\documentclass[12pt,twoside]{article}
\usepackage[dvips]{graphicx}
\usepackage{float}
\usepackage{latexsym}
\usepackage{amsmath,mathrsfs,bm,amssymb,color}%suzuki

\catcode`\@=11
\def\newsymbol#1#2#3#4#5{\let\next@\relax%
 \ifnum#2=\@ne\else%
 \ifnum#2=\tw@\let\next@\msyfam@\fi\fi%
 \mathchardef#1="#3\next@#4#5}
\def\mathhexbox@#1#2#3{\relax%
 \ifmmode\mathpalette{} {\m@th\mnnathchar"#1#2#3}
 \else\leavevmode\hbox{$\m@th\mathchar"#1#2#3$}\fi}
\font\tenmsy=msbm10
\font\sevenmsy=msbm7
\font\fivemsy=msbm5
\newfam\msyfam
\textfont\msyfam=\tenmsy
\scriptfont\msyfam=\sevenmsy
\scriptscriptfont\msyfam=\fivemsy
\edef\msyfam@{\hexnumber@\msyfam}
\def\Bbb#1{\fam\msyfam\relax#1}
\catcode`\@=\active

\newtheorem{theorem}{Theorem}[section]
\newtheorem{proposition}[theorem]{Proposition}
\newtheorem{lemma}[theorem]{Lemma}
\newtheorem{corollary}[theorem]{Corollary}
\newtheorem{definition}[theorem]{Definition}
\newtheorem{example}[theorem]{Example}
\newtheorem{remark}[theorem]{Remark}

\newtheorem{assumption}[theorem]{Assumption}
\newcommand{\sx}
{\sum_\s \!  \int \!  dx\,}
\newcommand{\meas}{{\cal X}}
\newcommand{\qs}{\mathscr{Q}}
\newcommand{\bd}[1]{\begin{definition}\label{#1}}
\newcommand{\ed}{\end{definition}}
\newcommand{\la}{\lambda}
\newcommand{\bl}[1]{\begin{lemma}\label{#1}}
\newcommand{\el}{\end{lemma}}
\newcommand{\bc}[1]{\begin{corollary}\label{#1}}
\newcommand{\ec}{\end{corollary}}
\newcommand{\bt}[1]{\begin{theorem}\label{#1}}
\newcommand{\et}{\end{theorem}}
\newcommand{\bp}[1]{\begin{proposition}\label{#1}}
\newcommand{\ep}{\end{proposition}}
\newcommand{\br}[1]{\begin{remark}\label{#1}}
\newcommand{\er}{\end{remark}}
\newcommand{\zz}{{\Bbb Z}_2}
\newcommand{\QQQ}{L^2(\QEE)}
\newcommand{\qqf}{L_{\rm fin}^2(\QSS)}
\newcommand{\ttt}{\tau}
\newcommand{\xtes}{X_t(\e,s)}
\newcommand{\zten}{Z_t^0(\e)}
\newcommand{\xte}{Z_t(\e)}
\newcommand{\XX}{X_t(\e)}

\newcommand{\XXx}{Z_t^x(\e)}
\newcommand{\zzzx}[1]{Z_t^x(#1)}
\newcommand{\zzzz}[1]{Z_t^0(#1)}

\newcommand{\Xn}[1]{Y_t^n({#1})}
\newcommand{\Xne}[1]{Y_t^n({#1},\e)}
\newcommand{\YY}[1]{Y_t({#1})}
\newcommand{\YYm}[1]{Y_t^{(m)}({#1})}

\newcommand{\YYE}[1]{Y_t({#1},\e)}
\newcommand{\iii}{I}
\newcommand{\XXX}[2]{X_{#1}(\e,#2)}
\newcommand{\xten}{X_t^n(\e)}
\newcommand{\xtem}{X_t^{(m)}(\e)}
\newcommand{\XP}{X_t^\perp(\e)}

\newcommand{\D}{\mathscr{H}_{\rm d}}
\newcommand{\OD}{\mathscr{H}_{\rm od}}
\newcommand{\QSS}{\qs}
\newcommand{\QEE}{\qs_{\rm E}}
\newcommand{\calb}{\Pi}

\newcommand{\QQ}{L^2(\QSS)}
\newcommand{\las}{j_s\la}
\newcommand{\ot}{\oplus_{\zz}}
\newcommand{\crr}{C_0^\infty(\BR)}
\newcommand{\ssss}{\sum_{\s\in\zz}}
\newcommand{\eq}[1]{\begin{equation}\label{#1}}
\newcommand{\en}{\end{equation}}
\newcommand{\eqn}{\begin{eqnarray*}}
\newcommand{\enn}{\end{eqnarray*}}
\newcommand{\eqnn}{\begin{eqnarray}}

\newcommand{\ennn}{\end{eqnarray}}
\newcommand{\proof}{{\noindent {\sc Proof}: \,}}
\newcommand{\qed}{\hfill {\bf qed}\par\medskip}
\newcommand{\BR}{{{\Bbb R}^3 }}
\newcommand{\bi}{\begin{description}}
\newcommand{\ei}{\end{description} }
\newcommand{\CC}{{{\Bbb C}}}
\newcommand{\RR}{{\Bbb R}}
\newcommand{\jjj}{\sum_{j=\pm1}}

\newcommand{\muu}{\sum_{\mu=1}^3}
\newcommand{\sh}{\beta}
\newcommand{\gh}{\psi_{\e}}

\newcommand{\slim}{{\rm s}\!\!-\!\!\lim}
\newcommand{\slimn}{\slim_{n\rightarrow\infty}}
\newcommand{\slime}{\slim_{\e\rightarrow0}}

\newcommand{\limn}{\lim_{n\rightarrow\infty}}
\newcommand{\limm}{\lim_{m\rightarrow\infty}}

\newcommand{\lime}{\lim_{\e\rightarrow0}}

\newcommand{\wick}[1]{{:\!\! #1 \!\!:}}
\newcommand{\kak}[1]{(\ref{#1})}

\newcommand{\tz}{Z}
\newcommand{\SSS}{\mathscr{S}}
\newcommand{\sss}{\mathscr{S}}

\newcommand{\DEE}{\D^{\rm E}}

\newcommand{\ODEE}{\OD^{\rm E}}

\renewcommand{\AA}{\mathscr{A}}
\newcommand{\BB}{\mathscr{B}}

\newcommand{\AAA}{\mathscr{A}^{\rm E}}
\newcommand{\BBB}{\mathscr{B}^{\rm E}}

\newcommand{\aaa}{{[a,b]}}
\newcommand{\ccc}{{[c,d]}}

\newcommand{\tot}{P^{\rm tot}}
\newcommand{\ab}[1]{\langle#1\rangle_\beta}
\newcommand{\abb}[1]{\langle#1\rangle_0}

\newcommand{\LR}{{L^2(\BR)}}
\newcommand{\hp}{H_{\rm p}}
\newcommand{\LRZ}{{\mathcal H}_{\rm b}}
\newcommand{\LRZZ}{L^2(\BR\!\times\!\zz)}

\newcommand{\LRS}{{L^2(\BR;\CC^2)}}
\newcommand{\om}{\omega_{{\rm b},m}}
\newcommand{\OOb}{\Omega}
\newcommand{\Ob}{1_{\QEE}}
\newcommand{\Obb}{1_{\QSS}}
\newcommand{\mue}{\mu_{\rm E}}
\newcommand{\wo}{\Omega}
\newcommand{\bn}{{{\omega}}}
\newcommand{\Nb}{N_{\rm b}}
\newcommand{\ob}{\omega_{\rm b}}
\newcommand{\LRT}{{L^2(\RR^{3 +1})}}
\newcommand{\e}{\varepsilon}
\newcommand{\ee}{\varepsilon/2}

\newcommand{\frm}{\s_{\rm F}}

\newcommand{\LRTr}{L_{\rm real}^2(\RR^{3+1})}
\newcommand{\EE}{{\Bbb E}^{x,\s}}
\newcommand{\E}{\EE}
\newcommand{\EEEE}{\eep }
\newcommand{\EEp}{{\Bbb E}_{\Omega}}
\newcommand{\EEEEE}{{\Bbb E}^{0,\s}}
\newcommand{\eep}{{\Bbb E}_{\rm P}}

\newcommand{\fff}{{\mathscr{F}}}
\newcommand{\fffs}{{\Sigma}}
\newcommand{\ffff}{\fff_{\rm fin}}
\newcommand{\is}{\inf\sigma}
\newcommand{\os}{S}
\newcommand{\ov}[1]{\overline{#1}}
\newcommand{\f}{^{-1}}
\newcommand{\lk}{\left(}
\newcommand{\rk}{\right)}
\newcommand{\lkk}{\left\{}
\newcommand{\rkk}{\right\}}

\renewcommand{\d}{\displaystyle}
\newcommand{\UU}{U}
\newcommand{\add}{a^{\dagger}}
\newcommand{\ass}{a^\sharp}
\newcommand{\MMM}[4]
{\left[ \!\!\!\begin{array}{cc}#1&#2\\
#3&#4\end{array}\!\!\!\right] }

\newcommand{\av}{{A}}
\newcommand{\bv}{{B}}
\newcommand{\PF}{H_{\rm PF}}

\newcommand{\PFs}{\hat H_{\rm PF}}
\newcommand{\PFp}{{H_{\rm PF}^\perp}}

\newcommand{\hf}{H_{\rm rad}}
\newcommand{\pf}{{P_{\rm f}}}

\newcommand{\hff}{H_{\rm rad}^{\mathscr{F}}}
\newcommand{\PFF}{H_{\rm PF}^{\mathscr{F}}}
\newcommand{\pff}{{P_{\rm f}^{\mathscr{F}}}}
\newcommand{\hhh}{{\mathcal H}^{\mathscr{F}}}
\newcommand{\pffm}{{P_{{\rm f}\mu}^{\mathscr{F}}}}

\newcommand{\half}{\frac{1}{2}}
\newcommand{\han}{{1/2}}
\newcommand{\npt}{N_p(t,U)}
\newcommand{\wpf}{ {H_{\rm PF}^0}}
\newcommand{\wpfe}{H_{\rm PF}^{0\,\e}}
\newcommand{\zpe}{\tz_t(\phi,\e)}

\newcommand{\ott}{\oplus^3}
\newcommand{\ZX}{Y}

\newcommand{\wwpf}{K_{\rm PF}}

\newcommand{\vvv}[1]
 {\left[
 \!\!\!\begin{array}{c}#1\end{array}\!\!\!\right]}
\newcommand{\mmm}[4]
{\left[ \!\!\!\begin{array}{cc}#1&#2\\
#3&#4\end{array}\!\!\!\right]}

\newcommand{\s}{\sigma}
\newcommand{\hhhh}{{\mathcal H}}
\newcommand{\hhhhh}{\hhhh_0}

\newcommand{\vp}{{\hat \varphi}}
\newcommand{\non}{\nonumber}

\makeatletter
\@addtoreset{equation}{section}
\makeatother
\def\theequation{\arabic{section}.\arabic{equation}}
\title
{Functional integral representations of the Pauli-Fierz model with
spin $\han$}
\author{
Fumio Hiroshima\thanks{e-mail: hiroshima@ math.kyushu-u.ac.jp} and
J\'ozsef L\H orinczi}

\begin{document}

\makeatletter \@addtoreset{equation}{section} \makeatother
\def\theequation{\arabic{section}.\arabic{equation}}
\title
{\sc Functional Integral Representation of the Pauli-Fierz Model
with Spin $\han$} \vspace{1.5cm}

\author{
\small Fumio Hiroshima\\[0.1cm]
{\small\it Department of Mathematics, University of Kyushu}    \\[-0.7ex]
{\small\it  6-10-1, Hakozaki, Fukuoka, 812-8581,  Japan}      \\[-0.7ex]
 {\small  {\tt hiroshima@math.kyushu-u.ac.jp}   }\\[0.5cm]
\small J\'ozsef L\H{o}rinczi\\[0.1cm]
{\it \small Zentrum Mathematik, Technische Universit\"at M\"unchen} \\[-0.7ex]
{\it \small Boltzmannstr. 3, 85747 Garching bei M\"unchen, Germany} \\[-0.7ex]
{\small {\tt  lorinczi@ma.tum.de}} \\[-0.4ex]
{\small and} \\
{\it \small School of Mathematics, Loughborough University} \\[-0.7ex]
{\it \small Loughborough LE11 3TU, United Kingdom} \\[-0.7ex]
{\small {\tt  J.Lorinczi@lboro.ac.uk}} \\[-0.7ex]}

\vspace{2cm}
\date{}
\pagestyle{myheadings} \markboth{The Pauli-Fierz model with spin}
{The Pauli-Fierz model with spin}
 \setlength{\baselineskip}{18pt}
\maketitle

\vspace{2cm}
\begin{abstract}
\noindent A Feynman-Kac-type formula for a L\'evy and an infinite
dimensional Gaussian random process associated with a quantized
radiation field is derived. In particular, a functional integral
representation of $e^{-t\PF}$ generated by the Pauli-Fierz
Hamiltonian with spin $\han$ in non-relativistic quantum
electrodynamics is constructed. When no external potential is
applied $\PF$ turns translation invariant and it is decomposed as a
direct integral $\PF = \int_\BR^\oplus \PF(P) dP$. The functional
integral representation of $e^{-t\PF(P)}$ is also given. Although
all these Hamiltonians include spin, nevertheless the kernels
obtained for the path measures are scalar rather than matrix
expressions. As an application of the functional integral
representations  energy comparison inequalities are derived.
\end{abstract}
\newpage

\section{Introduction}

Functional integration proved to be a useful approach in various
applications to quantum field theory. For the case of a quantum
particle linearly coupled to a scalar boson field, the so called
Nelson model, it gives a tool to proving existence or absence of a
ground state in Fock space \cite{sp98, LMS02a}. Furthermore, ground
state properties can be derived in terms of path measure
expectations \cite{bhlms}, and the question how the model
Hamiltonian and its ground state behave under lifting the so called
infrared and ultraviolet cutoffs can also be treated by the same method
\cite{LMS02b,GL07a,GL07b}. Another problem studied by this approach
is that of the effective mass \cite{bs,sp5}. Some of these results
have been obtained by functional integration only, thus sometimes it
offers a complementary method rather than a mere alternative.

In contrast with Nelson's model, the Pauli-Fierz model describes a
minimal coupling of a particle to the quantized radiation field. The
spectrum of the Pauli-Fierz Hamiltonian has been extensively studied
by a number of authors also using analytic methods. In particular,
the bottom of the spectrum of the Pauli-Fierz Hamiltonian is
contained in the absolutely continuous spectrum, no matter how small
the coupling constant is. Nevertheless, a ground state exists for
arbitrary values of the coupling constant without any infrared
cutoff \cite{bfs, gll, lilo}. Functional integration is also useful
in studying the spectrum of the Pauli-Fierz Hamiltonian which was
addressed in the spinless case so far \cite{bh, h10, h26,hilo}.

The spinless Pauli-Fierz Hamiltonian is written as
\begin{equation}
\PFs := \half (-i\nabla - e \AA)^2 + V + \hf \label{PF}
\end{equation}
on $\LR\otimes \QQ$, where the former is the particle state space
and the latter is the state space of the quantum field, $\AA$ stands
for the vector potential, $\hf$ for the photon field, and $V$ is an
external potential acting on the electron. These objects will be
explained in the following section in detail. The $C_0$-semigroup
$e^{-t\PFs}$ is defined through spectral calculus. A functional
integral representation of the semigroup $e^{-t\PFs}$ can be
constructed on the space $C([0,\infty);\BR) \times \QEE$, involving
a process consisting of $3$-dimensional Brownian motion $(B_t)_{t\geq0}$
for the particle, and an infinite dimensional Ornstein-Uhlenbeck process
on a function space $\QEE$ for the field \cite{ffg,ha, h4}. One immediate
corollary for the functional integral representation is the diamagnetic
inequality \cite{ahs,h4}
 \eq{in}
 \is(-(\han)\Delta +V+\hf)\leq\is(\PFs).
 \en
Using the fact that a path measure exists was also applied to
proving self-adjointness of $\PFs$ for arbitrary values of the
coupling constant $e$ \cite{h12,h19}. Furthermore, whenever $\PFs$
has a ground state, the path measure can be used to prove its
uniqueness \cite{h10} as an alternative to the methods making use of
ergodic properties of the semigroup in \cite{gr,gj1}. Other
applications for the study of the ground state include
\cite{bh,hilo}.

The path measure of the coupled Brownian motion and
Ornstein-Uhlenbeck process can be written in terms of a mixture of
two measures as the specific form of the coupling between particle
and field allows an explicit calculation of the Gaussian part. The
so obtained marginal over the particle is a Gibbs measure on
Brownian paths with densities dependent on the twice iterated It\^o
integral of a pair potential function describing the effective field
resulting from the Gaussian integration \cite{sp5,h10,bh,GL07a}.

Previous applications of rigorous functional integration to quantum
field theory covered, as far as we know, only cases when no spin was
present in the model. In this paper our main concern is to study by
means of a Feynman-Kac-type formula the Pauli-Fierz operator with
spin $\han$. (\ref{PF}) is in this case replaced by
\begin{equation}
 \PF := \half\lk  \vec\sigma \cdot (-i\nabla - e \AA)\rk ^2 +
 V + \hf,
\label{SPF}
\end{equation}
where $\vec \sigma = (\sigma_1,\sigma_2,\sigma_3)$ are the Pauli matrices
standing for the spin (see details in the next section). The random
process  of the particle modifies to a $3+1$ dimensional joint
Wiener and jump process $(\xi_t)_{t\geq 0} = (B_t,\s_t)_{t\geq 0}$,
where the effect of the spin appears in the process $\s_t =
\s(-1)^{N_t}$ hopping between the two possible values of the spin
variable $\s$, driven by a Poisson process $(N_t)_{t\geq 0}$. Our
approach owes a debt to the ideas in \cite{als}, where a path
integral representation of a $C_0$-semigroup generated by Pauli
operators in quantum mechanics was obtained by making use of an
$\BR\times \zz$-valued process, with $\zz$ the additive group of
order two. As we will see in the next subsection, the Pauli operator
is of a similar form as $\PF$, in fact both operators describe
minimal interactions. While in \cite{als} only a path integral
representation of operators with non-vanishing off-diagonal elements
was constructed, we improve on this here since this part of the spin
interaction in general may have zeroes.

Another model considered in the present paper is the so called
translation invariant Pauli-Fierz Hamiltonian which is the case of
$\PF$ above with zero external potential $V$. Translation invariance
yields a fiber decomposition $\PF = \int_\BR^\oplus \PF(P) dP$
with respect to total momentum $\tot$, where the fiber Hamiltonian
is given by
 \eq{jo}
  \PF(P) := \half \lk \vec \s\cdot (P-\pf -e\AA(0))\rk^2 + \hf,
  \quad P\in\BR.
 \en
Here $\pf$ denotes the momentum operator of the field. While the
translation invariant Hamiltonian does not have any point spectrum,
$\PF(P)$ under some conditions does \cite{fr,th}. In \cite{h26} the
functional integral representation of $e^{-t\hat \PF(P)}$ for the
spinless fiber Hamiltonian is constructed, where
 \eq{jojo}
  \hat H _{\rm PF}(P) := \half \lk P-\pf -e\AA(0)\rk^2 + \hf,
  \quad P\in\BR.
 \en
Furthermore, uniqueness of the ground state of $\hat H_{\rm PF}(0)$ as
well as the energy comparison inequality
 \eq{in2}
 \is(\hat H_{\rm PF}(0)) \leq \is(\hat H_{\rm PF}(P))
 \en
are shown.

Our main purpose in this paper is to extend the results on the
spinless Hamiltonians mentioned above to those with spin, i.e.,
\begin{enumerate}
\item[(1)]
construct a functional integral representation of $e^{-t\PF}$ and
$e^{-t\PF(P)}$ with {\it a scalar} kernel;
\item[(2)]
derive some energy comparison inequalities for $\PF$ and $\PF(P)$.
\end{enumerate}
We stress that $\PF$ and $\PF(P)$ include spin $\han$, nevertheless
the kernels of their functional integrals obtained here are scalar.
(1) is achieved in Theorems \ref{main} and \ref{main2}, and (2) in
Corollaries \ref{rotation} and \ref{main3} below.

Here is an outline of the key steps of proving (1) and (2). First we
assume that the form factor $\vp$ is a sufficiently smooth function
of compact support. Then we will see that there exists a Pauli
operator $\wpf(\phi)$, $\phi\in\QSS$, on $\LRZZ$,
 which can be used
to define
 \eq{tror}
 \wpf := \int_{\QSS}^\oplus \wpf(\phi)  d\mu(\phi).
 \en
As it will turn out, for {\it arbitrary} values of the coupling
constant $e$,
 \eq{tro}
 \PF = \wpf\,\,  \dot + \,\, \hf
 \en
holds as an equality of self-adjoint operators ($\dot + $ denotes
quadratic form sum). Although for weak couplings this results by the
Kato-Rellich Theorem, it is non-trivial for arbitrary values of $e$.
Thus it will suffice to construct a functional integral
representation of the right hand side of \kak{tro}. However, as was
mentioned before, the off-diagonal part of $\wpf(\phi)$ may have in
general zeroes or a compact support. In order to prevent the
off-diagonal part vanish we change $\wpf(\phi)$ for $\wpfe(\phi) $
by adding a term controlled by a small parameter $\e>0$. Then we
work with
 \eq{will2}
 \PF^\e:= \wpfe\,\,  \dot
 + \,\, \hf
 \en
and obtain the original Hamiltonian by $\lime e^{-t \PF^ \e} =
e^{-t\PF}$, where in fact
$$\wpfe :=\int_\qs ^\oplus \wpfe(\phi)
d\mu(\phi).$$
In particular, instead of for the semigroup $e^{-t
\PF}$, we construct the functional integral representation of
$e^{-t\PF^\e}$. By the Trotter-Kato product formula we write
 \eq{tlk}
 e^{-t\PF^\e} = \slim_{n\rightarrow\infty} (e^{-(t/n)\wpfe}
e^{-(t/n)\hf})^n \en and derive the functional integral of the
Pauli-operator $e^{-t \wpfe(\phi)}$ by using that the form factor
$\vp$ is chosen to be bounded and sufficiently smooth, with non-zero
off-diagonals. By making use of a hypercontractivity argument for
second quantization and the Markov property of projections, we are
able to construct the functional integral representation of
$e^{-t\PF^\e}$. An approximation argument on $\vp$ leads us then to
our main Theorem~\ref{main} for reasonable form factors.

The functional integral representation of $e^{-t\PF(P)}$ is further
obtained by a combination of that of $e^{-t\PF}$ and \cite{h26}.
Since the functional integral kernels are scalar, we can estimate
$|(F, e^{-t\PF}G)|$ and $|(F, e^{-t\PF(P)}G)|$ directly, and derive
some energy comparison inequalities.

%%%%%%%%%%%%%%%%%%%%%%%%%%%%%%%%%%%%%%%%%%%%%%%%%%%%%%%%%%%%%%%%%

%%%%%%%%%%%%%%%%%%%%%%%%%%%%%%%%%%%%%%%%%%%%%%%%%%%%%%%%%%%%%%%%%%%%
Our paper is organized as follows. In Section 2 we discuss the Fock
space respectively Euclidean representations of the Pauli-Fierz
Hamiltonian with spin $1/2$ in detail. Section 3 is devoted to
discussing L\'evy  processes and functional integral representations
of Pauli operators. In Section 4 by using results of the previous
section and hypercontractivity properties of second quantization we
construct the functional integral representation of $e^{-t\PF}$ and
derive comparison inequalities for ground state energies. In Section
5 we derive the functional integral of $e^{-t\PF(P)}$ and obtain
energy inequalities for this case. In Section 6 we comment on the
multiplicity of ground states of a model with spin. Section 7 is an
appendix containing details on Poisson point processes and a related
It\^o formula adapted to our context.

\section{Function space representation of the Pauli-Fierz model with spin}
% and statements of the results}
\subsection{Pauli-Fierz model with spin $1/2$ in Fock space}
We begin by defining the Pauli-Fierz Hamiltonian as a self-adjoint
operator.

\medskip \noindent \emph{Fock space} \quad
Let $\LRZ:=L^2(\BR\!\times\!\{-1,1\})$ be the Hilbert space of a
single photon, where $\BR\times\{-1,1\}\ni(k,j)$ are its momentum
and polarization, respectively. Denote $n$-fold symmetric tensor
product by $\bigotimes_{\rm sym}^n$, with $\bigotimes_{\rm
sym}^0\LRZ:=\CC$. The Fock space describing the full photon field is
defined then as the Hilbert space
\eq{fhi}
\fff:=\bigoplus_{n=0}^\infty
\left[
\bigotimes_{\rm sym}^n \LRZ\right]
\en
with scalar product \eq{sca}
(\Psi,\Phi)_{\fff}: = \sum_{n=0}^\infty(\Psi^{(n)},\Phi^{(n)})_
{\otimes^n_{\rm sym}\LRZ}, \en and $\Psi = \bigoplus_{n=0}^\infty
\Psi^{(n)}$, $\Phi = \bigoplus_{n=0}^\infty \Phi^{(n)}$.
Alternatively, $\fff$ can be identified as the set of
$\ell_2$-sequences $\{\Psi^{(n)}\}_{n=0}^\infty$ with $\Psi^{(n)}\in
\bigotimes_{\rm sym}^n \LRZ$. The vector $\OOb=\{1,0,0,...\}\in\fff$
is called Fock vacuum. The finite particle subspace $\ffff$ is
defined by
$$
\ffff:=\left\{\{\Psi^{(n)}\}_{n=0}^\infty\in\fff \,|\,\exists M \in
{\Bbb{N}}: \, \Psi^{^{(m)}} = 0, \; \forall m \geq M\right\}.
$$

\bigskip \noindent \emph{Field operators} \quad
With each $f\in\LRZ$ a photon creation and annihilation operator is
associated. The creation operator $\add(f):\fff\rightarrow \fff$ is
defined by
$$
(\add(f)\Psi)^{(n)}=\sqrt n
S_n(f\otimes \Psi^{(n-1)}),\ \ \ n\geq 1,
$$
where $S_n(f_1\otimes \cdots \otimes f_n)= (1/n!) \sum_{\pi \in
\Pi_n}f_{\pi(1)} \otimes \cdots \otimes f_{\pi(n)}$ is the
symmetrizer with respect to the permutation group $\Pi_n$ of degree
$n$. The domain of $\add(f)$ is maximally defined by
$$
D(\add(f)):= \left\{\{\Psi^{(n)}\}_{n=0}^\infty \,
 \left| \,
\sum_{n=1}^\infty n\| S_n(f\otimes \Psi^{(n-1)})\|^2<\infty\right.
\right\}.
$$
 The annihilation operator $a(f)$ is
introduced as the adjoint $a(f) = (\add(\bar f))^\ast$ of $\add(\bar
f)$ with respect to scalar product \kak{sca}.
 %Note that $\add(f)$ and $a(f)$ are linear in $f\in\LRZ$.
$\add(f)$ and $a(f)$ are closable operators, their closed extensions
will be denoted by the same symbols. Also, they leave $\ffff$
 invariant and obey the canonical commutation relations on
$\ffff$:
$$
[a(f), \add(g)]=(\bar f, g)1,\ \ \ [a(f),a(g)]=0,\ \ \ [\add(f),\add(g)]=0.
$$

\bigskip \noindent \emph{Second quantization and free  field Hamiltonian} \quad
Although the free field Hamiltonian
$$
\hff = \jjj \int |k| \add(k,j) a(k,j)dk
$$
is usually given in terms of formal kernels of creation and
annihilation operators, we define it as the infinitesimal generator
of a one-parameter unitary group since this definition has
advantages in studying functional integral representations. We use
the label ${\rm F}$ for objects defined in Fock space.
 This unitary
group is constructed through a functor $\Gamma$. Let ${\mathscr{C}} (X\to
Y)$ denote the set of contraction operators from $X$ to $Y$. Then
$\Gamma: \mathscr{C}(\LRZ\rightarrow \LRZ) \rightarrow \mathscr{C}
(\fff\rightarrow \fff)$ is defined as
$$
\Gamma(T):=\bigoplus_{n=0}^\infty [\otimes^n T]
$$
for $T\in \mathscr{C}(\LRZ\rightarrow \LRZ)$, where the tensor product for
$n=0$ is the identity operator. For a self-adjoint operator $h$ on $\LRZ$,
$\Gamma(e^{ith})$, $t\in\RR$, is a strongly continuous one-parameter
unitary group on $\fff$. Then by Stone's Theorem there exists a
unique self-adjoint operator $d\Gamma(h)$ on $\fff$ such that $
\Gamma(e^{ith}) = e^{itd\Gamma(h)}$, $t\in\RR. $ $d\Gamma(h)$ is
called the second quantization of $h$. The second quantization of
the identity operator,  $N:=d\Gamma(1)$ gives the photon number
operator. Let $\ob $ be the multiplication operator $f\mapsto \ob(k)
f(k,j)=|k|f(k,j)$, $k\in\BR$, $j=\pm 1$ on $\LRZ$. The operator
$\hff:=d\Gamma(\ob )$ is then the free field Hamiltonian.

\bigskip \noindent \emph{Polarization vectors} \quad
Two vectors $e(k,+1)$ and $e(k,-1)$, $k\not=0$, are polarization
vectors whenever $e(k,-1),e(k,+1),k/|k|$ form a right-handed system
in $\BR$ with
 (1) $e(k,-1)\times e(k,+1)=k/|k|$,
 (2) $e(k,j)\cdot e(k,j')=\delta_{jj'}$,
 (3)  $e(k,j)\cdot k/|k|=0$.
We have
 \eqn
 &&
 \jjj e_\mu (k,j) e_\nu(k,j) =
 \delta_{\mu\nu}-\frac{k_\mu k_\nu}{|k|^2},
 \enn
independently of the specific choice of these vectors. One can
choose the polarization vectors at convenience since the
Hamiltonians $\PFF$ defined below are unitary equivalent up to this
choice \cite{sa}.

\bigskip \noindent \emph{Quantized radiation field} \quad
Note that $\ass(f)$ is linear in $f$, where $a^\sharp=a,\add$, thus
formally $a^\sharp(f)=\jjj\int f(k,j)a^\sharp(k,j) dk$. The
quantized radiation field with ultraviolet cutoff function (form
factor) $\vp$ is defined through the vector potentials
$$
A_{\mu}(x):=\frac{1}{\sqrt2}\jjj \int e_\mu(k,j) \left
(\frac{\vp(k)}{\sqrt{\ob (k)}}\add(k,j)e^{-ik\cdot x} +
\frac{\vp(-k)}{\sqrt{\ob (k)}}a(k,j)e^{ik\cdot x} \right)dk.
$$
Here $\vp$ is Fourier transform of $\varphi$. A standing assumption
in this paper is
\begin{assumption}
 \label{as}
We take $\ov{\vp(k)}=\vp(-k)=\vp(k)$ and $\sqrt{\ob}\vp, \, \vp/\ob
\in\LR$.
\end{assumption}
Under Assumption \ref{as} $A_\mu(x)$ is a well-defined symmetric
operator in $\fff$. By $k\cdot e(k,j)=0$, the Coulomb gauge
condition
$$
\sum_{\mu=1}^ 3 [\partial_{x_\mu}, A_\mu(x)]=0,
$$
holds on $\ffff$. By the fact that $\sum_{n=0}^\infty \|A_{\mu}(x)^n
\Phi\|/n!<\infty$ for $\Phi\in\ffff$, and Nelson's analytic vector
theorem \cite[Th.X.39]{rs2} it follows that
$A_{\mu}(x)\lceil_{\ffff}$ is essentially self-adjoint. We denote
its closure $\ov{A_{\mu}(x)\lceil_{\ffff}}$ by the same symbol
$A_{\mu}(x)$.

\bigskip \noindent \emph{Electron state space and Schr\"odinger
Hamiltonian} \quad The Hilbert space describing the electron is
$\LRS$. Let $\s_1,\s_2,\s_3$ be the $2\times 2$ Pauli matrices
$$
\s_1:=\mmm 0 1 1 0, \quad \s_2:=\mmm 0 {-i} i 0 , \quad \s_3:=\mmm 1
0 0 {-1}.
$$
We have $\s_\alpha\s_\beta = \delta_{\alpha\beta} + i
\sum_{\gamma=1}^3 \epsilon^{\alpha \beta \gamma }\s_\gamma $, where
$\epsilon^{\alpha \beta \gamma }$ is the totally antisymmetric
tensor with $\epsilon^{123}=1$. Then the electron Hamiltonian on
$\LRS$ with external potential $V$ is given by
\eq{num}
\hp = \half \sum_{\mu=1}^3\left(\s_\mu(-i\nabla_\mu)\right)^2 + V.
\en
Here $V$ acts as a multiplication operator and in some statements below
it will be required to satisfy one or both of the following
conditions:
\begin{assumption}
Let $V$ be
 \label{as2}
 \begin{itemize}
 \item[(1)]
 relatively bounded with respect to $(-1/2)\Delta$ with a bound
 strictly less than 1;
 \item[(2)]
 $\sup_{x\in\BR}{\Bbb E}^x \left[e^{-2\int_0^t V(B_s) ds}\right]
 < \infty$, for all  $t \in (0,\infty)$.
 \end{itemize}
\end{assumption}
(1) above is a usual ingredient for self-adjointness of
Schr\"odinger operators. In (2) the expectation ${\Bbb E}^x$
is meant under Wiener measure for $3$-dimensional Brownian motion
$(B_s)_{s\geq 0}$ starting at $x$. It is in particular satisfied by
Kato-class potentials which includes Coulomb potential.

\bigskip \noindent \emph{Pauli-Fierz Hamiltonian} \quad
The state space of the joint electron-field system is
 \eq{willsee}
 \hhh=\LRS\otimes\fff. \en
The non-interacting system is described by the total free
Hamiltonian $\hp \otimes 1+1\otimes \hff$. To define the quantized
radiation field $A$ we identify $\hhh$ with the set of
$\CC^2\otimes \fff$-valued $L^2$ functions on $\BR$, i.e.,
$\hhh\cong \int_\BR^\oplus (\CC^2 \otimes \fff) dx.$ Then we have by
definition $A_{\mu}= \int_\BR^\oplus (1\otimes A_{\mu}(x)) dx$.
Hence $(A_{\mu}F)(x) = A_{\mu}(x)F(x)$ for $F(x)\in D(A_{\mu}(x))$
and $A_{\mu}$ is self-adjoint. Taking into account the minimal
interaction $-i\nabla_\mu \mapsto -i \nabla_\mu -e \av_\mu$, we
obtain the Pauli-Fierz Hamiltonian
\begin{equation}
\PFF := \half\lk \sum_{\mu=1}^3 \s_\mu (-i \nabla_\mu \otimes 1 -
eA_{\mu})\rk ^2 + V \otimes 1 + 1 \otimes \hff
\end{equation}
with coupling constant $e\in\RR$, i.e.,
\begin{equation}
\PFF = \half(-i\nabla-e A)^2 + V + \hff - \frac{e}{2}\muu \s_\mu
B_\mu,
\end{equation}
where we omit the tensor product for convenience and write
$$
B_\mu(x)= - \frac{i}{\sqrt2}\jjj \int (k \times e(k,j))_\mu
\frac{\vp(k)}{\sqrt{\ob(k)}}\left(\add(k,j)e^{-ik\cdot x} -
a(k,j)e^{ik\cdot x}\right) dk.
$$
In fact, $B_\mu(x) = (\nabla \times A(x))_\mu$, however, we regard
$\av$ and $\bv$ as independent operators in this paper.

\bigskip

A first natural question is whether $\PFF$ is a self-adjoint
operator.

\bp{22}
 Under Assumption \ref{as} $\PFF$ is self-adjoint on $D(-\Delta )
 \cap D(\hff)$ and bounded from below. Moreover, it is essentially
 self-adjoint on any core of $\hp +\hff$.
 \ep
 \proof
 See \cite{h12,h19}.
\qed

A special case considered in this paper is the translation invariant
Pauli-Fierz Hamiltonian obtained under $V=0$. Then
$$
e^{it P^{\rm tot}_\mu} \PFF  e^{-it P^{\rm tot}_\mu} = \PFF, \quad
t\in\RR, \; \mu=1,2,3,
$$
where $P^{\rm tot}$ denotes the total
electron-field momentum
$$
 P^{\rm tot}_\mu : = -i\nabla_\mu \otimes 1 +
 1\otimes \pffm
$$
and $\pffm = d\Gamma(k_\mu)$ is the momentum of the field. By
translation invariance the Hilbert space $\hhh$ and the Hamiltonian
$\PFF$ can both be decomposed with respect to the spectrum of $\tot$
as $\int_\BR^\oplus \hhh(P) dP$ and $\PFF := \int_\BR^\oplus K(P)
dP$, with a self-adjoint operator $K(P)$ labeled by $P$ on
$\hhh(P)$. It is seen that $K(P)$ and $\hhh(P)$ are isomorphic with
a self-adjoint operator resp. a Hilbert space. Define thus on
$\CC^2\otimes\fff$ the Pauli-Fierz operator at total momentum $P \in
\BR$ by
\begin{equation}
\PFF(P) := \half(P-\pff-eA(0))^2+ \hff-\frac{e}{2}\muu
\s_\mu\bv_\mu(0).
\end{equation}
Then we have
\bp{lms}

Under Assumption \ref{as} $\PFF(P)$, $P\in\BR$, is self-adjoint on
the domain $D(\hff) \bigcap_{\mu=1}^3 D((\pffm)^2)$, and essentially
self-adjoint on any core of the self-adjoint operator $\half
\sum_{\mu=1}^3 (\pffm)^2+\hff$. Moreover, $\hhh\cong\int_\BR^\oplus
\CC^2\otimes \fff dP$ and $\PFF \cong \int_\BR^\oplus \PFF(P) dP$
hold.
 \ep
 \proof
See \cite{h27,lms}.
 \qed

\noindent

Here is an incomplete list of results on the spectral properties of
the Pauli-Fierz Hamiltonian. The existence of the ground state of
$\PF$ is established in \cite{bfs,gll,lilo} and that of $\PF(P)$ in
\cite{fr,th,hh}. The multiplicity of the ground state is estimated
in \cite{h10,hisp2,bfp,h27}, a spectral scattering theory and
relaxation to ground states are studied in \cite{ar4, sp97,fgs1}.
The perturbation of embedded eigenvalues is reduced to investigating
resonances \cite{bfs2,bfs1}. Energy estimates are obtained in
\cite{fe, ffg,lilo2} and the effective mass is studied in
\cite{sp5, caha, hisp3, th2,bcfs, hk}. Related works on particle
systems interacting with quantum fields include
\cite{ge,bdg, agg, lms,sa}.

\subsection{Stochastic representation and spin variables in function space}
\subsubsection{Stochastic representation}
In this section we prepare the necessary items for a $Q$-representation of
$\PFF$ and explain how to accommodate spin in this framework.

To introduce a $Q$-representation, we define a bilinear form
and construct a Gaussian random process with mean zero and
covariance given in terms of this form. Define the field operator
$A_\mu(\hat f)$ by
$$
A_\mu(\hat f):=\frac{1}{\sqrt2}\jjj\int e_\mu(k,j) \left(\hat f(k)
\add(k,j)+\hat f(-k) a(k,j)\right) dk
$$
and the $3\times 3$ matrix $D(k)$, $k\not=0$, by
$$
D(k) := \left(\delta_{\mu\nu}-\frac{k_\mu k_\nu}{|k|^2}\right)_
{1\leq \mu,\nu\leq 3}.
$$
Consider the bilinear form $q_0: \oplus^3 \LR\times \oplus^3\LR
\rightarrow \CC$ given by the scalar product
$$
q_0(f,g):=\sum_{\mu,\nu=1}^3
 (A_\mu(f)\OOb, A_\nu(g)\OOb)_\fff=\half
\int_\BR \ov{\hat f(k)}\cdot D(k) \hat g(k) dk.
$$
Similarly to the representation of a Euclidean free field in terms
of path integrals over the free Minkowski field in constructive
quantum field theory \cite[Th.III.6]{si}, we introduce another
bilinear form $q_1$ to define an additional Gaussian random process.
Let $q_1:\oplus^3 \LRT\times\oplus^3 \LRT\rightarrow \CC$ be
$$
q_1(F,G) := \half \int_{\RR^{3+1}} \ov{\hat F(k,k_0)}\cdot D(k) \hat
G(k,k_0) dkdk_0.
$$
Note that $D(k)$ is independent of $k_0$ in the definition of $q_1$.
%In a moment we discuss $q_0$ and $q_1$ simultaneously. So from now on
Use the label ${\sh}$ for $0$ or~$1$, let $\SSS(\RR^{3+{\sh}})$ be
the set of real-valued Schwartz test functions on $\RR^{3+\sh}$ and
put ${\SSS}_\sh :=\ott \SSS(\RR^{3+{\sh}}) $.
The properties (1) $\sum_{i,j=1}^n \bar z_i z_j
\exp(-{q_{\sh}(f_i-f_j,f_i-f_j)}) \geq 0$ for arbitrary $z_i\in\CC$
and  $i=1,...,n$, $\forall n = 1,2,...$; (2) $\exp(-{q_{\sh}(g,g)})$
is strongly continuous in $g\in\oplus^3L^2(\RR^{3+\sh })$; (3)
$\exp(-{q_{\sh}(0,0)})=1$ can be checked directly.

Let $\qs_{\sh}:={\SSS}_\sh'$, where ${\SSS}_\sh'$ is the dual space
of ${\SSS}_\sh$, and denote the pairing between elements of
$\qs_{\sh}$ and $ {\SSS}_\sh$ by $\ab{\phi,f}\in\RR$.
By the three properties listed above and the Bochner-Minlos Theorem
there exists a probability space $(\qs_{\sh}, \BB_{\qs_{\sh}},
\mu_{\sh})$ such that $\BB_{\qs_{{\sh}}}$ is the smallest
$\s$-field generated by $\{\ab{\phi, f}, f\in{\SSS}_\sh \}$ and
$\ab{\phi,f}$ is a Gaussian random variable with mean zero and
covariance given by \eq{gauss} \int_{\qs_{{\sh}}} e^{i\ab{\phi, f}}
d\mu_{{\sh}}(\phi) = e^{-q_\sh (f,f)},\ \ \ f\in{\SSS}_\sh .
 \en
Although $\ab{\phi,\oplus_\mu^3 \delta_{\mu\nu}f}$ is a
$Q$-representation of the quantized radiation field with the
ultraviolet cutoff function  $f\in\SSS(\BR)$, we have to extend
$f\in {\SSS}_\sh$ to a more general class since our cutoff is
$(\vp/\sqrt\omega)^\vee\in\LR$. This can be done in the following
way. For any $f=f_{\rm Re}+i f_{\rm Im}\in \ott \sss(\RR^{3+{\sh}})$
we set $\ab{\phi,f} := \ab{\phi, f_{\rm Re}} + i\ab{\phi, f_{\rm
Im}}$. Since $\SSS (\RR^{3+{\sh}})$ is dense in
$L^2(\RR^{3+{\sh}})$
 and the inequality
$$
\int_{\qs_{\sh}}
 |\ab{\phi, f}|^2 d\mu_{\sh} (\phi)
  \leq \|f\|_{\ott
 L^2(\RR^{3+{\sh}})}^2
$$
holds by \kak{gauss}, we can define $\ab{\phi, f}$ for $f\in \ott
L^2(\RR^{3+{\sh}})$ by $\ab{\phi, f}=\slimn \ab{\phi, f_n}$ in
$L^2(\qs_\sh)$, where $\{f_n\}_{n=1}^\infty \subset \ott
\SSS(\RR^{3+{\sh}})$ is any sequence such that $\slimn f_n=f$ in
$\ott L^2(\RR^{3+{\sh}})$. Thus we define the multiplication
operator
\eq{defmulti}
\lk \AA^{\sh}(f) F\rk (\phi) := \ab{\phi, f} F(\phi),\quad
\phi\in \qs_{\sh},
\en
labeled by $f\in  \ott L^2(\RR^{3+{\sh}})$ in $L^2(\qs_\sh)$, with
domain
$$
D(\AA^{\sh}(f)) := \left\{F\in L^2(\qs_{\sh})
\left|
\int_{\qs_{\sh}}
|\ab{\phi, f} F(\phi)|^2d\mu_{\sh}(\phi)<\infty\right.
\right\}.
$$
Denote the identity function in $L^2(\qs_{\sh})$ by $1_{\qs_{\sh}}$
and the function $\AA^{\sh}(f) 1_{\qs_{\sh}}$ by $\AA^{\sh}(f)$ unless
confusion may arise. It is known that $L^2(\qs_{\sh}) = \bigoplus_{n=0}^
\infty L_n^2(\qs_{\sh})$, with
$$
L_n^2(\qs_{\sh})= \ov{{\rm L.H.} \{\wick{\AA^{\sh}(f_1)\cdots
\AA^{\sh}(f_n)}| f_j\in \oplus^3 L^2(\RR^{3+{\sh}}), \,
j=1,2,...,n\}}.
$$
Here $L_0^2(\qs_{\sh}) = \{\alpha 1_{\qs_{\sh}} | \alpha\in\CC\}$
and $\wick{X}$ denotes Wick product recursively defined by
 \eqn &&
 \wick{\AA^{\sh}(f)} = {\AA^{\sh}(f)},\\
 &&
 \wick{\AA^{\sh}(f) \AA^{\sh} (f_1) \cdots \AA^{\sh}
 (f_n)} = \AA^{\sh}(f)\wick{ \AA^{\sh} (f_1)
 \cdots\AA^{\sh}(f_n)}\\
 &&
 \hspace{5cm}
 -\sum_{j=1}^n q_{\sh}(f,f_j) \wick{\AA^{\sh}(f_1) \cdots
 \widehat{\AA^{\sh}(f_j)}\cdots \AA^{\sh}(f_n)},
 \enn
where $\widehat X$ denotes removing $X$.

Next we define the second quantization $\Gamma_{\sh\sh'}$ in
$Q$-representation as the functor
$$
\Gamma_{\sh\sh'}: \mathscr{C}\left(L^2(\RR^{3+\sh})\rightarrow
L^2(\RR^{3+\sh'})\right) \;\;\rightarrow\;\; \mathscr{C}\left
(L^2(\qs_\sh) \rightarrow L^2(\qs_{\sh'})\right).
$$
With $T\in \mathscr{C}(L^2(\RR^{3+\sh})\rightarrow L^2(\RR^{3+\sh'}))$,
$\Gamma_{{\sh}{\sh}'}(T)\in \mathscr{C}( L^2(\qs_\sh)\rightarrow
L^2(\qs_{\sh'}))$ is defined by
$$
\Gamma_{{\sh}{\sh}'}(T)1_{\qs_{\sh}} =1_{\qs_{{\sh}'}}, \quad
\Gamma_{\sh}(T) \,\wick{\AA^{\sh}(f_1)\cdots \AA^{\sh}
(f_n)}\,=\,\wick{\AA^{{\sh}'}(T f_1) \cdots \AA^{{\sh}'}
(Tf_n)}.
$$
For notational simplicity we use $\Gamma_{\sh}$ for
$\Gamma_{{\sh}{\sh}}$. For each self-adjoint operator $h$ in
$L^2(\RR^{3+\sh})$, $\Gamma_{\sh}(e^{ith})$ is a one-parameter
unitary group. Then $\Gamma_{\sh}(e^{ith}) =
e^{itd\Gamma_{\sh}(h)}$, $t\in\RR$, for the unique self-adjoint
operator $d\Gamma_\sh (h)$ in $L^2(\qs_\sh)$. We write
  \eq{hi}
  \QSS:=\qs_0,\ \ \  \QEE:=\qs_1,\ \ \
  \mu:=\mu_0,\ \ \ \mu_{\rm E}:=\mu_1,\ \ \ \AA:=\AA^0,\ \ \ \
  \AAA:=\AA^1
  \en
in what follows, using the label ${\rm E}$ for
 ``Euclidean" objects to distinguish from Fock space objects.
 Thus it is seen that $\fff$, $A_\mu(\hat f)$ and $d\Gamma(h)$ are
isomorphic to $L^2(\QSS)$, $\AA (\oplus_{\nu=1}^3
\delta_{\mu\nu}f)$ and $d\Gamma_0 (\hat h )$, respectively, where
$\hat h ={\rm F} h {\rm F}^{-1}$ and ${\rm F}$ denotes Fourier
transform on $\LR$. That is, there exists a unitary operator ${\Bbb
U}: \fff\to \QQ$ such that
\begin{itemize}
\item[(1)]
${\Bbb U}\OOb = \Obb$,
\item[(2)]
${\Bbb U} A_\mu(\hat f) {\Bbb U}^{-1} = \AA (\oplus_{\nu=1}^3
\delta_{\mu\nu} f)$,
\item[(3)]
${\Bbb U} d\Gamma(h) {\Bbb U}\f = d\Gamma_0(\hat h )$.
\end{itemize}
The isomorphism ${\cal U}:=1\otimes{\Bbb U}: \hhh\rightarrow
\LRS\otimes\QQ$ maps  $\PFF$ to a self-adjoint operator on
$\LRS\otimes\QQ$. Let
 \eq{Lambda}
 \la:=(\vp/\sqrt\ob)^\vee,
 \en
where {\it \v f} denotes inverse Fourier transform of $f$. Set
$\AA_\mu (\la(\cdot-x)) := \AA(\oplus_{\nu=1}^3 \delta_{\mu\nu}
\la(\cdot-x))$ and $ \hf := d\Gamma_0(\hat \ob)$ on $\QQ$.

Finally we define $\PF$, the main object in this paper,  by
 \eq {PFP}
 \PF:=
  \half(-i\nabla -e\AA)^2+V+\hf
       -\frac{e}{2}\muu \s_\mu \BB_\mu,
 \en
where $\AA_\mu  := \int_\BR^\oplus \AA _\mu(\la(\cdot-x))dx$ and
$\BB_\mu := \int_\BR^\oplus \BB_\mu (\la(\cdot-x)) dx$, with
$$
 \BB_\mu(\la(\cdot-x)) = \AA(\oplus_{\nu=1}^3
 \delta_{\nu\mu}(\nabla_x\times\la(\cdot-x))_\mu).
$$
Here the self-adjoint operator $\PF$ is the $Q$-representation of
$\PFF$, obtained through the map ${\cal U}\PFF{\cal U}\f = \PF$. In
this representation $A_\mu$ and $B_\nu$ turn into the multiplication
operators $\AA _\mu$ and $\BB_\nu$, respectively.

\subsubsection{Spin variables in function space}
In order to reduce $\kak{PFP}$ to a {\it scalar} operator, we
introduce a two-valued variable $\s$. Let $ \zz := {\Bbb Z}/2{\Bbb
Z} $ and $[z]_2$ denote the equivalence class of $z\in{\Bbb Z}$.
Use the affine map $x \mapsto 2x-1$ to arrive at the conventional
variables $\{-1,+1\} \cong \zz$. Addition modulo 2 gives
$(+1)\ot(+1)=+1$, $(+1)\ot (-1)=-1$, $(-1)\ot(-1)=+1$. Define
$$
\LRZZ := \left\{f:\BR\times \zz \rightarrow \CC \,\left| \,
\|f\|_{\LRZZ}^2 := \sum_{\s\in\zz} \|f(\cdot,\s)\|_\LR^2 < \infty
\right.\right\}.
$$
The isomorphism between $\LRS$ and $ L^2(\BR\times \zz)$ is given by
$$
\LRS \ni \vvv{u(x, +1)\\ u(x, -1)}\mapsto u(x, \s) \in L^2(\BR\times
\zz).
$$
Let $F=\vvv {F({+1})\\  F({-1})}\in \hhh$ with $F(\pm 1) \in
 \LR\otimes \QQ$. Then since
$$
\PF =  \half(-i\nabla -e\AA)^2+V+\hf
       -\frac{e}{2}\mmm {\BB _3} {\BB_1-i \BB_2} {\BB_1+i \BB _2}
       {- \BB_3},
$$
our Hamiltonian can be regarded as the self-adjoint operator on
 \eq{hil}
 \hhhh := \LRZZ \otimes \QQ
 \en
defined by
 \eq {5}
 (\PF  F)(\s) = \lk \half(-i\nabla -e\AA)^2+V+ \hf +
  \D(\s) \rk F(\s)+ \OD(-\s) F(-\s)
 \en
for $\s\in\zz$, where $\D $ and $\OD$ denote the diagonal resp.
off-diagonal parts of the spin interaction explicitly given by
  \eqnn
  &&
  \D :=\D (x,\s):=-\frac{e}{2}\s\BB_3(\la(\cdot-x)),\\
  &&
  \OD:=\OD(x,-\s)= -\frac{e}{2} \left( \BB_1(\la(\cdot-x))-
  i\s\BB_2(\la(\cdot-x))\right).
 \ennn

To investigate the translation invariant case let $\pf: =
d\Gamma_0(-i\nabla)$. The translation invariant Pauli-Fierz
Hamiltonian $\PFF(P)$ can also be mapped into a self-adjoint
operator on $\ell_2(\zz)\otimes\QQ$ defined by
 \eq{pfp}
 (\PF(P) F)(\s) = \left( \half(P-\pf -e\AA(0))^2
 +\hf  +\D (0) \right) F(\s) + \OD(0) F(-\s),
  \en
where $F(\pm 1)\in \QQ$, $\AA _\mu(0):=\AA_\mu(\la(\cdot-0))$, $\D
(0)=\D (0,\s)$ and $\OD(0)=\OD(0,-\s)$. In the following we will
construct functional integral representations for \kak{5} and
\kak{pfp}.

\section{A Feynman-Kac-type formula for jump processes}
\subsection{Pauli operators}
In this section we consider the functional integral representation
of the Pauli operator in the context of quantum mechanics. The spin
will be described in terms of a $\zz$-valued Poisson point process.
We start by reconsidering the path integral representation of the
Pauli operator  established in \cite{als}. We turn the results of De
Angelis, Jona-Lasinio and Sirugue into precise statements and
proofs, and add extensions and comments.

For a vector potential $a$ we define the Pauli operator on $\LRS$ by
\begin{equation}
h(a,b) :=
\half(-i\nabla-a)^2 +V-\half \muu  \s_\mu  b_\mu.
\end{equation}
Usually for Pauli operators $b=\nabla\times a$. However, for the
remainder of this section we treat $a$ and $b$ as not necessarily
dependent vectors. We require them to satisfy the following
conditions:
\begin{assumption}
 \label{tak}
Let $a=(a_1,a_2,a_3)$ and $b=(b_1,b_2,b_3)$ be real valued with
$a_\mu\in C_{\rm b}^2(\BR)$ and $b_\nu\in L^\infty(\BR)$, for
$\mu,\nu=1,2,3$.
\end{assumption}
Under Assumptions \ref{as2} and \ref{tak} $h(a,b)$ is self-adjoint
on $D(\Delta)$ and bounded from below, moreover it is essentially
self-adjoint on any core of $-(\han)\Delta $ as a consequence of the
Kato-Rellich Theorem. In a similar manner to the previous section,
$h(a,b)$ can also be reduced to the self-adjoint operator $\tilde
h(a,b)$ on $\LRZZ$ to obtain
 \eq{ab}
 (\tilde h(a,b) f)(\s) := \left( \half(-i\nabla -a)^2 +V
 - \half \s b_3 \right) f(\s)- \half (b_1-i\s b_2) f(-\s).
 \en

\subsection{A $3+1$ dimensional  jump  process}
In order to construct a Feynman-Kac formula for $e^{-t\tilde
h(a,b)}$, in addition to the Brownian motion we need a Poisson point
process to take the spin into account. For a summary of basic
definitions and facts as well as notations we refer to the Appendix.

Let $(B_t)_{t\geq 0}= (B_{t}^\mu)_{t\geq 0, \; 1\leq \mu\leq 3}$ be
three dimensional Brownian motion on $(W,\BB_W, P_W^x)$ with
the forward filtration ${\cal F}_t = \s(B_s,s\leq t)$, $t\geq0$,
where $W=C([0,\infty);\BR)$ and $P_W^x$ is Wiener measure with
$P_W^x(B_0 = x) = 1$.
Let, moreover, $(\os, \fffs, P_{\rm P})$ be a probability space with
a right-continuous increasing family of sub-$\s$-fields
$(\fffs_t)_{t\geq0}$, and $\eep $ denote expectation with respect to
$P_{\rm P}$. Fix a  measurable space $({\cal M}, B_{\cal M})$. Let
$p:(0,\infty)\times S\rightarrow {\cal M}$ be a stationary
$(\fffs_t)$-Poisson point process, and $D(p)\subset(0,\infty)$
denote its domain. Note that $\#D(p)$ is finite for each $\tau\in
S$. The intensity of $p$ is given by $\Lambda(t,U) := \eep [\npt] =
tn(U)$ for some measure $n$ on ${\cal M}$, where
 $N_p$ denotes counting measure on
 $((0,\infty) \times {\cal M}, \BB_{(0, \infty)} \times B_{\cal M})$ given by
$$
\npt := \#\left\{s\in D(p)\;|\; s\in (0,t], \, p(s)\in U \right\},
\quad t>0, \; U \in B_{{\cal M}},
$$
with $N_p[0, U] = 0$, and  $\BB_{(0,\infty)}$ is the Borel
$\s$-field of $(0,\infty)$. Then
$$
\eep [\npt = N] = \frac{\Lambda(t)^N}{N!} e^{-\Lambda(t)}.
$$
Assume that $n({\cal M})=1$. Write
 \eq{count}
 dN_t := \int_{{\cal M}} N_p(dtdm).
 \en
Hence
 \eq{po}
 \int_0^{t+}f(s,N_s) dN_s = \sum_{\stackrel{r\in D(p)} {0< r \leq t}}
 f(r,N_{r}).
 \en
Since $\#\{s\in D(p)\,|\,0 < s \leq t\} < \infty$, for each $\tau
\in S$ there exists $N = N(\tau)\in{\Bbb N}$ and $0 < s_1 =
s_1(\tau),...,s_N = s_N(\tau) \leq t$ such that
$$
\int_0^{t+}f(s,N_s) dN_s = \sum_{j=1}^N f(s_j,N_{s_j}) =
\sum_{j=1}^N f(s_j,j).
$$
Since $\eep [N_t] = t$ and $\eep [N_t = N] = t^N e^{-t}/N!$, the
expectation of \kak{po} reduces to Lebesgue integral:
 \eqn
 \eep  \left[\int_0^{t+} f(s,N_s) dN_s \right]
 = \eep  \left[\int_0^{t} f(s,N_s) ds\right] =
 \int_0^t \sum_{n=0}^\infty f(s,n)\frac{s^n}{n!}e^{-s} ds.
 \enn

Write $(\wo, \BB_\wo, P_\wo) := (W\times \os, \BB_W\times
\fffs, P_W\otimes P_{\rm P})$ and $\bn: = w \times \ttt\in W\times
\os $. For $\omega = w \times \ttt$, we put $B_t(\bn):=B_t(w)$ and
$p(s,\bn) := p(s,\ttt)$.

 \bd{zt}
The $\zz$-valued random process $\s_t:\zz\times \wo\rightarrow \zz$
is defined by
 $$
 \s_t := \s\ot [N_t]_2 = \s(-1)^{N_t}, \quad \s\in\zz.
 $$
 \ed
Here we have the paths $[N_t]_2$ with values $\pm 1\in\zz$
corresponding to the equivalence classes. The electron and spin
processes together give us finally the ($3+1$)-dimensional
$\BR\times\zz$-valued random process
$$
(\xi_t)_{t\geq 0} := \left(B_t,[N_t]_2\right)_{t\geq 0} =
(B_t,\s_t)_{t\geq 0}
$$
on $(\wo, \BB_\wo, P_\wo)$. Let  $\wo_t = {\cal F}_t\times
\fffs_t$, $t\geq0$. For notational convenience, we write
$$
\E[f(\xi_\cdot )] := \int_\wo f(x+B_\cdot, \s\ot [N_\cdot]_2 )dP_\wo
= \int_\wo f(x+B_\cdot, \s_\cdot )dP_\wo
$$
as well as $\EEp[f]=\int_\Omega f dP_\Omega$, ${\Bbb E}^x
[f(B_\cdot)] =\int_Wf(x+B_\cdot) dP_W^0 = \int_W f(B_\cdot) dP_W^x$,
${\Bbb E}^\s[g(\s_\cdot)] = \int_{S} g(\s_\cdot) dP_{\rm P}$, and
$\sx f(x,\s) := \ssss \int_\BR dx f(x,\s)$.

\subsection{Generator and a Feynman-Kac formula for $\xi_t$}

Next we compute the generator of the process $\xi_t$ and derive a
version of the Feynman-Kac formula.

Let $\frm$ be the fermionic harmonic oscillator defined by
 \eq{fermion}
 \frm := \half (\s_3+i\s_2)(\s_3-i\s_2)- \half.
 \en
Note that $\frm = - \s_1$. A direct computation yields
 \eq{gen}
 (f, e^{-t(-(\han)\Delta + \epsilon \frm)}g)
  = \sx \E[\bar f(\xi_0) g(\xi_t) \epsilon^{N_t}].
 \en
Thus the generator of $\xi_t$ is given by
 $$
 -\half \Delta + \frm
 $$
 and by making use of the two-valued variable $\s$,
$$
 \lk ( -\half \Delta + \epsilon \frm)f \rk (\s) = \half
 \Delta f(\s)-\epsilon  f(-\s)
$$
follows.
 \bp{88}{\rm[\textbf{De Angelis, Jona-Lasinio, Sirugue}]}
Suppose
 \eq{F}
 \int_0^t ds \int_\BR (2\pi s)^{-3/2}
 \left |\log \half \sqrt{b_1(y)^2+b_2(y)^2}
 \right | e^{-|y-x|^2/(2s)}dy <\infty
 \en
for all $(x,t) \in \BR \times [0,\infty)$. Then
 \eq{posdoc}
 \left(e^{-t\tilde h(a,b)}g\right)(x,\s) =
 e^t \EE [e^{\tz_t}g(\xi_t)].
 \en
Here
 \eqn
  \tz_t&=&
  -i\sum_{\mu=1}^3 \int_0^t a_\mu(B_s)\circ dB_{s}^\mu -
  \int_0^t V(B_s) ds\\
  &&\hspace{4cm}
  -\int_0^t \left(-\half\right)\s_sb_3(B_s) ds + \int_0^{t+}
  W(B_s,-\s_{s-}) dN_s,
 \enn
$\displaystyle \int_0^t a_\mu(B_s) \circ dB_{s}^\mu$ denoting
Stratonovich integral and
 \eqn
   W(x,-\s) := \log\left(\half  (b_1(x)-i\s b_2(x)) \right).
 \enn
 \ep
\begin{remark}
 \rm{
We will prove Proposition \ref{88} by making use of the It\^o
formula. In order that It\^o's formula applies, however, the
integrand in $\int_0^{t+}\ldots dN_s$ must be predictable with
respect to the given filtration. $\s_s$ is, though, right continuous
in $s$ for each $\omega\in\Omega$, so we define
$\s_{s-}=\lim_{\epsilon \uparrow 0} \s_{s-\epsilon}$.
 Then $\s_{s-}$
is left continuous and $W(B_s,-\s_{s-})$ is predictable, i.e.,
$W(B_s,-\s_{s-})$ is $\Omega_s$ measurable and left continuous in
$s$ for each $\omega\in\Omega$. This allows then an application of
It\^o's formula to $\int_0^{t+}W(B_s,-\s_{s-})dN_s$, for more
details see the Appendix.
 }
\end{remark}

Before turning to the proof of Proposition \ref{88}, we consider a
simplified model. Let $\UU(\cdot, \s)$ and $W(\cdot, -\s)$ be
multiplication operators on $\LRZZ$. Define the operator $K: \LRZZ
\rightarrow \LRZZ$ by
 \eq{18}
 (Kf)(x, \s): = \UU(x,\s) f(x, \s)-e^{W(x, -\s)} f(x,-\s).
 \en
First we construct a functional integral for $e^{-tK}$.
 \bp{7}
Let $\UU(x,\s)$ and $W(x,-\s)$ be continuous bounded functions in
$x\in \BR$, for each $\s=\pm 1$, such that
$\ov{\UU(x,\s)}=\UU(x,\s)$, $\ov{W(x,-\s)}=W(x,+\s)$. Then $K$ is
self-adjoint and
 \eq{14}
 (e^{-tK}g)(x,\s)= e^t \E \left[g(x,\s_t) e^{-\int_0^t \UU(x,\s_s)
 ds + \int_0^{t+} W(x,-\s_{s-})dN_s}\right].
 \en \ep
 \proof
The proof of the self-adjointness of $K$ is trivial. Write
 $$
 K_tg(x,\s) := \E\left[g(x,\s_t) e^{-\int_0^t \UU(x,\s_s) ds+\int_0^{t+}
 W(x,-\s_{s-})dN_s}\right].
 $$
Note that for each $(x,\bn)\in \BR\times \wo$,
 \eq{13}
 \left|\int_0^{t+}W(x, -\s_{s-})dN_s\right|\leq M\int_0^t dN_s=M N_t,
 \en where
$M = \sup_{x\in\BR,\s\in\zz}|W(x,-\s)|$. Then
 $$
 \|K_tg\| \leq \|g\| \,e^{tM'}\, \E[e^{M N_t}]= \|g\| \,e^{tM'}\,
 e^{t(e^M-1)},
 $$
where $M' = \sup_{x\in\BR,\s\in\zz}\EE [e^{-\int_0^t\UU(x,\s_s) ds}]$,
and $K_t$ is bounded. For each $(x, \bn)\in \BR\times \wo$ it is
seen that $\int_0^{t+}W(x, -\s_{s-}) dN_s$ is continuous in a
neighborhood of $t=0$, since $\#\{0<s<\epsilon\,|\,s\in D(p)\}=0$
for sufficiently small $\epsilon>0$, and then
 $$
 \int_0^{t+}W(x,-\s_{s-})dN_s=\sum_{\stackrel{s\in D(p)}
 {0<s\leq t}}W\left(x, -\s (-1)^{N_{s-}} \right)=0
 $$
for small enough $t$.
 Hence for $g\in C_0^\infty(\BR\times\zz)$,
 \eqn
 \lefteqn{
 \lim_{t\rightarrow 0} \|g-K_tg\|^2 } \\
 &&
 \leq \lim_{t\rightarrow 0}
 \sx\E\left[|g(x,\s)-g(x,\s_t) e^{-\int_0^t \UU(x,\s_s)ds+\int_0^{t+}
 W(x,-\s_{s-})dN_s}|^2\right] = 0
 \enn
by dominated convergence. Since $C_0^\infty(\BR\times\zz)$ is dense
in $\LRZZ$, it follows that $K_t$ is strongly continuous at $t=0$.
Also, $K_t$ has the following semigroup property. Since $N_s$ is a
Markov process, for each $(x,\s)\in\BR\times\zz$, we have
 \eqn
 \lefteqn{
 (K_sK_t g)(x, \s)} \\
 &&
 = \E\left[
 e^{-\int_0^s \UU(x, \s_r) dr+\int_0^{s+} W(x, -\s_{r-})dN_r}
 {\Bbb E}^{x, \s_s} \left[e^{-\int_0^t \UU(x, \s_{l}) dl +
 \int_0^{t+} W(x, -\s_{l-})dN_{l}}g(x, \s_t)\right]\right]\\
 &&
 = \E\left[\frac{}{}
 e^{-\int_0^s \UU(x, \s_r) dr+\int_0^{s+}
 W(x, -\s_{r-})dN_r}\right.\\
 &&\hspace{4.1cm}
 \times \E \left.
 \left[\left. e^{-\int_s^{s+t} \UU(x, \s_{l}) dl+
 \int_s^{(s+t)+} W(x, -\s_{l-})dN_{l}}
 g(x, \s_{s+t})\right|{\wo_s}\right]\right]\\
 &&
 = \E\left[e^{-\int_0^s \UU(x,\s_r) dr+\int_0^{s+} W(x,-\s_{r-})dN_r}
  e^{-\int_s^{s+t} \UU(x,\s_{l}) dl
  +\int_s^{(s+t)+} W(x,-\s_{l-})
 dN_{l}} g(x,\s_{s+t})\right]\\
 &&
 =(K_{s+t}g)(x,\s).
 \enn
$K_t$ is thus a $C_0$-semigroup, hence the Hille-Yoshida Theorem
says that there is a closed operator $h$ in $\LRZZ$ such that $K_t=
e^{-th}$, $t\geq 0$. We show that $h=K+1$.

Put $d X_t:=X_t-X_0$.
By It\^o's formula, see  Proposition~\ref{40} below, we have $d\s_t
= \int_0^{t+}(-2\s_{s-}) dN_s$ and $dg(x,\s_t) = \int_0^{t+}
\left(g(x,-\s_{s-})-g(x,\s_{s-}) \right) dN_s.$ Let
 $$
 \ZX_t := -\int_0^t \UU(x,\s_s) ds + \int_0^{t+} W(x,-\s_{s-})dN_s.
 $$
Then it follows that
 $$
 de^{\ZX_t}= -\int_0^t e^{\ZX_s}\UU(x,\s_s)ds +\int_0^{t+}
 e^{\ZX_{s-}}(e^{W(x,-\s_{s-})}-1) dN_s.
 $$
By using the product rule we get
 \eqn
 &&
 d\left(e^{\ZX_t}g(x,\s_t)\right)\\
 &&
 = -\int_0^t g(x,\s_{s})e^{\ZX_s}\UU(x,\s_s)ds +\int_0^{t+} g(x,\s_{s-})
 e^{\ZX_{s-}} (e^{W(x,-\s_{s-})}-1)dN_s\\
 &&
 \hspace{0.5cm}
 +\int_0^{t+}e^{\ZX_{s-}} (g(x,-\s_{s-})-g(x,\s_{s-}) dN_s\\
 &&
 \hspace{0.5cm}
 +\int_0^{t+}(g(x,-\s_{s-})-g(x,\s_{s-})) e^{\ZX_{s-}}(e^{W(x,-\s_{s-})}-1)
 dN_s\\
 &&
 = -\int_0^t g(x,\s_{s})e^{\ZX_s}\UU(x,\s_s)ds+\int_0^{t+}e^{\ZX_{s-}}
 \left( g(x,-\s_{s-})e^{W(x,-\s_{s-})}-g(x,\s_{s-})\right) dN_s.
 \enn
Therefore
 \eq{15}
 \E\left[e^{\ZX_t}g(x,\s_t)-e^{\ZX_0}g(x,\s_0)\right] =
 \int_0^t \E[G(s)]ds,
 \en
where $G(s)=G(x,\s,s)$ is defined by
 $$
 G(s):=\lkk\begin{array}{ll} -e^{\ZX_s}g(x,\s_{s})\UU(x,\s_s)
 +e^{\ZX_{s-}}(g(x,-\s_{s-})e^{W(x,-\s_{s-})}-g(x,\s_{s-})),&s>0,\\
 & \\
 -g(x,\s)\UU(x,\s)+ g(x,-\s)e^{W(x,-\s)}-g(x,\s),&s=0.
\end{array}\right.
 $$
Thus for each $(x,\bn)\in \BR\times \wo$, $G(s)$ is continuous in
$s$ at $s=0$ and is bounded as $|G(s)|\leq e^{MN_s}M'|g(x,\s)|$,
with constants $M$ and $M'$. Dominated convergence gives then
 $$
 \lim_{s\rightarrow 0+} \sx  \E[G(s)] = \sx \E[G(0)].
 $$
Hence
 \eqn
 &&
 \lim_{t\rightarrow 0}\frac{1}{t}(f,(K_t g-g))\\
 &&=
  \lim_{t\rightarrow 0}\frac{1}{t} \sx  \ov{ f(x,\s)}
 \E[e^{\ZX_t}g(x,\s_t)-e^{\ZX_0} g(x,\s)] \\
 &&=
  \lim_{t\rightarrow 0}\frac{1}{t} \int_0^t ds \sx
 \ov{ f(x,\s)} \E[G( s)] \\
 && =
  \sx  \ov{ f(x,\s)} \E[G(0)]\\
 && =
  \sx  \ov{f(x,\s)} \left(-\UU(x,\s)g(x,\s)+g(x,-\s)
 e^{W(x,-\s)} -g(x,\s)\right) \\
 &&=
  (f, -(K+1)g).
 \enn
Since $C_0^\infty(\BR\times\zz)$ is a core of $K$, $h=K+1$ follows.
\qed

\medskip
{\sc Proof of Proposition \ref{88}}: \quad  We put $\UU(x,\s) =
-(\han) \s b_3(x)$ and $W(x,-\s) = \log[(\han)(b_1(x)-i\s b_2(x))]$.
Recall that
 $$ \tz_t=-i\muu  \int_0^t
 a_\mu(B_s)\circ
 dB_s^\mu -\int_0^t \UU(B_s,\s_s) ds+\int_0^{t+}W(B_s,-\s_{s-}) dN_s-\int_0^t V(B_s) ds.
 $$
$W(B_s,-\s_{s-})$ is predictable and first we have to check that
$|\int_0^{t+} W(B_s,-\s_{s-}) dN_s| $ is finite for almost every
$\omega\in\Omega$ in order to apply It\^o's formula.
 Indeed,
 \eqn
 \lefteqn{
 \left|\EE\left[ \int_0^{t+} W(B_s, -\s_{s-}) dN_s\right]\right|}
 \\&&
 \leq
 \EE\left[\int_0^t
 \left|\log\left(\half \sqrt{b_1(B_s)^2+b_2(B_s)^2}\right)\right|
 dN_s\right]\\
 &&
 = 2 \int_0^t  ds \int_\BR (2\pi s)^{-3/2}
 e^{-|y-x|^2/(2s)} \left
 |\log \left(\half\sqrt{b_1(y)^2+b_2(y)^2}\right)\right| dy
 \enn
is finite by the assumption, hence $|\int_0^{t+} W(B_s, -\s_{s-})
dN_s| < \infty$, for almost every $\omega\in\Omega$.

Define $S_t:\LRZZ\rightarrow \LRZZ$ by
$$
S_tg(x,\s) = \EE\left[
e^{\tz_t}g(B_t,\s_t)\right].
$$
It can be seen that
$$
\|S_t g\|\leq V_M^\han e^{M't} e^{(M-1)t/2}\|g\|,
$$
where $M'=\sup_{x\in\BR} |b_3(x)/2|$,  $M=\sup_{x\in\BR}
(b_1^2(x)+b_2^2(x))/4$ and
 \eq{vm}
 V_M :=\sup_{x\in\BR} {\Bbb E}^x[e^{-2\int_0^t V(B_s) ds}],
 \en
which is finite by Assumption \ref{as2}. Thus $S_t$ is bounded.
Since $\tz_t$ is continuous at $t=0$ for each $\bn\in\wo$, dominated
convergence yields
$$
\|S_t g-g\|\leq \sx \EE [|g(x,\s)-g(B_t,\s_t)e^{\tz_t}|]\rightarrow 0
$$
as $t\rightarrow 0$. The semigroup property of $S_t$ follows from
the Markov property of the process $(B_t,N_t)$, which is shown in a
similar way as that of $K_t$ in Proposition \ref{7}. Thus $S_t$ is a
$C_0$-semigroup. Denote the generator of $S_t$ by the closed
operator $h$. We will see below that $S_t = e^{-th} =
e^{-t(h(a,b)+1)}$. From  Proposition \ref{40} it follows that
 \eqn
 d g(B_t, \s_t)
 &=&
 \muu \int_0^t \partial _{x_\mu} g(B_s,\s_s)dB_s^\mu +\half \int_0^t
 \Delta_x g (B_s, \s_s)ds\\
 &&
 +\int_0^{t+}\lk g(B_s, -\s_{s-})-g(B_s, \s_{s-})\rk dN_s,
 \enn
and
 \eqn
 de^{\tz_t}
 &=&
 \muu\int_0^t e^{Z_s} (-ia_\mu(B_s))\circ dB_s^\mu
 +\int_0^t e^{Z_s}(-V(B_s))ds\\
 &&
 +
 \half \int_0^t e^{Z_s}\lk
 (-i\nabla\cdot a)(B_s)+(-ia(B_s))^2\rk  ds\\
 &&
  +
  \int_0^t e^{Z_s}
 (-\UU(B_s,\s_s))ds
  +\int_0^{t+}\left( e^{Z_{s-} +W(B_s,-\s_{s-})}-e^{Z_{s-}}\right) dN_s.
 \enn
By the product rule and the two identities above we have
 \eqn
 d (e^{Z_t} g(B_t,\s_t))
 &=&
 \int_0^t e^{Z_s}
 \left[ \half \Delta_x g(B_s,\s_s) +(-ia(B_s))\cdot
 (\nabla_x g)(B_s,\s_s)    \right.\\
 &&
 \left.\hspace{1cm}  + \lk \half (-ia(B_s))^2  -V(B_s)
 -\UU(B_s,\s_s)\rk g(B_s,\s_s) \right] ds \\
 &&
 +\muu \int_0^t e^{Z_s}
  \left( \partial_{x_\mu} g(B_s,\s_s) +(-ia_\mu (B_s))
 g(B_s,\s_s)\right)  \cdot dB_s^\mu \\
 &&
 + \int_0^{t+} e^{Z_{s-}}
 \left[ \frac{}{}\lk g(B_s, -\s_{s-})-
 g(B_s, \s_{s-})\rk  \right.\\
 &&
 \hspace{2cm}
 +(g(B_s, -\s_{s-})-g(B_s, \s_{s-}))(e^{W(B_s,-\s_{s-})}-1) \\
 &&
 \hspace{2cm}
 \left.
 +g(B_s,\s_{-s})(e^{W(B_s, -\s_{s-})}-1)\frac{}{}\right] dN_s.
 \enn
Take expectation on both sides above. The martingale part vanishes
and by \kak{8} we obtain that
 \eqn
 &&
 \EE [e^{\tz_t}g(B_t,\s_t)-g(x,\s)]
 = \int_0^t \EE [G(s)]ds,
 \enn
where
 \eqn G(s) &:= & e^{Z_s}\left[ \half \Delta_x g(B_s,\s_s)
 +(-ia(B_s))\cdot (\nabla_x g)(B_s,\s_s) \right.\\
 &&
 \hspace{1cm}
 \left.+\lk \half (-ia(B_s))^2  -V(B_s)
 -\UU(B_s,\s_s)\rk g(B_s,\s_s) \right] \\
 &&
 \hspace{1cm}
 + e^{Z_{s-}}\left((g(B_s,-\s_{s-})e^{W(B_s,-\s_{s-})}-
 g(B_s,\s_{s-})\right),
 \enn
with $s>0$, and
 \eqn
 &&
 G(0):= \lkk \half \Delta_x-ia(x)\cdot \nabla_x+\half
 (-ia(x))^2-V(x)-\UU(x,\s)-1 \rkk g(x,\s) \\
 &&
 \hspace{1.5cm}
 + e^{W(x,-\s)}g(x,-\s)\\
 &&
 \hspace{1cm}
 = -(h(a,b)+1)g (x,\s).
 \enn
We see that $G(s)$ is continuous at $s=0$, for each $\bn\in\wo$,
whence
 \eqn
 \lim_{t\rightarrow 0} \frac{1}{t}(f,(S_t-1) g)
 &=&
 \lim_{t\rightarrow 0} \frac{1}{t}\int_0^t ds \sx \ov{f(x,\s)}
 \EE [G(s)] \\
 &=&
 \sx  \bar f(x,\s) \EE [G(0)]\\
 &=&
 (f, -(h(a,b)+1) g).
 \enn
Since $C_0^\infty(\BR\times\zz)$ is a core of $h(a,b)$, \kak{posdoc}
follows.
 \qed

\medskip
Note that \kak{F} is a sufficient condition making sure that
  \eq{FF}
  \int_0^{t+} |W(B_s,-\s_{s-})| dN_s<\infty, \quad {\rm a.e.}\
  \omega\in\Omega.
  \en
When, however, $b_1(x)-i\s b_2(x)$ vanishes for some $(x,\s)$,
\kak{FF} is not clear. This case is relevant and Proposition
\ref{88} must be improved since we have to construct the path
integral representation of $e^{-t\tilde h(a,b)}$ in which the
off-diagonal part $b_1-i\s b_2$ of $\tilde h(a,b)$ has zeroes or a
compact support. Since the generator of $\xi_t$ is
$-(\han)\Delta+\frm$, as was seen above, this then becomes singular.
Take $\epsilon\rightarrow 0$ on both sides of
 \eq{gen2}
 (f, e^{-t(-(\han)\Delta + \epsilon \frm)}g)
  = \sx \E[\bar f(\xi_0) g(\xi_t) \epsilon^{N_t}].
 \en
Then the right hand side of \kak{gen2} converges to $\sx {\Bbb E}^x
[\bar f(x,\s) g(B_t,\s)]$, see Remark \ref{kaiketu} below. The
off-diagonal part of $h(a,b)$, however, in general may have zeroes.
For instance, $a_\mu$ for all $\mu=1,2,3$ have compact support, and
so does the off-diagonal part in the case of $b = \nabla \times a$.
Therefore, in order to avoid that the diagonal part vanishes, we
introduce
 \eqnn
 \tilde h^\e(a,b)f(\s)
 &:=&
 \lk \half(-i\nabla-a)^2+V-\half \s b_3\rk f(\s) \non \\
 && \label{off}
 +\lk
 -\half (b_1-i\s b_2)+\e\gh\lk
 -\half (b_1-i\s b_2)\rk
 \rk f(-\s),
 \ennn
where $\gh$ is the indicator function
 \eq{ie}
 \gh(x):=\lkk \begin{array}{ll} 1,& |x|<\ee,\\
 0,& |x|\geq \ee.\end{array}\right.
 \en
We define $\gh(K)$ for a self-adjoint operator $K$ by the spectral
theorem. In particular, the identity
$$
\gh(K) = (2\pi)^{-\han}\int_\RR \hat \gh (k) e^{ik K} dk
$$
holds.
Thus $|-\half (b_1 - i\s b_2)
+ \e \gh(-\half (b_1-i \s
b_2))| > \ee$, which does not vanish for any $\e>0$.

\bp{888}
 We have
 \eq{eps}
 \left(e^{-t\tilde h^\e (a,b)}g\right)(\s, x)
 = e^t \EE[e^{\tz_t^\e }g(\xi_t)],
 \en
and
 \eq{eps2}
 \left(e^{-t\tilde h(a,b)}g\right)(\s, x) =
 \lime e^t \EE [e^{\tz_t^\e }g(\xi_t)],
 \en
where
 \eqn
 \tz_t^\e
 &=&
 -i\muu  \int_0^t a_\mu(B_s)\circ dB_s^\mu  -
  \int_0^t V(B_s) ds\\
 &&\hspace{4cm}
 -\int_0^t \left(-\half\right)\s_sb_3(B_s) ds + \int_0^{t+}
 W_\e(B_s,-\s_{s-}) dN_s,
 \enn
and
 \eqn
 W_\e (x,-\s):=
 \log\left( \half  (b_1(x)-i\s b_2(x))
 -\e \gh\left(-\half (b_1(x)-i\s b_2(x))\right)\right).
 \enn
 \ep
 \proof
\kak{eps} is derived as in Proposition \ref{88}. Since $e^{-t\tilde
h^\e(a,b)}$ converges strongly to $e^{-t\tilde h(a,b)}$ as
$\e\rightarrow 0$, \kak{eps2} follows.
 \qed
 \begin{remark}\label{kaiketu}
\rm{ We have the following cases.
\begin{enumerate}
\item[(1)]
Let the measure of
 $$
 {\cal O}_\e = \left\{(x,\s)\in \BR\times\zz \;|\;
 |(\han)(b_1(x)-i\s b_2(x))|<\ee \right\}
 $$
be zero for some $\e>0$. Then Proposition \ref{88} stays valid.

\item[(2)]
In case when the off-diagonal part identically vanishes, we have
 \eqn
 \lefteqn{
 \lime \E \left[e^{\tz_t^\e}g(\xi_t)\right]} \\
 &&=
  \lime e^t \E\left [e^{-i\muu \int_0^t a_\mu (B_s)\circ dB_s^\mu
 -\int_0^t V(B_s) ds
 -\int_0^t (-\half)\s_sb_3(B_s)  ds} \e^{N_t} g(\xi_t)\right]\\
 &&=
  {\Bbb E}^x \left [e^{-i\muu \int_0^t a_\mu (B_s)\circ dB_s^\mu
 -\int_0^t V(B_s)
 ds-\int_0^t (-\half) \s_s b_3(B_s)  ds}  g(B_t,\s)\right]\\
 &&=  e^{-t\lk
 \half (-i\nabla-a)^2 + V -\half \s_3 b_3\rk} g(x,\s).
 \enn
Here we used that as $\e\rightarrow 0$ the functions on $K_t :=
\{\omega\in\Omega \,|\, N_t(\omega)\geq 1\}$ vanish and those on
$K_t^c := \{\omega\in\Omega \,|\, N_t(\omega)=0\}$ stay different
from zero. Note that for $\omega\in K_t^c$, $N_s (\omega) = 0$
whenever $0\leq s\leq t$, as $N_t$ is counting measure. Clearly,
then the right hand side in the expression above describes the
diagonal Hamiltonian.

\item[(3)]
Since the diagonal part $-(\han) \s b_3(x)$ acts as an external
potential up to the sign $\s=\pm$, heuristically we have the
integral $\int_0^t (-\han) \s_s b_3(B_s) ds$ in $\tz_t$. This
explains why $\int_0^t \log[(\han)(b_1(B_s) -i\s_s b_2(B_s))]dN_s$
appears in $\tz_t$. Consider $T_tF (x,\s):= \E [F(B_t,\s_t)
e^{\int_0^t W(B_s, -\s_{s-}) dN_s}]$. Take, for simplicity, that $W$
has no zeroes. Compute the generator $-K$ of $T_t$ by It\^o's
formula for L\'evy processes to obtain
 \eqnn
 d \left(e^{\int_0^{t+} W(B_s, -\s_s)dN_s}\right)
 &=&
 \lk e^{\int_0^{t+} W(B_s, -\s_{s-})dN_s + W(B_t, -\s_t)}-
 e^{\int_0^{t+} W(B_s, -\s_{s-}) dN_s} \rk dN_t \non
 \\
 \label{ww}
 &=&
 e^{\int_0^{t+} W(B_s, -\s_{s-}) dN_s}(e^{W(B_t, -\s_t)}-1)dN_t.
 \ennn
On the other hand, we have
 \eq{v}
 d\lk e^{-\int_0^t V(B_s) ds}\rk  = e^{-\int_0^t V(B_s) ds}
 (-V(B_t)) dt.
 \en
From this we obtain that $e^{-t (-(\han)\Delta+V)}f(x) = {\Bbb E}^x
[e^{-\int_0^t V(B_s) ds} f(B_t)]$. Comparing \kak{ww} and \kak{v},
it is seen that It\^o's formula gives the differential for
continuous processes and the difference for discontinuous ones. From
\kak{ww} it follows that the generator $K$ of $T_t$ is given by
$$
Kf(\s)= \lk
-\half \Delta -e^{W(x, -\s)}+1 \rk f(-\s).
$$
Thus $e^{-tK}F(x, \s)= e^t {\Bbb E}^\s  [F(x, \s_t) e^{\int_0^t
W(x,-\s_{s-})dN_s}]$ giving rise to the special form of the off-diagonal
part.
\end{enumerate}
 }
\end{remark}

\section{Functional integral representation of $e^{-t\PF}$}
\subsection{Hypercontractivity and Markov property}
In this section we discuss hypercontractivity and turn to the
functional integral representation of $e^{-t\PF}$. Also, we derive a
comparison inequality for ground state energies.

Let  $ \|F\|_p = \lk \int_{\qs_\sh } |F(\phi)|^p d\mu_\sh
(\phi)\rk^{1/p}$ be $L^p$-norm on $(\qs_\sh,\mu_\sh)$ and
$(\cdot,\cdot)_2$ the scalar product on $L^2(\qs_\sh)$.
As explained in Section 2, $\Gamma_{\sh}(T)$ for $\|T\|\leq 1$ is a
contraction on $L^2(\qs_{\sh})$.
It has also the strong property of
\emph{hypercontractivity}, i.e., for a bounded operator
$K:L^2(\RR^{3+\beta})\rightarrow L^2(\RR^{3+\sh'})$ such that
$\|K\|<1$, $\Gamma_{\sh\sh'}(K)$ is a bounded operator from
$L^2(\qs_\sh)$ to $L^4(\qs_\sh)$. Nelson proved the sharper result
below.
 \bp{nelson}
Let $1\leq q\leq p$ and $\|T\|^2\leq (q-1)(p-1)\f\leq 1$. Then
$\Gamma_{\sh}(T)$ is a contraction operator from $L^q(\qs_{\sh})$ to
$L^p(\qs_{\sh})$, i.e., for $\Phi\in L^q(\qs_{\sh})$,
$\Gamma_{\sh}(T)\Phi\in L^p(\qs_{\sh})$ and
$\|\Gamma_{\sh}(T)\Phi\|_p \leq \|\Phi\|_q$.
 \ep
 \proof
See \cite{ne}.
 \qed
We factorize $e^{-t\hf}$ as is usually done. Let $j_t:\LR\rightarrow
\LRT$, $t\geq 0$, be defined by
$$
\widehat {j_tf}(k,k_0) := \frac{e^{-itk_0}}{\sqrt\pi}
\sqrt{\frac{\ob(k)}{\ob(k)^2+|k_0|^2}}\hat f(k), \quad (k,k_0)\in
\BR\times \RR.
$$
The range of $j_t$, $a\leq t\leq b$, defines the $\s$-field
$\Sigma_\aaa$ of $\QEE$,
 and the projection $E_\aaa$ to
the set of $\Sigma_\aaa$-measurable functions can be represented as
the second quantization of a contraction operator.
 By using the
 Markov property of the family of projections $E_{[\cdots]}$
 and hypercontractivity
 of $E_\aaa E_\ccc$ with $\aaa\cap \ccc = \emptyset$,
 it can be shown
  that $\int_{\QEE} |J_a F| | J_b G| |\Phi| d\mue <\infty$ for
$F,G\in\QQ$ and $\Phi\in L^1(\QEE)$.
 We will  prove this for the
 {\it massless} case in Corollary \ref{h20}.

The isometry $j_t$ preserves realness and $j_t^\ast j_s =
e^{-|t-s|\ob(-i\nabla)}$, $s,t\in\RR$, follows. Define
$$
J_t :=\Gamma_{01}(j_t), \quad J_t: \QQ \rightarrow \QQQ.
$$
Hence $J_t^\ast J_s = e^{-|t-s|\hf}$ on $\QQ$. The operator $e_t :=
j_tj_t^\ast$ is the projection from $\LRTr$ to ${\rm  Ran} j_t$.
Define
$$
U_\aaa := \ov{{\rm L.H.}\{f\in \LRTr \;|\; f\in {\rm Ran} j_t \;\;
\mbox{for some $t \in \aaa$}\}}
$$
and let $e_\aaa : \LRTr\rightarrow {U_\aaa}$ denote orthogonal
projection. Define the projections on $\QQQ$ by $E_t := J_t J_t^\ast
= \Gamma_1(e_t)$ and $E_\aaa:=\Gamma_1(e_\aaa)$. Let $\Sigma_\aaa$
be the minimal $\s$-field generated by $\{\AAA(f) \in \QQQ\;|\;f\in
{U_\aaa}\}$ and denote the set of $\Sigma_\aaa$-measurable functions
in $\QQQ$ by ${\cal E}_\aaa$. The projection $E_{[a,b]}$ has the
properties below:
 \bl{h18}
Let $a\leq b \leq t \leq c\leq d$. Then (1) $e_ae_be_c=e_ae_c$, (2)
$e_\aaa e_t e_\ccc=e_\aaa e_\ccc$, (3) ${\rm Ran} E_\aaa = {\cal
E}_\aaa$, (4) $E_\aaa E_t E_\ccc=E_\aaa E_\ccc$.
 \el
 \proof
 See \cite{si,h4}.
 \qed
Lemma \ref{h18} implies that $E_{[a,b]}$ is the projection from
$\QQQ$ onto ${\cal E}_\aaa$. The fact that $E_\aaa E_t E_\ccc =
E_\aaa E_\ccc$ is called {\it Markov property} of the family $E_s$.
Let $\om = \sqrt{|k|^2+m^2}$ with $m\geq 0$. Define $j_t^{(m)}$,
$J_t^{(m)}$, $e_\aaa^{(m)}$, $e_t^{(m)}$, $E_\aaa^{(m)}$,
$E_t^{(m)}$  and  ${\cal E}_\aaa^{(m)}$ by $j_t$, $J_t$, $e_\aaa$,
$e_t$, $E_\aaa$, $E_t$ and ${\cal E}_\aaa$ with $\ob$ replaced by
$\om$, respectively. Then Lemma \ref{h18} stays true for $e_\aaa$
and $E_\aaa$ replaced by $e_\aaa^{(m)}$ and $E_\aaa^{(m)}$,
respectively. Note that $\Gamma_{01}(e^{-t\om})$, $m>0$, is
hypercontractive but it fails to be so for $m=0$.
 \bl{h19}
 Let $a \leq b < t < c\leq d$, $F\in {\cal  E}_{\aaa}^{(m)}$ and
 $G\in {\cal E}_\ccc^{(m)}$. Take $1\leq r<\infty$, $1<p$,
 $1<q$, $r<p$ and $r<q$. Suppose that $e^{-2m(c-b)}\leq (p/r-1)
 (q/r-1)\leq 1$ and $F\in L^p(\QEE)$ and $G\in L^q(\QEE)$. Then
 $FG\in L^r(\QEE)$ and $\|FG\|_r\leq \|F\|_p\|G\|_q$. In particular,
 for $r$ such that
 $$
 r\in [1, \frac{2}{1+e^{-m(c-b)}}]\cup[\frac{2}{1-e^{-m(c-b)}},\infty),
 $$
 we have $\|FG\|_r\leq \|F\|_2\|G\|_2$.
 \el
 \proof
Let $F_N=\lkk \begin{array}{ll}F,&|F|<N,\\0,&|F|\geq N,
\end{array}\right.$ and $G_N = \lkk \begin{array}{ll}G, &|G|<N, \\0,
&|G| \geq N.\end{array}\right.$ Then $|F_N|^r\in {\cal
E}_\aaa^{(m)}$, $|G_N|^r\in {\cal E}_\ccc^{(m)}$, and it follows
that
$$
\int_{\QEE}  |F_N|^r |G_N|^r d\mue  = \left(E_\aaa^{(m)}|F_N|^r,
E_\ccc^{(m)}|G_N|^r\right)_2 = \left(|F_N|^r, \Gamma_1 (e_\aaa^{(m)}
e_\ccc^{(m)}) |G_N|^r\right)_2.
$$
Note that $T_e := e_\aaa^{(m)}e_\ccc^{(m)}$ satisfies
 \eqn
 \|T_e\|^2
 &=&
 \|e_\aaa^{(m)}e_b^{(m)}e_c^{(m)}e_\ccc^{(m)}\|^2
 \leq \|j_b^{(m)\ast} j_c^{(m)}\|^2\\
 &=&
 \|e^{-|c-b|\om}\|^2 \leq e^{-2m(c-b)}\leq(p/r-1)(q/r-1).
 \enn
Thus by H\"older inequality,
 \eq{yu1} \|F_N G_N \|^r_r\leq \||F_N|^r \|_{q/r}
 \|\Gamma_1(T_e)|G_N|^r\| _s,
 \en
where $\displaystyle 1 = \frac{1}{s} + \frac{r}{q}$. Since
$\|T_e\|^2 \leq (p/r-1)(q/r-1) = (p/r-1)(s-1) \f \leq 1$, by
Proposition \ref{nelson} it is seen that $\|\Gamma_1(T_e)|G_N|^r
\|_s\leq \||G_N| ^r\|_{p/r}$. Together with \kak{yu1} this yields
 \eq{hh36}
 \|F_N G_N \|_r\leq \|F_N \|_q\|G_N \|_p \leq \|F\|_q\|G\|_p.
 \en
Taking the limit $N\rightarrow \infty$ on both sides of \kak{hh36},
by monotone convergence the lemma follows. \qed

An immediate consequence is
 \bc{h20}
 Let $\Phi\in L^1(\QEE)$ and $F,G\in \QQQ$. Then, for $a \not = b$,
 $(J_a F) \Phi (J_bG )\in L^1(\QEE)$ and \eq{hh22}
 \int_{\QEE}|(J_a F)\Phi(J_bG )|d\mue \leq \|\Phi \|_1\|F\|_2\|G\|_2.
 \en \ec
 \proof
Let $a < b$, and $ \displaystyle r^{(m)}=\frac{2}{1-e^{-m(b-a)}}$
and $s^{(m)}>1$ be such that $\displaystyle \frac{1}{r^{(m)}} +
\frac{1}{s^{(m)}} = 1$, i.e., $s^{(m)} = r^{(m)}/(r^{(m)}-1)$.
Without loss of generality we can assume that $\Phi$ is a
real-valued function. Truncate $\Phi $ as
$$
\Phi _N := \lkk \begin{array}{cl} N, & \Phi >N,\\ \Phi, & |\Phi
|\leq N,\\ -N, & \Phi <-N.
\end{array}
\right.
$$
By Lemma \ref{h19}
 \eqn
 |(J_a^{(m)}F,\Phi_N J_b^{(m)} G)_2|
 &\leq&
 \int_{\QEE} |(J_a^{(m)}F)| |\Phi_N| |(J_b^{(m)}G)| d\mue \\
 &\leq&
 \|\Phi_N\|_{s^{(m)}}\|(J_a^{(m)}F) (J_b^{(m)} G)\|_{r^{(m)}}\\
 &=& \|\Phi_N\|_{s^{(m)}} \|J_a^{(m)} F\|_2\|J_b^{(m)} G\|_2\\
 &=& \|\Phi _N\|_{s^{(m)}} \| F\|_2\| G\|_2.
 \enn
 Since
 $\slim_{m\rightarrow 0} J_t^{(m)} = J_t$ in $\QQQ$ by
 $\slim_{m\rightarrow 0} j_t^{(m)}=j_t$ in $\LRT$,
and  $\Phi _N$ is a bounded multiplication operator, we have
 \eq{hh1}
 (|J_aF|, |\Phi_N| | J_b G|)_2 \leq \|\Phi_N\|_1\|F\|_2\|G\|_2
  \leq \|\Phi \|_1\|F\|_2\|G\|_2.
 \en
Since $|\Phi _N|\uparrow |\Phi|$ as $N\rightarrow \infty$, by
monotone convergence $|J_aF||\Phi||J_bG| \in L^1(\QEE)$ and
\kak{hh22} follow. This completes the proof. \qed

\subsection{Functional integral}
As explained in Section 1, a key idea of constructing a functional
integral representation of $e^{-t\PF}$ is to use the identity
 \eq{hh21}
 \hhhh =  \int_{\QSS}^\oplus \LRZZ d\mu(\phi).
 \en
We define the  Pauli operator $\wpf(\phi)$  in \kak{pauli} for each
fiber $\phi\in \QSS$ and set
 \eq{46}
 \wwpf := \hf\, \, \dot + \, \,
 \int_{\QSS}^\oplus \wpf(\phi) d\mu (\phi),
  \en
where $\dot{+}$ denotes quadratic form sum. It is seen that
$\PF=\wwpf$ as a self-adjoint operator. Using the path integral
representation of Pauli operators discussed in Section 3, we can
construct the functional integral representation of
$e^{-t\wpf(\phi)}$ for each $\phi\in \QSS$. From this the path
integral representation of $e^{-t\PF}$ can be derived through the
identity $\PF=\wwpf$ and the Trotter product formula for quadratic
form sums \cite{km}.

\medskip
Define the Pauli operator $\wpf(\phi)$ on $\LRZZ$ by
 \eq{pauli}
  (\wpf(\phi)f)(\s) := \left( \half (-i\nabla -e\AA(\phi))^2 + V
 +\D (\phi) \right) f(\s) + \OD(\phi) f(-\s),
 \en
where
 \eqn
 &&
 \D (\phi)=\D (x,\s,\phi) = -\frac{e}{2}\s \BB_3(\phi),\\
 &&
 \OD(\phi)=\OD(x,-\s,\phi)=-\frac{e}{2}(\BB_1(\phi)-i\s\BB_2(\phi)).
 \enn
To avoid that the off-diagonal part $\OD(\phi)$ vanishes, we
introduce $\wpfe (\phi)$ in a similar manner as in $\tilde
h^\e(a,b)$ above by
 \eqnn
 && (\wpfe (\phi)f)(\s) := \lk \half (-i\nabla -e\AA(\phi))^2+V
 +\D (\phi)\rk f(\s) \label{48} \\
 &&\non
 \hspace{7cm}
 + \lk \OD(\phi) + \e \gh(\OD(\phi)) \rk  f(-\s),
 \ennn
where $\gh$ is the indicator function given by \kak{ie}. Since $
|\D(\phi) + \e \gh(\D (\phi))|\geq \ee$ for all $(x,\s)\in
\BR\times\zz$, we can define
 \eqn
 W_{\phi}^\e (x,-\s) := \log\lk
 -\OD(x,-\s, \phi) -
 \e\gh(\OD(x,-\s, \phi)) \rk.
 \enn
 \bl{24}
Assume that $\la\in C_0^\infty(\BR)$. Then for each $\phi\in \QSS$,
$\wpfe (\phi)$ is self-adjoint on $D(-\Delta)\otimes\zz$ and for
$g\in\LRZZ$,
 \eqn (e^{-t\wpfe  (\phi)}g)(x,\s)=
 \EE[ e^{-\int_0^t V(B_s) ds} e^{\zpe }g(\xi_t)],
 \enn
where
 \eqn
 \zpe
 &=&
 -i\muu
 \int_0^t \AA_\mu (\la(\cdot-B_s),\phi) dB_s^\mu \\
 &&
 -\int_0^t \D(B_s,\s_s,\phi) ds + \int_0^{t+}
 W_{\phi}^\e (B_s,-\s_{s-})dN_s.
 \enn \el
 \proof
Since $\la\in \crr$, we have
$$
\AA_\mu(\phi) = \AA_\mu(\la(\cdot-x),\phi) := \langle {\phi,
\oplus_{\nu=1}^3 \delta_{\mu\nu} \la (\cdot-x)} \rangle_0 \in C_{\rm
b}^\infty(\RR^3_x), \quad  \phi\in\QSS.
$$
Then $\wpfe  (\phi)$ is the  Pauli operator with a sufficiently
smooth bounded vector potential $\AA(\phi)$, and the off-diagonal
part is perturbed by the bounded operator $\e\gh(\OD(\phi))$. Hence
it is self-adjoint on $D(-\Delta)\otimes\zz$ and the functional
integral representation follows  by Proposition~\ref{88}.
 \qed
 Next we define the operator $\wwpf^\e$ on $\hhhh$ through $\wpfe
(\phi)$ and the constant fiber direct integral representation
\kak{hh21} of $\hhhh$. Assume that $\la\in \crr$. Define the
self-adjoint operator $\wpfe $ on $\hhhh$ by
 $$
\wpfe  := \int_{\QSS}^\oplus \wpfe  (\phi) d\mu(\phi),
 $$
that is, $ (\wpfe   F)(\phi)=\wpfe  (\phi)F(\phi)$ with domain
 $$
D(\wpfe )= \left\{F\in \hhhh \,\left|\, \int_{\QSS} \|(\wpfe
F)(\phi)\|^2_{\LRZZ} d\mu(\phi)<\infty\right. \right\}.
 $$
Set
 \eq{high}
 \wwpf^\e := \wpfe \,\, \dot{+} \,\, \hf.
 \en
Let $\qqf := \bigcup_{m=0}^\infty\{ \bigoplus_{n=0}^m L_n^2(\QSS)
\bigoplus_{n=m+1}^\infty \{0\}\}$ and define the dense subspace
 \eq{dense subspace}
 \hhhhh := C_0^\infty(\BR\times\zz)
 \,\hat\otimes \,\qqf,
 \en
where $\hat\otimes$ denotes algebraic tensor product. Also, define
 \eq{this}
 \PF^\e := \PF +  \MMM 0
 { \e \gh(- \frac{e}{2}(\BB_1-i\BB_2))}
 { \e \gh(- \frac{e}{2}(\BB_1+i\BB_2))}
 {0 }.
 \en
 \bl{common}
Let $\la\in C_0^\infty(\BR)$. Then
 \eq{25}
 (F, e^{-t\PF}G) = \lime(F, e^{-t\wwpf^\e}G).
 \en
 \el
 \proof
It is seen that $\wwpf^\e = \PF^\e$ on $\hhhhh$, implying that
$\wwpf^\e = \PF^\e$ as a self-adjoint operator since $\hhhhh$ is a
core of $\PF^\e$ \cite{h12,h19}. Moreover, $\PF^\e\rightarrow \PF$
on $\hhhhh$ as $\e \rightarrow 0$ and $\hhhhh$ is a common core of
the sequence $\{\PF^\e\}_{\e\geq 0}$. Thus $\slime e^{-t\PF^\e} =
e^{-t\PF}$, whence \kak{25} follows.
 \qed
By \kak{25} it suffices to construct a functional integral
representation for the expressions at its right hand side and then
use a limiting procedure. Set
 \eqnn
 &&\DEE(x,\s,s)=-\frac{e}{2}\s\BBB_3(j_s\la(\cdot-x)),\\
&&\ODEE(x,-\s,s)=-\frac{e}{2}\lk
\BBB_1(j_s\la(\cdot-x))-i\s\BBB_2(j_s\la(\cdot-x))\rk.
 \ennn

 \bl{projection1}
As a bounded multiplication operator on $\QQ$, for each $(x,\s)\in
\BR\times\zz$
 \eq{asa}
 J_s \gh(\OD(x,-\s))J_s^\ast = E_s \gh(\ODEE(x,-\s,s))E_s.
 \en
 \el
 \proof
Note that $\gh(\OD(x,-\s))$ is a function of the Gaussian random
variable $\Phi := \OD(x,-\s) = (-e/2)(\BB_1(x)-i\s\BB_2(x))$ of mean zero
and covariance
 \eq{mean}
 \rho := \int_{\QSS}\Phi^2 d\mu =
 \frac{e^2}{4}
 \int_{\QSS}(\BB_1(x)^2+\BB_2(x)^2) d\mu =
 \frac{e^2}{8}  \int
 \frac{|\vp(k)|^2}{\ob(k)}|k|^2
 \left(2-\frac{|k_1|^2+|k_2|^2}{|k|^2}\right)dk,
 \en
since \eqn
 \jjj (k\times e(k,j))_\mu (k\times e(k,j))_\nu=|k|^2
 \left(\delta_{\mu\nu}-\frac{k_\mu k_\nu}{|k|^2}\right).
 \enn
In general, for a given function $g\in L^2(\RR)$, $g(\Phi)$ is
approximated by
  \eq{app}
  g_n(\Phi) = (2\pi)^{-\han} \int_\RR  \hat g_n(k) e^{i k \Phi}dk
  \en
in $\QQ$, where $g_n\in \SSS(\RR)$ is such that $g_n\rightarrow
g$ as $n\rightarrow \infty$ in $L^2(\RR)$. This follows from
  \eq{tou}
  \|g(\Phi)-g_n(\Phi)\|_2^2\leq (2\pi\rho )^{-\han}
  \int_\RR  |g(x)-g_n(x)|^2  dx.
  \en
For the vector
$$
F = \int f(k_1,...,k_n) e^{-i \sum_{j=1}^n \abb{\phi, h_j}}
dk_1\cdots dk_n
$$
with $f \in \SSS(\RR^n)$ and $h_j\in \oplus^3 \LR$, we have
$\limn g_n(\Phi) F = g(\Phi)F$ strongly by \kak{tou}. Since the set
of vectors of form $F$ are dense in $\QQ$, as bounded multiplication
operators $g_n(\Phi)$ strongly converge to $g(\Phi)$ as
$n\rightarrow\infty$. Thus there is a sequence
$\{\gh^n(\Phi)\}_{n=1}^\infty$ such that
 \eq{413}
 \gh^n(\Phi) =
(2\pi)^{-\han}\int_\RR \hat\gh^n(k) e^{ik\Phi} dk \en
 with
$\hat\gh^n\in\SSS(\RR)$ and $\limn \gh^n(\Phi) = \gh(\Phi)$ in
strong sense. By \kak{413}
 \eqn
 \lefteqn{
  J_s \gh^n(-\OD(x,-\s)) J_s^\ast
  =
 (2\pi)^{-\han}\int_\RR \hat\gh^n(k) J_s e^{ik\Phi}J_s^\ast dk }\\
  &&
  =
  (2\pi)^{-\han}\int_\RR \hat\gh^n(k) E_s e^{ik\Phi_s} E_s dk
  = E_s\gh^n(-\ODEE(x,-\s,s))E_s,
 \enn
where $\Phi(s)=(-e/2)(
\BBB_1(j_s\la(\cdot-x))-i\s\BBB_2(j_s\la(\cdot-x)))$,
and $\gh^n(\ODEE(x,-\s,s))$ converges strongly to $\gh
(\ODEE(x,-\s,s))$ with $n\rightarrow\infty$ as a bounded
multiplication operator on $\QQQ$, yielding \kak{asa}.
 \qed
The next statement is our key lemma.
 \bl{26}
Let $\la \in \crr$, $F\in {\cal E}_\aaa$ and $s \not \in \aaa$.
Then
 \eq{500}
 (F, J_s e^{-t \wpfe }J_s^\ast G) = e^t \sx
 \EE \left[ e^{-\int_0^t V(B_{r}) dr} \int_{\QEE}
 \ov{F(\xi_0)} e^{ \xtes } E_s G(\xi_t) d\mue  \right].
 \en
Here
 \eqnn
 \xtes & = & -ie \sum_{\mu=1}^3 \int_0^t \AAA_\mu
 (\las (\cdot-B_{r})) dB_{r}^\mu
 \label{310}\\
 &&
 -\int_0^t \DEE (B_{r},\s_{r},s)dr+\int_0^{t+}
 W^\e(B_{r}, -\s_{r-},s) dN_{r}, \non
 \ennn
and
 \eqnn
 \label{28}
  W^\e(x,-\s,s) :=\log\left(-\ODEE(x,-\s,s) -
  \e\gh(\ODEE(x,-\s,s))\right)
 \ennn
 \el
 \proof
First notice that the right hand side of \kak{500} is bounded. By
Corollary \ref{h20}, $F(x,\s) = J_l J_l^\ast F(x,\s)$ for some
$l\in[a,b]$ and $E_s G(B_t,\s_t)= J_s J_s^\ast G(B_t,\s_t)$. We
obtain
 \eq{501}
 |{\rm r.h.s.}\
 \kak{500}| \;\leq\;
 \EEp \left[ e^{-\int_0^t V(B_{r}) dr}
 \sx\|F(x,\s)\|_2 \|G(B_t+x,\s_t)\|_2
 \|e^{ \xtes }\|_1 \right].
 \en
We will prove in Lemma \ref{cclemon} below that there exists a
random variable $c = c(\omega)$ such that
\begin{itemize}
\item[(1)]
$\|e^{ \xtes }\|_1^2\leq c$, a.e. $\omega\in \Omega$,
\item[(2)]
$c$ is independent of $(x,\s)\in \BR\times\zz$,
\item[(3)]
$c$ is independent of  $B_t^\mu$, $\mu=1,2,3$,
\item[(4)]
$\EEp [c^\han]<\infty$.
\end{itemize}
By \kak{501},
 \eqnn
 \lefteqn{
 |{\rm r.h.s.}\ \kak{500}| } \non\\ &&
 \leq
 \EEp \left[ \lk \sx \|G(B_t+x,\s_t)\|_2^2\rk^\han \lk \sx
 \|F(x,\s)\|_2^2 e^{-2\int_0^t V(B_{r}+x) dr}c\rk^\han\right]
 \non\\
 &&
 \leq \|G\|_{\hhhh} \,\EEp \left[c^\han \lk \sx \|F(x,\s)\|_2^2
 e^{-2\int_0^t V(B_{r}+x) dr}\rk^\han\right]\non\\
 &&
 \leq \|G\|_{\hhhh}\, \EEp [c^\han] \, \EEp \left[\lk \sx \|F(x,\s)\|_2^2
 e^{-2\int_0^t V(B_{r}+x) dr}\rk^\han \right]\non\\
 &&
 \leq \|G\|_{\hhhh} \, \|F\|_{\hhhh} V_M^\han \EEp[c^\han] < \infty,
 \label{502}
 \ennn
where we used (1) above in the second line, (2) in the third line,
(3) in the fourth line, Assumption \ref{as2} and (4) in the fifth
line, and where $V_M$ is defined in \kak{vm}.

Next we prove \kak{500}. By Lemma \ref{24} we have
 \eqn
 \lefteqn{
 (J_s^\ast F,  e^{-t\wpfe }J_s^\ast G)}\\
 &&
 = \int_{\QSS}d\mu(\phi) (({J_s^\ast F})(\phi), e^{-t\wpfe  (\phi)}
 (J_s^\ast G)(\phi))_{\LRS}\\
 &&
 = \int_{\QSS}d\mu(\phi) \sx  \EE \left[e^{-\int_0^t V(B_{r})dr}
 \ov{({J_s^\ast F})(\phi,\xi_0)} e^{\zpe}
 (J_s^\ast G)(\phi,\xi_t) \right]\\
 &&
 = \sx \EE \left[ e^{-\int_0^t V(B_{r})dr} \int_{\QSS}d\mu(\phi)
 \ov{(J_s^\ast F)(\phi, \xi_0)} e^{ \zpe } (J_s^\ast G)(\phi, \xi_t)\right].
 \enn
Here we used Fubini's Theorem in the fourth line. Put
$$
\xte  = -ie\sum_{\mu=1}^3 \int_0^t \AA_\mu(\la(\cdot-B_{s}))
dB_{s}^\mu  -\int_0^t \D (B_{s},\s_{s}) ds +\int_0^{t+} W^\e (B_{s},
-\s_{s-}) dN_{s},
$$
with $W^{\e}(x,-\s) := \log \left(-\OD(x,-\s)- \e \gh(\OD(x,-\s))
\right)$. Pick $F,G\in \hhhhh$. Given that
 $J_s^\ast F\in \QQQ$ and $e^ {\xte }J_s^\ast G
(B_t,\s_t)\in \QQQ$, we rewrite as
$$
(J_s^\ast F,  e^{-t \wpfe  }J_s^\ast G) = \sx \EE \left[
e^{-\int_0^t V(B_{r} )dr}(F(\xi_0), J_s e^{ \xte } J_s^\ast
G(\xi_t))_{\QQQ}\right].
$$

The kernel $J_s e^{ \xte } J_s^\ast$ is computed as follows. Divide
it up into
 \eqnn
 &&
 J_s e^{ \xte }J_s^\ast =
 \underbrace{ J_s e^{-ie\sum_{\mu=1}^3 \int_0^t
 \AA_\mu(\la(\cdot-B_{r})) dB_{r}^\mu } J_s^\ast }_{:={\rm I}}
 \underbrace{
 J_s e^{-\int_0^t \D  (B_{r},\s_{r}) dr}J_s^\ast}_{:={\rm II}}
 \non\\
 &&\label{261}\hspace{5cm}
 \times \underbrace{ J_s e^{\int_0^{t+} W^\e (B_{r}, -\s_{r-})
 dN_{r}}J_s^\ast}_{:={\rm III}}.
 \ennn
We compute the three factors $\rm I,II, III$ separately. First, by
\cite{h4}
\eqn &&
  J_s \exp\lk -ie \sum_{\mu=1}^3 \int_0^t \AA_\mu (\la(\cdot-
B_{r})) dB_{r}^\mu \rk J_s ^\ast \\
&& \hspace{1cm}=
  E_s\exp\lk -ie \sum_{\mu=1}^ 3 \int_0^t \AAA_\mu(\las (\cdot-
B_{r})) dB_{r}^\mu \rk E_s. \enn
Secondly, for $\bn\in\wo$, there exist $N=N(\bn)\in{\Bbb N}$ and
 $s_1=s_1(\bn),...,s_N=s_N(\bn)\in (0,\infty)$ such that on $\hhhhh$
 \eqn
 \lefteqn{
 J_s \exp\lk {\int_0^{t+}W^\e (B_{r},-\s_{r-}) dN_{r}}\rk
 J_s^\ast } \\
 &&
 =
 J_s \prod_{i=1}^N \lk -\OD(B_{s_i}, -\s_{s_i-}) -
 \e\gh(-\OD(B_{s_i}, -\s_{s_i-}))\rk J_s^\ast\\
 &&=
 E_s \prod_{i=1}^N \lk -\ODEE(B_{s_i}, -\s_{s_i-},s)
 -\e\gh(-\ODEE(B_{s_i}, -\s_{s_i-},s))\rk E_s\\
 && =
 E_s\exp\lk {\int_0^{t+}W^\e  (B_{r},-\s_{r-},s) dN_{r}}\rk
 E_s,
 \enn
where we used that $J_s\AA(f_1)\cdots\AA(f_n)J_s^\ast = E_s
\AAA(j_sf_1) \cdots \AAA(j_sf_n)E_s$ as multiplication operators,
and that $J_s \gh ( \OD(B_{s_i}, -\s_{s_i-})) J_s^\ast = E_s \gh
(\ODEE(B_{s_i}, -\s_{s_i-},s))E_s$ by Lemma~\ref{projection1}.
Finally, it can be seen that, similarly to $\rm III$, factor $\rm
II$ is computed on $\hhhhh$ as
 \eqn
 \lefteqn{
 J_s \exp\lk -\int_0^t \D (B_{r},\s_{r}) dr\rk J_s^\ast
 = \limn J_s \prod_{i=0}^n \exp \lk \D (B_{it/n}, \s_{it/n})
 \frac{t}{n} \rk J_s^\ast} \\
 &&=
 \limn \prod_{i=0}^n E_s \exp \lk \DEE(B_{it/n}, \s_{it/n},s)
 \frac{t}{n} \rk E_s = \exp\lk -\int_0^t \DEE (B_{r},\s_{r},s)
 dr\rk E_s.
 \enn
Putting all this together we get
 \eq{hh123}
 (F, J_s e^{-t\wpfe }J_s^\ast G)
 = \sx \EE \left[e^{-\int_0^t V(B_{r})dr} \int_{\QEE} d\mue
 \ov{F(\xi_0)} e^{\xtes }E_s G(\xi_t) \right]
 \en
for $F,G\in \hhhhh$. By a limiting argument and the bound \kak{502}
it is seen that \kak{hh123} extends for $F,G\in\hhhh$, completing
the proof.
 \qed
 \bl{cclemon}
There exists a random variable $c=c(\omega)$ satisfying (1)-(4) in
the proof of Lemma \ref{26}.
 \el
 \proof
Note that
 $$
 \|e^{ \xtes }\|_1^2
 \leq \|e^{-\int_0^t \DEE(B_{r}, \s_{r},s) dr} \|_2^2\,
 \| e^{\int_0^t|W^\e (B_{r}, -\s_{r-},s)|dN_{r}}\|_2^2.
 $$
We estimate the right-hand side of this expression. Since
$$
\int_0^t \DEE(B_{r}, \s_{r},s) dr= \BBB_3\left(-\frac{e}{2}
\int_0^t\s_{r} \las (\cdot-B_{r})dr\right)
$$
and $\BBB_\mu(f)$ is a Gaussian random variable with mean zero and
covariance
 \eq{cov}
 \int_{\QEE} \BBB_\mu(f)\BBB_\nu(g)d\mue  = \half \int
 \ov{\hat f(k,k_0)} \hat g(k,k_0) |k|^2\left(\delta_{\mu\nu} -
 \frac{k_\mu k_\nu}{|k|^2}\right) dkdk_0,
 \en
we have
 \eqnn
 \lefteqn{
 \left\|e^{-\int_0^t \DEE(B_{r}, \s_{r}, s) dr}\right\|_2^2 =
 \left(\Ob, e^{-2\int_0^t \DEE (B_{r},\s_{r},s) dr}\Ob\right)\non} \\
 &&=
 \exp\lk 4 \half \lk\frac{e}{2}\rk^2 \half \int_0^t dr\int_0^t
 dl\s_{r}\s_{l} \int_\BR \frac{|\vp(k)|^2}{\ob(k)}
 e^{-ik\cdot(B_{r}-B_{l})}(|k_1|^2+|k_2|^2) dk\rk\non \\
 &&\label{matui}
 \leq \exp\lk \lk\frac{e}{2}\rk^2  t^2 \int_\BR \frac{|\vp(k)|^2}
 {\ob(k)} |k|^2 dk \rk := c_1 < \infty.
 \ennn
$c_1$ is thus independent of $(x,\s)\in \BR\times\zz$. Next consider
$\|e^{\int_0^t |W_\e(B_{r}, -\s_{r-},s)| dN_{r}}\|_2^2$. Set
$\BBB_\mu(t) := \BBB_\mu(\las (\cdot-B_t))$ for notational convenience.
For each $\bn\in\wo$, there exists $N = N(\bn)\in {\Bbb N}$ and
$s_1=s_1(\bn),...,s_N=s_N(\bn)\in(0,\infty)$ such that
 \eqnn
 \lefteqn{
 \left\|e^{\int_0^t|W^\e (B_{r},-\s_{r-},s)|dN_{r}}\right\|_2^2} \\
 &\leq&
 \left(\Ob,\exp\left(2\int_0^t \log\left[\frac{|e|}{\sqrt 2}
 \sqrt{\BBB_1(r)^2+\BBB_2(r)^2+\e^2} \right]dN_{r}\right)
 \Ob\right)_2\non \\
 &=&
 \left(\Ob, \exp \left(2\sum_{i=1}^N \log\left[\frac{|e|}{\sqrt 2}
 \sqrt{\BBB_1(s_i)^2+\BBB_2(s_i)^2+\e^2}\right]\right)
 \Ob\right)_2\non \\
 &=&
 \lk \frac{|e|}{\sqrt 2}\rk^{2N} \left(\Ob, \prod_{i=1}^N \lk
 \BBB_1(s_i)^2+\BBB_2(s_i)^2+\e^2\rk \Ob\right)_2\non \\
 &=&
 \lk \frac{|e|}{\sqrt 2}\rk^{2N}
 \sum_{m=0}^ N \e^{2(N-m)} \sum_{{\rm comb}_m}
 (\Ob, \underbrace{(\BBB_{\#})^2\cdots(\BBB_{\#})^2}_{\mbox{\tiny{$m$-fold}}}
 \Ob)_2
 \non \\
 &=&
 \lk \frac{|e|}{\sqrt 2}\rk^{2N}
 \sum_{m=0}^ N \e^{2(N-m)} \sum_{{\rm comb}_m} \|\underbrace{\BBB_{\#}
 \cdots \BBB_{\#}}_{\mbox{\tiny{$m$-fold}}}\Ob\|_2^2 \non \\
 &\leq&
 \lk \frac{|e|}{\sqrt 2}\rk^{2N}
 \sum_{m=0}^ N \e^{2(N-m)} 2^m \,(\sqrt2)^{2m}\, m!\,
 \|\sqrt{|k|}\vp\|^{2m }:=c_2,
 \label{matui2}
 \ennn
where $\sum_{{\rm comb}_m}$ denotes summation over the $2^m$ terms
in the expansion of the product $\prod_{i=1}^m (\BBB_1(s_i)^2 +
\BBB_2(s_i)^2)$, $\BBB_{\#}$ denotes one of $\BBB_\mu (s_i)$, $\mu
=1,2$, $i=1,...,N$, and we used that $|a+ib +\e|\leq \sqrt 2
\sqrt{a^2 + b^2 + \e^2}$, $a,b,\e\in\RR$, in the first line, and the
basic inequality $ \|\BBB_\mu(s_i)\Psi\|_2 \,\leq\, {\sqrt 2}
\|\sqrt{|k|}\vp\|\|\Nb^\han \Psi\|_2 $ in the sixth. Note that
$c_2(\omega)$ is independent of $(x, \s) \in\BR\times\zz$ and
$B_t^\mu$. Set
  \eq{cc}
  c(\omega) = c_1 c_2(\omega).
  \en
Then
 \eq{mot}
 \EEp [c^\han] \,\leq\, e^{ \half (|e|/2)^2 t^2
 \|\sqrt{|k|}\vp\|^2} \sum_{N=0}^\infty
 \left(\frac{|e|}{\sqrt 2}\right)^{N}
 \sum_{m=0}^ N \frac{\e^{N-m} \,\sqrt{m!}\, 2^m
 \, \|\sqrt{|k|}\vp\|^m}{N!} e^{-t} < \infty.
 \en
This completes the proof of claims (1)-(4) above.
 \qed

\medskip

Next we define the $\LRT$-valued stochastic integral $\int_0^t
j_s\la(\cdot-B_s) dB_s^\mu$ by a limiting procedure. Let
$\Delta_n(s)$ be the step function on the interval $[0,t]$ given by
 \eq{chi}
 \Delta_n(s) := \sum_{i=1}^n \frac{t(i-1)}{n} 1_{(t(i-1)/n,ti/n]}(s).
 \en
Define the sequence of the $\LRT$-valued random variable $\xi_n^\mu:
\Omega\rightarrow \LRT$ by
$$
\xi_n^\mu := \int _0^t j_{\Delta_n(s)}\la(\cdot-B_s) dB_s^\mu, \quad
\mu=1,2,3.
$$
This sequence converges, which is guaranteed by
 \eqn
 \EEp [ \|\xi_n^\mu- \xi_m^\mu\|^2]
 &=&
 \EEp \left[\int_0^t \|j_{\Delta_n(s)}\la (\cdot-B_s) -
 j_{\Delta_m(s)}\la (\cdot-B_s)\|^2 ds \right]\\
 &=&
 2 \EE \left[ \int _0^t \left( \|\la\|^2 - (\la(\cdot-B_s),
 e^{-|\Delta_n(s)-\Delta_m(s)| \ob} \la(\cdot-B_s))\right) ds \right]
 \rightarrow 0
 \enn
as $n, m\rightarrow \infty$.
 \bd{hiro}
We define
$$
\int_0^t j_s\la(\cdot-B_s) dB_s^\mu  := \slimn \xi_n^\mu, \quad
\mu=1,2,3,
$$
and set
$$
\int_0^t \AAA_\mu(j_s\la(\cdot-B_s))dB_s^\mu :=
\AAA_\mu\left(\int_0^t j_s \la(\cdot-B_s) dB_s^\mu\right).
$$
 \ed

Now we are in the position to state the main theorem of this
section.
 \bt{main}
For every $t \geq 0$ and all  $F,G\in\hhhh$
 \eq{hhmain}
 (F,e^{-t\PF^\e}G) =
 e^t \sx \EE \left[e^{-\int_0^t V(B_s) ds}
 \int_{\QEE} d\mue  \ov{J_0 F(\xi_0)}
 e^{{\XX }} J_t G(\xi_t)\right]
 \en
and
 \eq{hmain}
 (F, e^{-t\PF} G) =
 \lime e^t \sx \EE \left[ e^{-\int_0^t V(B_s) ds}
 \int_{\QEE}d\mue  \ov{J_0 F(\xi_0)} e^{{\XX}}
 J_t G(\xi_t)\right].
 \en
Here
 \eqn
 \lefteqn{
 \XX
 =
 -ie \sum_{\mu=1}^3 \int_0^t\AAA_\mu(\las (\cdot-B_s)) dB_s^\mu} \\
 &&
 - \int_0^t \DEE(B_s,\s_s,s) ds
 + \int_0^{t+} \log\left(-\ODEE(B_s,-\s_{s-},s) -
   \e\gh(\ODEE(B_s,-\s_{s-},s)) \right) dN_s.
 \enn
 \et
 \proof
Notice that $\BBB_\mu(j_s f)$, $f\in\LR$, $s\in\RR$, $\mu=1,2,3$, is
a Gaussian random variable with mean zero and covariance
$$
\int_{\QEE}\BBB_\mu(j_s f)\BBB_\nu (j_t g)d\mue  = \half\int_\BR
\ov{\hat f(k)} \hat g(k) |k|^2\left(\delta_{\mu\nu}-\frac{k_\mu
k_\nu} {|k|^2}\right) e^{-|t-s|\ob(k)} dk.
$$
Then similarly to \kak{502} we obtain $|{\rm r.h.s.}
\kak{hhmain}|\leq \|F\|_{\hhhh}\|G\|_{\hhhh} V_M^\han
\EE[c^\han]<C$, where $c$ is given by  \kak{cc} and $C$ is a
constant independent of $\e$. Since  $e^{-t\PF^\e}\rightarrow
e^{-t\PF}$ strongly as $\e\rightarrow 0$, \kak{hmain} follows from
\kak{hhmain}.

Now we turn to proving \kak{hhmain}. Take $\la =
(\vp/\sqrt{\ob})^\vee \in C_0^\infty(\BR)$. Then by \kak{502}
 $ \EE[
e^{-\int_0^t V(B_r) dr} e^{\xtes}G(\xi_t)] \in \hhhh$ for $G\in
\hhhh$, and
 \eqn
 &&
 \left\| \EE \left[ e^{-\int_0^t  V(B_r) dr}
 e^{\xtes }G(\xi_t)\right]_{\hhhh}\right\|
 \;\leq \;
 V_M^\han \, \EE[c^\han] \, \|G\|_{\hhhh}.
 \enn
Remember that $ \xtes $ was defined in \kak{310} and $V_M$ in
\kak{vm}. Define the bounded operator
$$
(S^\e_{t,s}G)(x,\s) := e^t \EE \left[e^{-\int_0^t V(B_u) du}
e^{\xtes }G(\xi_t)\right], \quad \hhhh\rightarrow \hhhh.
$$
 Set \eqn &&
\XXX{S,T}{s}=
 -ie\sum_{\mu=1}^3 \int_S^T
  \AA_\mu(j_s\la(\cdot-B_l)) dB_l^\mu\\
  &&\hspace{3cm}
   -\int_S^T  \D (B_l,\s_l,s) dl +
 \int_S^{T+}  W^\e (B_l, -\s_{l-},s) dN_l.
 \enn
 By making use of the Markov property of $\xi_t$ we get
 \eqnn
 \lefteqn{
 (S^\e_{t,r}S^\e_{s,l}G)(x,\s) \non} \\
 &&
 = e^{s+t} \EE \left[e^{-\int_0^{t}  V(B_u) du}
 e^{\XXX{0,t}{r}}{\Bbb{E}}^{B_{t},\s_t}
 \left[e^{-\int_0^s V(B_u) du} e^{ \XXX{0,s}{l}}
 G(\xi_s) \right] \right]\non \\
 &&
 = e^{s+t} \EE\left[e^{-\int_0^{t}V(B_u) du} e^{\XXX{0,t}{r}}
 \EE\left[e^{-\int_s^{s+t}V(B_u) du}e^{ \XXX{t,s+t}{l}}
 G(B_{s+t},\s_{s+t}) \,|\, \Omega_{t}\right]\right]\non \\
 &&
 =
 \label{318}
 e^{s+t}\EE\left[e^{-\int_0^{s+t}V(B_u) du}
 e^{\XXX{0,t}{r}+ \XXX{t,s+t}{l}} G(B_{s+t},\s_{s+t})\right].
 \ennn
Note that for $s_1\leq \cdots\leq s_n$,
 \eq{429}
 \exp\lk \XXX{0,t_1}{s_1} + \XXX{t_1,t_1+t_2}{s_2}
 + \cdots + \XXX{t_1+\cdots+t_{n-1}, t_1+\cdots+t_n}
 {s_n} \rk
 \in E_{[s_1,s_n]}\QQQ.
 \en
For operators $T_j$, $j=1,...,N$, write $\prod_{i=1}^n T_i:=
T_1T_2\cdots T_n$.
 By using the identity $\PF^\e =
 \hf\, \, \dot + \, \,
 \int_{\QSS}^\oplus \wpf(\phi) d\mu (\phi)$,
 we have
 \eqnn
 (F, e^{-t\PF^\e}G)
 &=&
 \left(F, e^{-t(\wpfe\,\dot+\, \hf)}G\right)\non \\
 &=&
 \limn \left(F, \lk
 e^{-(t/n) \wpfe} e^{-(t/n)\hf}\rk ^n G\right)\non \\
 &=&
 \limn \left(J_0F,  \lk \prod_{i=0}^{n-1} J_{it/n}
 e^{-(t/n) \wpfe}
 J_{it/n}^\ast\rk J_t G\right)\non \\
 &=&
 \limn \left(J_0F, \lk \prod_{i=0}^{n-1} E_{it/n}
 S^\e_{t/n,it/n} E_{it/n} \rk
 J_t G\right)\non \\
 &=&
 \limn \left(J_0F,
 \lk\prod_{i=0}^{n-1} S^\e_{t/n,it/n}\rk J_t G\right)\non \\
 &=&
 e^t \limn
 \sx \EE \left[ e^{ -\int_0^t V(B_{r}) dr} \int_{\QEE}d\mue
 \ov{J_0F(x,\s)} e^{\xten}J_t G(\xi_t)\right],\non\\
 &&
 \label{503}
 \ennn
where we applied the Trotter-Kato product formula \cite{km} to the
quadratic form sum in the second line, the equality $J_s^\ast J_t =
e^{-|t-s|\hf}$ in the third, Lemma \ref{26} in the fourth, \kak{429}
and the Markov property of the family of projections $E_{[\cdots]}$
in the fifth, and \kak{318} in the sixth line. Moreover $\xten =
\Xn{1}+\Xn{2}+\Xne{3},$ with
 \eqn
 \Xn{1}
 &:=&
 -ie \sum_{\mu=1}^3 \sum_{i=1}^n \int_{t(i-1)/n}^{ti/n}
 \AAA(j_{t(i-1)/n}\la(\cdot-B_s)) dB_s^\mu\\
 &=&
 -ie \AAA\left(\oplus_{\mu=1}^3 \int_0^t j_{\Delta_n(s)}
 \la(\cdot-B_s)dB_s^\mu\right),\\
 \Xn{2}
 &:=&
 -\sum_{i=1}^n \int_{t(i-1)/n}^{ti/n}\DEE(B_s,\s_s,t(i-1)/n)ds
 = -\int_0^t \DEE (B_s,\s_s,\Delta_n(s))ds,\\
 \Xne{3}
 &:=&
 \sum_{i=1}^n \int_{t(i-1)/n}^{ti/n+} W^\e
 (B_s, -\s_{s-}, {t(i-1)/n}) dN_s
  = \int_0^t W^\e (B_s,-\s_{s-},{\Delta_n(s)}) dN_s,
 \enn
and with $W^\e(x,-\s,r) $ defined in \kak{28} and step function
$\Delta_n(s)$ given by \kak{chi}. Furthermore, put
 \eqn
 &&
 \YY{1}:= -ie \AAA\lk \oplus_{\mu=1}^3
  \int_0^t  \las (\cdot-B_s)
 dB_s^\mu\rk,\\
 &&
 \YY{2}:=-\int_0^t \DEE(B_s,\s_s,s)ds,\\
 &&
 \YYE{3}:= \int_0^{t+} W^\e(B_s, -\s_{s-},s)dN_s.
 \enn
Then $\XX=\YY{1}+\YY{2}+\YYE{3}$. We claim that
 \eq{hh2}
 {\rm r.h.s .}\ \kak{503}  =
 e^t \sx \EE \left[e^{-\int_0^tV(B_s) ds} \int_{\QEE} d\mue
 \ov{J_0F(\xi_0)} e^{\XX  } J_tG(\xi_t)\right].
 \en
Note that
 \eqnn
 \lefteqn{
 \sx \EE \left[e^{-\int_0^tV(B_s) ds} \int_{\QEE}
 |J_0F(\xi_0)| \, | J_t G(\xi_t)| \,
 |e^{\xten }-e^{\XX  }|d\mue  \right]\non}\\
 &&\label{325}
  \leq \, \|G\|_{\hhhh} \,
 \EE\left[\lk \sx e^{-2\int_0^tV(B_s) ds} \|F(x,\s)\|_2^2
 \, \|e^{\xten }-e^{\XX  }\|_1^2
 \rk^\han\right]
 \ennn
and
 $$
 \|e^{\xten }\|_1^2 \,\leq\,
 \left(\Ob, |e^{\Xn{2}}|^2 \Ob\right)
 \left(\Ob,|e^{\Xne{3}}|^2\Ob\right).
 $$
We continue by estimating the right-hand side above. It readily
follows that
 \eqnn
 &&
 \hspace{-0.5cm}
 \left(\Ob, e^{2\Xn{2}}\Ob\right)\non \\
 &&
 \hspace{-0.5cm}
 = \exp\lk \frac{e^2}{4} \int _0^t ds\int_0^t dr \s_s \s_{r}
 \int_\BR \frac{|\vp(k)|^2}{\ob(k)} e^{-ik(B_s-B_{r})}
  (|k_1|^2+|k_2|^2) e^{-|\Delta_n(s)-\Delta_n(r)|\ob(k)}dk\rk \non \\
 &&
 \label{326}
 \leq \, \exp\lk \frac{e^2}{4} t^2 \int_\BR |\vp(k)|^2|k|
 dk \rk = c_1,
 \ennn
and the estimate of $\left \|e^{\int_0^t
W^\e(B_s,-\s_{s-},{\Delta_n(s)}) dN_s}\right \|_2^2$ goes as that of
$\left
 \|e^{\int_0^t  W^\e (B_{r},-\s_{r-},s) dN_{r}}\right \|_2^2$
explained in \kak{matui2}, with $\BBB_\mu(j_{s_i} \la
(\cdot-B_{s_i}))$ replaced by $\BBB_\mu(j_{\Delta_n(s_i)} \la
(\cdot-B_{s_i}))$. Then, for each $\bn\in\wo$, $\left \|
e^{\int_0^t W^\e (B_{s}, -\s_{s-}, {\Delta_n(s)}) dN_{s}}\right\|_2^2
\leq c_2(\omega)$, with $c_2(\omega)$ given in \kak{matui2}. Thus we
conclude that $\|e^{\xten}\|_1^2 < c(\bn)$, where $c(\bn)=c_1
c_2(\bn)$ and $\EE [c^\han] < \infty$. Similarly,
$\|e^{\XX}\|_1<C(\bn)$ and $\EE[C^\han] < \infty$ follows for a
random variable $C(\bn)$. Note that both $c$ and $C$ are independent
of $(x,\s)\in \BR\times\zz$, $B_t^\mu $ and $n$. Thus by \kak{325}
and dominated convergence, it suffices to show that for almost every
$\bn\in\wo$, $e^{\xten} \rightarrow e^{\XX}$ as $n\rightarrow
\infty$ in $L^1(\QEE)$. We have
 \eqnn
 e^{\xten }-e^{\XX  }
 &=&
 \underbrace{e^{\Xn{1}}e^{\Xn{2}}e^{\Xne{3}}-
 e^{\YY{1}}e^{\Xn{2}}e^{\Xne{3}}}_{\rm :=I}\non \\
 &&
 +\underbrace{e^{\YY{1}}e^{\Xn{2}}e^{\Xne{3}}-
 e^{\YY{1}}e^{\YY{2}}e^{\Xne{3}}}_{\rm :=II}\non\\
 &&
 \label{tohoku}
 + \underbrace{e^{\YY{1}}e^{\YY{2}}e^{\Xne{3}}-
 e^{\YY{1}}e^{\YY{2}}e^{\YY{3}}}_{\rm :=III}.
 \ennn
We estimate $\rm I,II$ and $\rm III$. %First we estimate $I$.
Notice that
  \eq{328}
 \|{\rm I}\|_1
 \leq
 \|e^{\Xn{1}}-e^{\YY{1}}\|_2 \,
 \|e^{\Xn{2}}e^{\Xne{3}}\|_2,
 \en
By a minor modification of \kak{matui} and \kak{matui2} it is seen
that there is $N=N(\omega)$ such that
 \eqnn
 \label{c3}
 \|e^{\Xn{2}}e^{\Xne{3}}\|_2^2
 &\leq&
 \||e^{\Xn{2}}|^2\|_2 \| |e^{\Xne{3}}|^2\|_2 \\
 &\leq &
 e^{4(e/2)^2  t^2 \|\sqrt {|k|} \vp\|^2}
 \underbrace{
 \left(\frac{|e|}{\sqrt 2}\right)^{4N} \, \sum_{m=0}^ {2N}
 \e^{2N-m} \,m!\,  2^{2m}\, \|\sqrt{|k|}\vp\|^{2m}}_{:=c_3}.
 \non
 \ennn
By the expression of $\YY{1}$ in Definition \ref{hiro}
$$
\left(e^{\Xn{1}},e^{\YY{1}}\right)_2 = \exp\lk
 - \frac{e^2}{2} q_1(\varrho_1^n,\varrho_1^n)\rk,
$$
with $\d \varrho_1^n = \oplus_{\mu=1}^3 \int_0^t (j_{\Delta_n(s)}
\la(\cdot-B_s)-j_s\la(\cdot-B_s))dB_s^\mu $. Moreover,
 \eqn \EE \left[q_1 ( \varrho_1^n,\varrho_1^n)\right]
 &\leq &
 \frac{3}{2}\EE \left[\int_0^t \| j_{\Delta_n(s)}
 \la(\cdot-B_s)-j_s\la(\cdot-B_s)\|^2 ds \right]\\
 &\leq&
 \frac{3}{2}\EE \left[\int_0^t \left(2\|\la\|^2 -
 2\Re(\la(\cdot-B_s), e^{-|\Delta_n(s)-s|\ob} \la (\cdot-B_s))
 \right) ds\right] \rightarrow 0
 \enn
as $n\rightarrow 0$. This implies that there exists a subsequence
$m$ such that for almost every $\omega\in \Omega$,
$\lim_{m\to\infty} (e^{Y_t^{m}(1)},e^{\YY{1}})_2 = 1$ and thus
$\|e^{Y_t^{m}(1)}-e^{\YY{1}}\|_2 \rightarrow 0$. We relabel this
subsequence by $n$. Then
 \eq{har1}
 \limn \|{\rm I}\|_1=0
 \en
follows by \kak{328} for almost every $\omega\in\Omega$.

Next we estimate $\rm II$. Since $|e^{\YY{1}}|=1$, we have
$$
 \|{\rm  II}\|_1 \leq  \|e^{\Xn{2}}-e^{\YY{2}}\|_2 \,
 \|e^{\Xne{3}}\|_2
$$
and $\|e^{\Xne{3}}\|_2\leq c_3(\omega)$, see \kak{c3}. A direct
computation yields
 \eqn
 \lefteqn{
 \|e^{\Xn{2}}\|_2^2} \\
 &&
 \hspace{-1cm}
 =\exp\lk \left(\frac{e}{2}\right)^2 \int_0^t ds \int_0^t ds
 \s_s \s_{r} \int dk \frac{|\vp(k)|^2}{\ob(k)} e^{-ik(B_s-B_{r})}
 (|k_1|^2+|k_2|^2) e^{-|\Delta_n(s)-\Delta_n(r)|\ob(k)} \rk\\
 &&
 \hspace{-0.5cm}
 \rightarrow \exp\lk \left(\frac{e}{2}\right)^2 \int_0^t ds
 \int_0^t dr \s_s\s_{r}\int dk \frac{|\vp(k)|^2}{\ob(k)}
 e^{-ik(B_s-B_{r})} (|k_1|^2+|k_2|^2)  e^{-|s-r|\ob(k)}\rk\\
 &&
 \hspace{-0.5cm}
 = \|e^{\YY{2}}\|_2^2
 \enn
and
 \eqn
 \lefteqn{
 (e^{\Xn{2}}, e^{\YY{2}})_2 }\\
 &&
 =
 \exp\lk \frac{1}{4} \left(\frac{e}{2}\right)^2
 \int_0^t ds \int_0^t dr \s_s\s_{r}\int dk
 \frac{|\vp(k)|^2}{\ob(k)} e^{-ik\cdot (B_s-B_{r})}
 (|k_1|^2+|k_2|^2) \right.\\
 &&
 \left.
 \frac{}{}\hspace{1.5cm}\times
 \lk
 e^{-|s-r|\ob(k)}+e^{-|s-\Delta_n(r)|\ob(k)}+
 e^{-|r-\Delta_n(s)|\ob(k)}
 +e^{-|\Delta_n(s)-\Delta_n(r)|\ob(k)}\rk
  \rk\\
 &&
 \rightarrow \exp\lk
 \left(\frac{e}{2}\right)^2 \int_0^t ds \int_0^t dr \s_s\s_{r}
 \int dk \frac{|\vp(k)|^2}{\ob(k)} e^{-ik\cdot (B_s-B_{r})}
 (|k_1|^2+|k_2|^2) e^{-|s-r|\ob(k)} \rk\\
 &&
 = \|e^{\YY{2}}\|_2^2
 \enn
as $n\rightarrow \infty$. Thus
 \eq{har2}
 \limn \|{\rm II}\|_1^2 \leq  \limn\lk \|e^{\Xn{2}}\|_2^2
 - 2\Re(e^{\Xn{2}}, e^{\YY{2}})_2 + \|e^{\YY{2}}\|_2^2\rk c_3^2  = 0
 \en
is obtained.

Finally, we deal with {\rm III}. Since
$$
\|e^{\YY{1}}e^{\YY{2}} e^{\Xne{3}} -
e^{\YY{1}}e^{\YY{2}}e^{\YYE{3}}\|_1 \leq \|e^{\YY{2}}\|_2 \,
\|e^{\Xne{3}} - e^{\YYE{3}}\|_2
$$
and $\|e^{\YY{2}}\|_2^2 \, \leq \, e^{4 ({e}/{2})  t^2 \|\sqrt{|k|}
\vp\|^2}$, it is enough to show that $e^{\Xne{3}}\rightarrow
 e^{\YYE{3}}$ in $\QQQ$. By the definition of $\Xne{3}$ we have
$$
e^{\Xne{3}}=\prod_{i=1}^n  \exp\lk {\int_{t(i-1)/n}^{ti/n+}
W^\e(B_s,-\s_{s-}, {t(i-1)/n})dN_s}\rk.
$$
For each $\bn\in\wo$ there exists $N=N(\bn)\in{\Bbb N}$ such that
$D(p)=\{ s_1, ..., s_N\}$, where $p$ is the point process defining
the counting measure $N_t$, see \kak{count}. For sufficiently large
$n$ the number of $s_k$ contained in the interval $(t(i-1)/n, ti/n]$
is at most one. Then by taking $n$ large enough and putting
$(n(s_i), n(s_i)+t/n]$ for the interval containing $s_i$,
$i=1,...,N$, we get
 \eq{ka4}
 e^{\Xne{3}}= \prod_{i=1}^N \lk -
 \ODEE(B_{s_i},-\s_{s_i-}, n(s_i)) - \e\gh
 (\ODEE(B_{s_i},-\s_{s_i-}, n(s_i)) \rk.
 \en
Clearly, $n(s_i)\rightarrow s_i$ as $n\rightarrow \infty$. We want
to show that
 \eq{ka1}
 \limm {\rm r.h.s.}\ \kak{ka4}  = \prod_{i=1}^N \lk
 -\ODEE(B_{s_i},-\s_{s_i-},s_i) -\e \gh(\ODEE(B_{s_i},-\s_{s_i-},s_i))\rk.
 \en
Since $\ODEE(B_{s_i},-\s_{s_i-},n(s_i))$ converges strongly to
$\ODEE(B_{s_i},-\s_{s_i-},s_i)$ as $n\rightarrow \infty$ in $\QQQ$,
we have by Lemma \ref{kawaa} below that in $\QQQ$
 \eq{kawa}
 \lim_{n\rightarrow 0}
 \gh(\ODEE(B_{s_i},-\s_{s_i-},n(s_i)))=
 \gh(\ODEE(B_{s_i},-\s_{s_i-},s_i)).
 \en
Set
$\iii (n,i) := \gh(\ODEE(B_{s_i}, -\s_{s_i-},n(s_i)))$, $\iii
(\infty,i) := \gh(\ODEE(B_{s_i}, -\s_{s_i-},s_i))$,
 $A(n,i) :=
\ODEE(B_{s_i}, -\s_{s_i-},n(s_i))$ and $A(\infty,i) :=
\ODEE(B_{s_i}, -\s_{s_i-},s_i)$. Since these are commutative as
operators, the right  hand side of \kak{ka4} can be expanded as a
finite sum of functions of the form $\displaystyle C(n):=
\prod_{k}\iii (n,\#) \prod_{N-k} A(n,\#)$, where $\#$ stands for one
of $1,...,N$. It suffices to show that each $C(n)$ converges to
$C(\infty)$ as $n\rightarrow \infty$ in $\QQQ$, where $C(\infty)$ is
$C(n)$ with $n(s_i)$ replaced by $s_i$, $i=1,...,N$. Take, for
example $C_0(n):= \iii (n,1)\cdots \iii (n,k) A(n,k+1)\cdots
A(n,N)$. Then
 \eqnn
 \label{ka2}
 \lefteqn{
 C_0(n)-C_0(\infty) =} \label{ka3} \\
 &&
 \iii (n,1)\cdots \iii (n,k) \left(A(n,k+1)\cdots A(n,N)
 -A(\infty,k+1) \cdots A(\infty,N)\right) \non \\
 &&\non
 + \left( \iii (n,1)\cdots \iii (n,k)
 - \iii (\infty,1)\cdots \iii (\infty,k) \right)
 A(\infty,k+1)\cdots A(\infty,N).
 \ennn
Since $\iii (n,i)$ is uniformly bounded in $n$, the first term at
the right hand side of \kak{ka3} goes to zero as $n\rightarrow
\infty$ in $\QQQ$. The second term can be estimated in this way.
First note that
 \eqn
 \lefteqn{
 \|\lk \iii (n,i) - \iii (\infty,i) \rk
 A(\infty,k+1)\cdots A(\infty,N)\|_2^2 =} \\
 &&
 \lk
 A(\infty,k+1)^2\cdots A(\infty,N)^2,  \iii (n,i)-\iii (\infty,i)
 \rk_2.
 \enn
Since $\limn \|(\iii (n,i)-\iii (\infty,i))^2\| = \limn \|\iii
(n,i)-\iii (\infty,i)\| = 0$ by \kak{kawa}, the second term of the
right hand side of \kak{ka2} also converges to zero. Then
$C_0(n)\rightarrow C_0(\infty)$ as $n\rightarrow \infty$ in $\QQQ$
follows, and hence \kak{ka1}. Since the right-hand side of \kak{ka1}
equals $e^{\YYE{3}}$, it is seen that $\limn \|e^{\Xne{3}} -
e^{\YYE{3}}\|_2 = 0$, and
 \eq{har3}
 \limn\|\rm III\|_1 = 0.
 \en
A combination of \kak{har1}, \kak{har2} and \kak{har3} implies
\kak{hh2}, and thus \kak{hhmain}.

Now we extend \kak{hmain} to form factors for which $\sqrt\ob \vp$,
$\vp/\sqrt\ob\in \LR$, through a limiting argument. Let $\vp_m\in
\crr$ satisfy $\vp_m/\sqrt\ob\rightarrow \vp/\sqrt\ob$ and $\sqrt\ob
\vp_m \rightarrow \sqrt\ob \vp$ strongly in $\LR$ as $m\rightarrow
\infty$. For each $\vp_m$, \kak{hmain} holds. Let ${\PF^\e }(m)$ be
$\PF^\e$ with $\vp$ replaced by $\vp_m$. Thus ${\PF^\e}(m)\rightarrow
\PF^\e$ as $m\rightarrow \infty$ on the common core $\hhhhh$. Then
$e^{-t{\PF^\e}(m)}\rightarrow e^{-t\PF^\e}$ strongly in $\hhhh$ as
$m\rightarrow \infty$. Define $\xtem$, $\YYm{1}$, $\YYm{2}$ and
$\YYm{3,\e}$ by $\XX$, $\YY{1}$, $\YY{2}$ and $\YY{3,\e}$ with $\vp$
replaced by $\vp_m$, respectively. It is enough to see that
$e^{\xtem}\rightarrow e^{\XX }$ in $L^1(\QEE)$. We divide
$e^{\xtem}-e^{\XX  }$ in the same way as \kak{tohoku} with $\Xn{i}$
replaced by $\YYm{i}$. Then it suffices to show that $e^{\YYm{i}}
\rightarrow e^{\YY{i}}$ strongly in $\QQQ$, for almost every $\omega
\in\Omega$ as $m\rightarrow\infty$. First, we have
$$
(e^{\YYm{1}},e^{\YY{1}})_2 = \exp\lk -\frac{e^2}{2}
q_1(\varrho_2^m,\varrho_2^m)\rk,
$$
where $\d \varrho_2^m=\oplus_{\mu=1}^3 \int_0^t (j_s \la_m
(\cdot-B_s) - j_s\la(\cdot-B_s))dB_s^\mu$
 and $\la_m=(\vp_m/\sqrt\ob)^\vee$.
Furthermore,
 \eqn
 \EE [q_1\lk \varrho_2^m,\varrho_2^m\rk ]
 &\leq&
 \frac{3}{2}\, \EE\left[\int_0^t \|j_s \la_m(\cdot-B_s)-j_s
 \la(\cdot-B_s)\|^2 ds \right]\\
 &\leq&
 \frac{3}{2} \, \|\vp_m/\sqrt\ob-\vp/\sqrt\ob\|\rightarrow 0
 \enn
as $m\rightarrow \infty$. Then there is a subsequence $l$ such that
$(e^{Y_t^{(l)}(1)},e^{\YY{1}})_2\rightarrow 1$ as $l \rightarrow
\infty$ for almost every $\omega\in \Omega$, and hence
 \eq{har4}
 \lim_{l\rightarrow \infty}
 \|e^{Y_t^{(l)}(1)}-e^{\YY{1}}\|_2=0.
 \en
We relabel $l$ as $m$ again. Secondly, we have
  \eqn
  \lefteqn{
  \|e^{\YYm{2}}\|_2^2} \\
  &&
  =\exp\lk \left(\frac{e}{2}\right)^2  \int_0^t ds
  \int_0^t dr \s_s\s_{r}\int dk \frac{|\vp_m(k)|^2}{\ob(k)}
  e^{-ik\cdot (B_s-B_{r})}(|k_1|^2+|k_2|^2) e^{-|s-r|\ob(k)}
  \rk,
  \enn
  \eqn
  \lefteqn{
  (e^{\YYm{2}}, e^{\YY{2}})_2} \\
  &&
  = \exp\lk\frac{1}{4}\left(\frac{e}{2}\right)^2 \int_0^t ds
  \int_0^t dr \s_s \s_{r}\int_\BR dk
  \frac{|\vp(k)+\vp_m(k)|^2}{\ob(k)} e^{-ik\cdot (B_s-B_{r})}
  \right. \\
  && \left. \hspace{9.2cm}
  \frac{}{}\times (|k_1|^2+|k_2|^2) e^{-|s-r|\ob(k)}\rk.
  \enn
From here
  \eq{har5}
  \limm \|e^{\YYm{2}}-e^{\YY{2}}\|_2 ^2 =
  \limm \left(\|e^{\YYm{2}}\|_2^2 -2\Re(e^{\YYm{1}}, e^{\YY{1}})_2
  +\|e^{\YY{2}}\|_2^2\right) = 0
  \en
follows. Finally we see that for each $\bn\in \wo$,
$e^{\YYm{3,\e}}\Ob \rightarrow e^{\YYE{3}}\Ob $ as $m\rightarrow
\infty$ in $\QQQ$. There exists $N=N(\bn)\in{\Bbb N}$,
$s_1=s_1(\bn),...,s_N(\bn)\in (0,\infty)$ such that
$$
e^{\YYm{3,\e}} = \prod_{i=1}^N \lk -\ODEE(B_{s_i},-\s_{s_i-},s_i, m)
-\e \gh\lk \ODEE(B_{s_i},-\s_{s_i-},s_i, m)\rk \rk ,
$$
where $\ODEE(B_{s_i},-\s_{s_i-},s_i, m)$ is  defined by
$\ODEE(B_{s_i},-\s_{s_i-},s_i)$ with $\vp$ replaced by $\vp_m$.
Since $\ODEE(B_{s_i},-\s_{s_i-},s_i, m)$ converges strongly to
$\ODEE(B_{s_i},-\s_{s_i-},s_i)$ as $m\rightarrow 0$ in $\QQQ$, by
Lemma \ref{kawaa} we obtain
 \eq{446}
 \lim_{m\rightarrow 0} \gh(\ODEE(B_{s_i},-\s_{s_i-},s_i, m)) =
 \gh(\ODEE(B_{s_i},-\s_{s_i-},s_i))
 \en
in $\QQQ$. Similarly to the proof of $\limn e^{\Xne{3}} =
e^{\YYE{3}}$, we argue that
 \eq{har6}
 \lim_{m\rightarrow \infty} \|e^{\YYm{3,\e}}-e^{\YYE{3}}\|_2 = 0.
 \en
From \kak{har4}, \kak{har5} and \kak{har6} we finally obtain
\kak{hh2}, completing the proof.
  \qed

It remains to show \kak{kawa} and \kak{446}.
 \bl{kawaa}
We have
 \eqnn
 &&
 \label{ka6}
 \limn \gh(\ODEE(B_{s_i},-\s_{s_i-},n(s_i)))
 = \gh(\ODEE(B_{s_i},-\s_{s_i-},s_i))\\
 &&
 \label{ka7}
 \lim_{m\rightarrow 0} \gh(\ODEE(B_{s_i},-\s_{s_i-},s_i,m)) =
 \gh(\ODEE(B_{s_i},-\s_{s_i-},s_i))
 \ennn
strongly in $\QQQ$.
 \el
 \proof
We show \kak{ka7}, the proof of \kak{ka6} is similar.
 Put $\eta_m
 = \ODEE(B_{s_i},-\s_{s_i-},s_i,m)$ and $\eta =
\ODEE(B_{s_i},-\s_{s_i-},s_i)$. Let $g_n\in\SSS(\RR)$ be such
that $g_n\rightarrow \gh$ as $n\rightarrow\infty$ in $L^2(\RR)$. We
have
$$
\|\gh(\eta)-\gh(\eta_m)\| \,\leq\,
\|\gh(\eta)-g_n(\eta)\|+\|g_n(\eta)-g_n(\eta_m)\| +
\|g_n(\eta_m)-\gh(\eta_m)\|.
$$
It is readily seen that
 \eq{kop}
 \|\gh(\eta)-g_n(\eta)\|^2 \leq  \int|\gh(x)-g_n(x)|^2
 (2\pi\rho)^{-\han}dx
 \en
and
 \eq{kopp}
 \|g_n (\eta_m)-\gh (\eta_m)\|^2\leq \int|\gh(x)-g_n(x)|^2
 (2\pi\rho_m)^{-\han} dx,
 \en
where $\rho$ is given by \kak{mean} and $\rho_m$ is obtained by
replacing $\vp$ by $\vp_m$. Since $\rho_m\rightarrow \rho$ as
$m\rightarrow 0$, the left hand sides of \kak{kop} and \kak{kopp}
are bounded by $C\|\gh-g_n\|^2$ with some constant $C$ independent
of $m$. Consequently, they both converge to zero uniformly in $m$.
We also see that
 \eq{koppp}
 \|g_n(\eta)-g_n(\eta_m)\|\leq (2\pi)^{-\han}\int_\RR |\hat g_n(k)|
 \|e^{ix\eta}-e^{ix\eta_m}\|dx.
 \en
Since $\|e^{ix\eta}-e^{ix\eta_m}\|\rightarrow  0$ as $m\rightarrow
0$ for each $n$, the left hand side of \kak{koppp} converges to zero
as $m\rightarrow 0$. This gives the lemma.
 \qed

\subsection{Energy comparison inequality}

Write
$$
\is(\PF) = E(\AA ,\BB _1,\BB _2,\BB _3)
$$
for the bottom of the spectrum of $\PF$. Then for the spinless
Pauli-Fierz Hamiltonian $\PFs$ we have $\is(\PFs) = E(\AA ,0,0,0)$
and the diamagnetic inequality $E(0,0,0,0) \,\leq\, E(\AA,0,0,0)$
is well-known to hold \cite{ahs, h4}. In this subsection we extend
this inequality to the case of the Hamiltonian with spin.

Define
 \eq{hir2}
 \PFp:=
 \hp+\hf - \mmm
 {\frac{e}{2}\BB_3}{\frac{|e|}{2} \sqrt{\BB_1^2+\BB_2^2}}
 {\frac{|e|}{2}\sqrt{\BB_1^2+\BB_2^2}}{- \frac{e}{2}\BB_3}.
 \en
Furthermore, to avoid zeroes of the off-diagonal part to occur we
also define
 \eq{sp}
 \PF^{\perp\e} := \PFp -\mmm{0}
 {\e\gh\lk  \frac{|e|}{2} \sqrt{\BB_1^2 +\BB_2^2} \rk }
 {\e\gh\lk  \frac{|e|}{2} \sqrt{\BB_1^2 +\BB_2^2} \rk }
 {0}.
 \en
Since the spin interaction is infinitesimally small with respect to
the free Hamiltonian $\hp+\hf$, $\PFp$ and $\PF^{\perp\e}$ are
self-adjoint on $D(-\Delta)\cap D(\hf)$ and bounded from below. Note
that $|\OD|=\frac{|e|}{2} \sqrt{\BB_1^2+\BB_2^2}$ and $\gh(\OD) =
\gh(|\OD|) = \gh (\frac{|e|}{2} \sqrt{\BB_1^2+\BB_2^2})$. The
functional integral representation of $e^{-t\PFp}$ is given by
 \eqn
 \lefteqn{
 (F, e^{-t\PFp} G) = \lime (F, e^{-t\PF^{\perp\e}} G)} \\
 &&
 = \lime \sx\EE \left[e^{-\int_0^t V(B_s) ds} \int_{\QEE} d\mue
 \ov{J_0 F(\xi_0)} e^{\XP} J_t G(\xi_t)\right],
 \enn
where
 \eqn
  &&
  \XP = -\int_0^t \D (B_s,\s_s,s) ds \\
  && \hspace{3cm} +
  \int_0^{t+} \log \left[|\ODEE(B_s,-\s_{s-},s)|+
  \e\gh(|\ODEE(B_s,-\s_{s-},s)|  )\right] dN_s.
  \enn
  \bc{rotation}
For all $t \geq 0$ and  $F,G\in\hhhh$ we have
  \eq{yuj}
  |(F,e^{-t\PF}G)| \,\leq\,
  \left(|F|, e^{-t\PFp}|G|\right)
  \en
and
  \eq{max}
  \max\lkk\begin{array}{l}E(0,\sqrt{\BB _1^2+\BB _2^2},0,\BB _3)\\
  E(0,\sqrt{\BB _3^2+\BB _1^2},0,\BB _2)\\
  E(0,\sqrt{\BB _2^2+\BB _3^2},0,\BB _1)
  \end{array}
  \rkk\leq E(\AA ,\BB _1,\BB _2,\BB _3).
  \en \ec
  \proof
Since $\PFp$ is unitary equivalent with the Hamiltonian obtained on
replacing $e$ by $-e$, we may assume that $e>0$ without loss of
generality. By the functional integral representation of $e^{-t\PF}$
we have
 \eqnn
 \lefteqn{
 |(F, e^{-t\PF} G)| =\lime|(F, e^{-t \PF^\e}G)|\non}\\
 &&
 \leq
 \lime \sx \EE \left[e^{-\int_0^t V(B_s)ds}\int_{\QEE} d\mue
 |J_0 F(\xi_0)| |J_t G(\xi_t)| e^{\XP}\right] \non \\
 &&
 \leq
 \lime \sx \EE \left[e^{-\int_0^t V(B_s)ds}\int_{\QEE} d\mue
 (J_0|F(\xi_0)|) (J_t |G(\xi_t)|) e^{\XP}\right],\non\\
 &&
 =
 \lime (|F|, e^{-t\PF^{\perp\e}}|G|)=(|F|, e^{-t\PF^\perp}|G|),\non
 \label{kuro}
 \ennn
where we used $|e^{\XX}|\leq e^{\XP}$ and the fact that $|J_t G|\leq
J_t |G|$ as $J_t$ is positivity preserving.  Thus \kak{yuj} follows.
From this, $E(0,\sqrt{\BB _1^2+\BB _2^2},0,\BB _3)\leq E(\AA , \BB
_1,\BB _2,\BB _3)$ is obtained. Since $ E(\AA ,\BB _1,\BB _2,\BB _3)
= E(\AA ,\BB _3,\BB _1,\BB _2) = E(\AA , \BB _2,\BB _3,\BB _1) $ by
symmetry, \kak{max} follows.
 \qed

\section{Translation invariant Hamiltonians}
In this  section we assume that $V=0$. In the previous section we
derived the functional integral representation of $e^{-t\PF}$ and
$e^{-t\PF^\e}$. By using them we can construct the functional
integral representation of the translation invariant Hamiltonian
$$
 \PF(P) = \half (P-\pf-e\AA(0))^2+\hf-\frac{e}{2} \muu
 \s_\mu  \BB_\mu (0).
$$
Before going to do this, we show translation invariance of the
operator $\PF^\e$ defined in \kak{this}.

 \bl{tra1}
$\PF^\e$ is translation invariant and it follows that
 $$
\PF^\e = \int_\BR^\oplus \PF^\e(P) dP,
 $$
where
 \eq{toutou}
 \PF^\e(P)=\PF(P)
 + \MMM 0 {\e \gh(-\frac{e}{2}(\BB_1(0)-i\BB_2(0)))}
 {\e\gh(-\frac{e}{2}(\BB_1(0)+i\BB_2(0)))} 0.
 \en
 \el
 \proof
Let $\Phi=\Phi(x)=(-e/2)(\BB_1(\la(\cdot-x))-i\BB_2(\la(\cdot-x)))$. Note
that
  $$
  \PF^\e = \PF +\MMM 0 {\e \gh(\Phi)}
  {\e\gh(\bar \Phi)} 0,
  $$
where $\bar\Phi$ denotes the complex conjugate of $\Phi$. The term
$\PF$ is translation invariant, therefore we only show that so is
$\gh(\Phi)$. We already know that there exists $\gh^n\in\SSS(\RR)$
such that $\gh^n(\Phi)\rightarrow \gh(\Phi)$ strongly as a
bounded multiplication operator when $n \rightarrow \infty$, where
$\gh^n(\Phi) = (2\pi)^{-\han}\int_\RR \hat \gh^n(k) e^{ik\Phi} dk$.
Thus $\gh^n$ is translation invariant, since $\Phi$ is. Hence
$\gh(\Phi)$ is also a translation invariant bounded multiplication
operator. The proof for $\gh(\bar\Phi)$ is similar.

Furthermore, $\PF+\gh^n(\Phi)$ is decomposed as
 $$
 \PF+ \mmm 0 {\gh^n(\Phi)}  { \gh^n(\bar \Phi)} 0 =
 \int_\BR^\oplus \left( \PF(P)+\mmm 0 {\e \gh^n(\Phi(0))}
 { \e \gh^n(\bar \Phi(0))} 0  \right)  dP.
 $$
Since $\gh^n(\Phi(0))$ and $\gh^n(\bar\Phi(0))$ converge strongly to
$\gh(\Phi(0))$ and $\gh(\bar \Phi(0))$, respectively, \kak{toutou}
follows.
 \qed

 \bt{main2}
For $t\geq 0$ and $\Phi,\Psi\in\zz\otimes\QQ$ we have
 \eq{yu3}
  (\Phi, e^{-t\PF^\e (P)}\Psi)%_{\CC^2\otimes \fff}
  = e^t \sum_{\s\in\zz} {\Bbb E}^{0,\s}\left[
  e^{iP\cdot B_t} \int_{\QEE}d\mue  \ov{J_0\Phi(\s)}
  e^{\XX  } J_t e^{-i  \pf\cdot B_t } \Psi(\s_t) \right]
  \en
and
  \eq{yu}
  (\Phi, e^{-t\PF(P)} \Psi)%_{\CC^2\otimes \fff} =
  =\lime e^t \sum_{\s\in\zz} {\Bbb E}^{0,\s}
  \left[ e^{iP\cdot B_t} \int_{\QEE} \ov{J_0\Phi(\s)}
  e^{\XX} J_t e^{-i\pf\cdot B_t}\Psi(\s_t) d\mue \right] .
  \en
  \et
  \proof
It suffices to show \kak{yu3}. The idea of proof is similar to that
of Theorem~3.3 in \cite{h27}. Set $ F_s(\s) =\rho_s   \otimes
\Phi(\s)$ and $G_{r}(\s) = \rho_{r} \otimes \Psi(\s)$, where
$\rho_s(x) = (2\pi s)^{-3/2} \exp(-|x|^2/(2s))$, $s>0$, is the heat
kernel, and $\Phi(\s),\Psi(\s) \in  \qqf$. We have by  Lemma
\ref{tra1}, for $\xi\in\BR$,
$$
(F_s, e^{-t\PF^\e}e^{-i\xi\cdot\tot} G_{r})_{\hhhh} = \int_\BR dP
((UF_s)(P), e^{-t\PF^\e(P)}e^{-i\xi\cdot P} (UG_{r})(P))_{\zz
\otimes \fff},
$$
where the unitary operator $U: \hhhh \rightarrow \hhhh$ is defined
by
$$
(UF_s)(P) = (2\pi)^{-3/2}\int_\BR e^{-ix\cdot P} e^{ix\cdot
\pf}\rho_s(x)\Psi(\s) dx.
$$
Hence we have
 \eq{suki}
 \lim_{s\rightarrow 0} (F_s,e^{-t\PF^\e }e^{-i\xi\cdot\tot}G_{r})_
 {\hhhh} = (2\pi)^{-3/2} \int_{\BR}dP (\Psi, e^{-t\PF^\e(P)}
 e^{-i\xi\cdot P}(UG_{r})(P))_{\zz \otimes \fff}.
 \en
On the other hand, we have through the functional integral
representation \kak{hmain},
$$
(F_s, e^{-t\PF^\e } e^{-i\xi\cdot \tot}G_{r})_{\hhhh}
 = \int_{\RR^3} \rho_s(x) \Upsilon(x) dx,
$$
where
$$
\Upsilon(x) = \sum_\s \EE \left[ \rho_{r}(B_t-\xi) \int_{\QEE}
\ov{J_0\Psi(\s)} e^{\XX} J_t e^{-i\xi\cdot \pf}\Phi(\s_t) d\mue
\right].
$$
In Lemma \ref{saigonosaigo} below we show that $\Upsilon$ is bounded
and is continuous at $x=0$. Thus further we obtain that
$$
\lim_{s\rightarrow 0} \int_{\RR^3} \rho_s(x)\Upsilon(x) dx = \Upsilon(0)=
\sum_\s {\Bbb E}^{0,\s} \left[\rho_{r}(B_t-\xi) \int_{\QEE}
\ov{J_0\Psi(\s)} e^{\XX} J_t e^{-i\xi\cdot \pf}\Phi(\s_t) d\mue
\right].
$$
Hence, together with \kak{suki} we have
  \eqnn
  \lefteqn{
  (2\pi)^{-3/2}\int_\BR dP e^{-i\xi\cdot P}
  (\Psi,e^{-t\PF^\e (P)}(UG_{r})(P))_{\zz \otimes \fff}\non }\\
  && \label{las}
  = \sum_{\s\in\zz} {\Bbb E}^{0,\s}
  [\rho_{r}(B_t-\xi)\ov{J_0\Psi(\s)} e^{\XX} J_t e^{-i\xi\cdot \pf}
  \Phi(\s_t) ].
  \ennn
Since $(\Psi, e^{-t\PF^\e (\cdot)} (UG_{r})(\cdot ))_{\zz \otimes
\fff}\in \LR$, by taking inverse Fourier transform on both sides of
\kak{las} we arrive at
  \eqnn
  \lefteqn{
  \hspace{-1cm}
  \left(\Psi,e^{-t\PF^\e (P)}(U G_{r})(P)\right)_
  {\zz \otimes \fff}\label{las2}}\\
  &&
  \hspace{-1cm} \non
  = (2\pi)^{-3/2}
  \sum_{\s\in \zz}{\Bbb E}^{0,\s}\left[\int _\BR d\xi
  e^{i\xi\cdot P}\rho_{r}(B_t-\xi)\int_{\QEE} \ov{J_0\Psi(\s)}
  e^{\XX} J_t e^{-i\xi\cdot \pf}\Phi(\s_t) d\mue \right]
  \ennn
for almost every $P\in \BR$. Since both sides of \kak{las2} are
continuous in $P$, the equality holds for all $P\in\BR$. Taking $r
\rightarrow 0$ on both sides of \kak{las2}, we get the desired
result.
  \qed
We conclude by showing the lemma used above.
 \bl{saigonosaigo}
 $\Upsilon$ is bounded and is continuous at $x=0$.
 \el
 \proof
The boundedness is trivial, we proceed to show continuity. We
have
 \eq{up}
 |\Upsilon(x)-\Upsilon(0)|\leq \sum_\s {\Bbb E}^{0,\s}
 \left[\|\Psi(\s)\|_2 \|\Phi(\s_t)\|_2\|
 e^{\XXx}-e^{Z_t^0(\e)}\|_1\right],
 \en
with
 \eqn
 &&
 \XXx
 =
 \underbrace{-ie \sum_{\mu=1}^3 \int_0^t\AAA_\mu(\las (\cdot-B_s-x))
 dB_s^\mu}_{:=\zzzx{1}}
  \underbrace{-\int_0^t \D(B_s+x,\s_s,s)ds}_{:=\zzzx{2}}\\
  &&
   +\underbrace{\int_0^{t+} \log\left [-\OD(B_s+x,-\s_{s-},s)
   -\e\gh(\OD(B_s+x,\s_{s-},s))\right] dN_s}_{:=\zzzx{3,\e}}.
 \enn
By \kak{up} it is enough to show that
 \eq{exs}
 \lim_{x\rightarrow 0} {\Bbb E}^{0,\s}[\| e^{\XXx} - e^{\zten}\|_1] =
 0,
 \en
similarly to the proof of Theorem \ref{main}. We estimate ${\rm
I,II,III}$ below:
 \eqnn
 e^{\XXx}-e^{\zten}
 &=&
 \underbrace{e^{\zzzx{1}}e^{\zzzx{2}}e^{\zzzx{3,\e}}-
 e^{\zzzz{1}}e^{\zzzx{2}}e^{\zzzx{3,\e}}}_{\rm :=I}\non \\
 &&
 + \underbrace{ e^{\zzzz{1}}e^{\zzzx{2}}e^{\zzzx{3,\e}}-
 e^{\zzzz{0}}e^{\zzzz{2}}e^{\zzzx{3,\e}}}_{\rm :=II}\non\\
 &&
 \label{tttt}
 + \underbrace{ e^{\zzzz{1}}e^{\zzzz{2}}e^{\zzzx{3,\e}}-
 e^{\zzzz{1}}e^{\zzzz{2}}e^{\zzzz{3,\e}}}_{\rm :=III}.
 \ennn
We have $\|e^{\zzzx{2}}e^{\zzzx{3,\e}}\|_2\leq e^{4(e/2)^2
t^2\|\sqrt{|k|}\vp\|^2} c_3(\bn):=c_4(\bn)$, where $c_3(\bn)$ is
given in \kak{c3}, and
$$
\|e^{\zzzx{1}}- e^{\zzzz{1}}\|_2^2=2 -2 \Re(e^{\zzzx{1}},
e^{\zzzz{1}}) = 2 - 2\exp \lk -\frac{e^2}{2}
q_1(\varrho_3^x,\varrho_3^x) \rk ,
$$
where $\d \varrho_3^x=\oplus_{\mu=1}^3\int_0^t j_s(\la(\cdot-B_s-x)
- \la(\cdot-B_s))dB_s^\mu$. Moreover,
$$
\EEEEE[q_1(\varrho_3^x,\varrho_3^x)] \leq \frac{3}{2}\,
\EEEEE\left[\int_0^t\|\la(\cdot-B_s-x)-\la(\cdot-B_s)\|^2 ds\right]
\rightarrow 0
$$
as $x\rightarrow 0$. Thus
 \eqn
 \lim_{x\rightarrow 0}
 \EEEEE \| I\|_1
 &\leq&
  \lim_{x\rightarrow 0}
  \EEEEE \|e^{\zzzx{1}}-e^{\zzzz{1}}\|_2
 \|e^{\zzzx{2}}e^{\zzzx{3,\e}}\|_2\\
 &\leq& \lim_{x\rightarrow 0}
 \EEEEE \|e^{\zzzx{1}}-e^{\zzzz{1}}\|_2
 \EEEEE [c_4^\han]\\
 & \leq&
  \lim_{x\rightarrow 0}
\EEEEE[1-e^{-(e^2/2)
q_1(\varrho_3^x,\varrho_3^x)}] \EEEEE
[c_4^\han]\\
 &\leq& \lim_{x\rightarrow 0}
  \EEEEE[ (e^2/2)
 q_1(\varrho_3^x,\varrho_3^x)]
\EEEEE [c_4^\han]
 = 0. \enn
Next we estimate ${\rm II}$. We have
 \eqn
 \lefteqn{
 (e^{\zzzx{2}}, e^{\zzzz{2}})_2} \\
 &&= \exp\lk \frac{e^2}{2}\int_0^t ds\int_0^tdr\s_s\s_{r}\int dk
 \frac{|\vp(k)|^2}{\ob(k)} e^{-ik(B_s-B_{r}-x)}(|k_1|^2+|k_2|^2)
 e^{-|s-r|\ob(k)}\rk\\
 &&
 \rightarrow \|e^{\zzzz{2}}\|_2^2
 \enn
as $x\rightarrow 0$. Then from $\|e^{\zzzx{2}}- e^{\zzzz{2}}\|_2^2 =
2\|e^{\zzzz{2}}\|_2^2 -2\Re (e^{\zzzx{2}}, e^{\zzzz{2}})\rightarrow
0$ it follows that
$$
\lim_{x\rightarrow 0} \|{\rm II}\|_1^2 \,\leq\, c_3
\lim_{x\rightarrow 0} \|e^{\zzzx{2}}- e^{\zzzz{2}}\|_2^2 =0
$$
for almost every $\bn\in\Omega$. Finally we estimate ${\rm III}$.
For each $\omega\in \Omega$, there  exist $N = N(\bn)\in {\Bbb N}$
and $s_1=s_1(\bn),...,s_N(\bn)\in (0,\infty)$ such that
$$
e^{\zzzx{3,\e}} = \prod_{i=1}^N \lk -\ODEE(x+B_{s_i},-\s_{s_i-},s_i)
-\e \gh\lk \ODEE(x+B_{s_i},-\s_{s_i-},s_i)\rk\rk.
$$
Since $\ODEE(x+B_{s_i},-\s_{s_i-},s_i)$ converges strongly to
$\ODEE(B_{s_i},-\s_{s_i-},s_i)$ as $x\rightarrow 0$ in $\QQQ$, we
see that $\lim_{x\rightarrow 0} \gh(\ODEE(x+B_{s_i}, -
\s_{s_i-},s_i)) = \gh(\ODEE(B_{s_i},-\s_{s_i-},s_i))$ in $\QQQ$.
This can be proven in the same way as Lemma \ref{kawaa}. Hence
 \eqnn
 \lefteqn{
 \lim_{x\rightarrow 0}
  \prod_{i=1}^N \lk -\ODEE(x+B_{s_i},-\s_{s_i-},s_i)-\e
 \gh\lk
 \ODEE(x+B_{s_i},-\s_{s_i-},s_i)\rk  \rk \non} \\
 &&
 \label{sssaigo2}
 = \prod_{i=1}^N \lk -\ODEE(B_{s_i},-\s_{s_i-},s_i)
 -\e \gh \lk \ODEE(B_{s_i},-\s_{s_i-},s_i)\rk \rk
 \ennn
follows. Thus we obtain $\lim_{x\rightarrow 0} \|e^{\zzzx{3,\e}} -
e^{\zzzz{3,\e}}\|_2 =0$ as well as $\lim_{x\rightarrow 0} \|{\rm
III}\|_1\leq \lim_{x\rightarrow 0}
\|e^{\zzzx{3,\e}}-e^{\zzzz{3,\e}}\|_2 \|e^{\zzzz{2}}\|_2=0$ for
almost every $\bn\in \Omega$, proving  \kak{exs}.
  \qed

\medskip
From \kak{yu}, we can derive energy inequalities in a similar manner
to Corollary~\ref{rotation}. Write
$$
\is (\PF(P)) = E(P,\AA , \BB _1, \BB _2, \BB _3),
$$
and define
$$
 \PFp(P)=\half(P-\pf)^2 +\hf - \mmm
 {\frac{e}{2}\BB_3(0)}{\frac{|e|}{2} \sqrt{\BB_1(0)^2+\BB_2(0)^2}}
 {\frac{|e|}{2}\sqrt{\BB_1(0)^2+\BB_2(0)^2}}{- \frac{e}{2}\BB_3(0)}.
$$
 \bc{main3}
For $t\geq 0$
 \eq{hum}
 |(\Phi, e^{-t\PF(P)} \Psi)| \leq
 \left(|\Phi|, e^{-t\PFp(0)}|\Psi|\right)
 \en
and
 \eq{hum2}
 \max\lkk \begin{array}{l} E(0, 0,\sqrt{\BB _1^2+\BB _2^2},0,\BB _3)\\
 E(0, 0,\sqrt{\BB _3^2+\BB _1^2},0,\BB _2) \\
 E(0, 0,\sqrt{\BB _2^2+\BB _3^2},0,\BB _1)
 \end{array}\rkk\leq
 E(P,\AA , \BB _1, \BB _2, \BB _3).\en
 \ec
 \proof
Clearly, $|e^{-i\pf\cdot B_t}\Psi|\leq e^{-i\pf\cdot B_t}|\Psi|$.
Therefore
 \eqn
{
 |(\Phi, e^{-t\PF(P)} \Psi)|}
  &\leq& e^t \lime \sum_{\s\in\zz} {\EE}\left[\int_{\QEE}
 (J_0 |\Phi(\s)|) e^{\XP} (J_t e^{-i\pf\cdot B_t} |\Phi(\s_t)|)
 \right]d\mue   \\
 &&
 ={\rm r.h.s.}\
 \kak{hum}.
 \enn
\kak{hum2} is immediate from \kak{hum}. \qed

\section{Concluding remarks}

It is known that $\PF$ has degenerate ground states for weak enough
couplings \cite{hisp2,h27}. In this subsection we comment on the
breaking of ground state degeneracy of a toy model by using the
functional integral obtained in Theorem \ref{main}.

Consider the self-adjoint operator on $\hhhh$ with the spin
interaction replaced by the fermion harmonic oscillator
\kak{fermion} in $\PF$:
  $$
  H(\epsilon)=\half (-i\nabla-e\AA)^2+V+\hf+\epsilon \s_{\rm F}.
  $$
Whenever $\epsilon=0$, the ground state of $H(0)$ is degenerate at
any coupling. In this case
 \eqn
 (F, e^{-tH(0)}G)
  &=&
  e^t  \lime \sx \EE \left[e^{-\int_0^t V(B_s) ds}
  ({J_0 F(\xi_0)}, e^{-iA} \e^{N_t} J_t G(\xi_t))\right]\\
  &=&
  e^t\sx {\Bbb E}^x \left[e^{-\int_0^t V(B_s) ds}
  ({J_0 F(x,\s)},e^{-iA} J_t G(B_t,\s))\right],
  \enn
where $A = \AAA (\oplus_ {\mu=1}^3 \int_0^t j_s\la(\cdot-B_s)
dB_s^\mu)$.
We show, however, that the ground state of $H(\epsilon)$ becomes
unique for arbitrary values of coupling constants as soon as
$\epsilon\not=0$. Since the fermion harmonic oscillator $\frm$ is
identical to $-\s_1$, the off-diagonal part of $H(\epsilon)$ is the
non-zero constant $-\epsilon$. Then we have the functional integral
representation of $e^{-tH(\epsilon)}$ with the exponent $X_t(0)$ in
\kak{hmain} replaced by
$$
-i e A +\int _0^t \log \epsilon dN_s.
$$
 Thus
$$
 (F, e^{-tH(\epsilon)} G) = e^t \sx \EE [ \epsilon^{N_t}
 e^{-\int_0^t V(B_s) ds}({J_0 F(\xi_0)}, e^{-ie  A} J_t
 G(\xi_t))].
$$
Take the unitary operator $\theta = e^{-i(\pi/2)N}$. In \cite{h10}
it was seen that $T_t:=J_0^\ast \theta \f e^{-iA}\theta J_t$ is
positivity improving. This implies
 \bc{phase}
$\theta\f e^{H(\epsilon)}\theta$ is positivity improving for
$\epsilon>0$ and, in particular, the ground state of $H(\epsilon)$,
$\epsilon \not = 0$, is unique whenever it exists.
 \ec
 \proof
Note that $H(\epsilon)$ and $H(-\epsilon)$ are isomorphic, therefore
we only take $\epsilon>0$. By a direct computation and the definition
of $T_t$, we have
  \eqn
 \lefteqn{
  (F, \theta \f e^{-tH(\epsilon)}\theta  G) }\\
 &&
 = e^t\sx {\Bbb E}^x \left[e^{-\int_0^t V(B_s) ds}\frac{}{} \times
 \right.\\
 &&
 \left. \frac{}{} \times \lk (F(x,\s), T_t  G(B_t,\s)) \cosh \epsilon t +
 ( F(x,\s), T_t G(B_t,-\s))\sinh \epsilon t\rk \right].
 \enn
Then for non-zero $0\leq F,G\in L^2(\BR\times\zz\times\QSS)$ we see
that the right-hand side above is strictly positive, i.e., $(F,
\theta\f e^{-tH(\epsilon)} G)>0$. This means that
$e^{-tH(\epsilon)}$ is positivity improving. The uniqueness of the
ground state follows by an application of the Perron-Frobenius
theorem \cite{gj1,gr}. \qed

The translation invariant version of the model is given by
$$
H(\epsilon, P) := \half(P-\pf-e \AA(0))^2+ \hf + \epsilon\frm.
$$
The ground state of $H(0,P)$ is degenerate, whenever it exists,
however in this case too the degeneracy is broken. By Theorem
\ref{main2}, the functional integral representation of
 $e^{-tH(\epsilon,P)}$ is given by
 \eq{pb}
(\Psi,e^{-tH(\epsilon,P)}\Phi)
 = e^t\sum_{\s\in\zz} {\Bbb E}^{0,\s}
 \left [\epsilon^{N_t} e^{iP\cdot B_t}
 ({J_0\Phi(\s)},  e^{-i A} J_t e^{-i \pf \cdot B_t }
 \Psi(\s_t) )\right].
 \en
If $P=0$, the phase $e^{iP\cdot B_t}$ vanishes. Then, since
$e^{-i\pf \cdot B_t}$ is positivity preserving in
$Q$-representation, similarly to Corollary \ref{phase} we see that
for $P=0$ and $\epsilon>0$, $\theta \f e^{-tH(\epsilon,0)} \theta$
is positivity improving. This yields
 \bc{phase2}
Let $P=0$ and $\epsilon\not=0$. Then $\theta\f e^{-tH(\epsilon,0)}
\theta$ is positivity improving and the ground state of $H(\epsilon,
0)$ is unique, whenever it exists.
 \ec
\begin{remark}
 \rm{
The spin-boson model is defined by
 $$
 H_{\rm SB}= \s_1\otimes 1+1\otimes H_{\rm f}+\alpha\s_3\otimes \phi(f),
 \; \alpha\in\RR,
 $$
on $\CC^2\otimes {\cal F}(\LR)$, where $H_{\rm f}$ is the free field
Hamiltonian of ${\cal F}(\LR)$ and $\phi(f)$ is the field operator
labeled by $f\in \LR$. We can also construct the functional integral
representation of $e^{-tH_{\rm SB}}$ by making use of the
$\zz$-valued jump process $\s_t$. The functional integral can then
be used to prove uniqueness of the ground state whenever it exists
\cite{sp3, hi1,hi2,hihi}.
 }
\end{remark}

\section{Appendix: It\^o formula for L\'evy  processes}

In this appendix we recall and discuss some basic facts on Poisson
processes and related It\^o formulas to make this paper sufficiently
self-contained. A general reference on this subject is
 \cite{iw,dv}.

Let $(\os, \fffs, P_{\rm P})$ be a complete probability space with a
right-continuous increasing family of sub-$\s$-fields
$(\fffs_t)_{t\geq0}$, where each $\fffs_t$ contains all $P_{\rm  P}$-null
sets. Also, let $(\meas ,\BB_\meas )$ be a measurable space and $\varpi$
the set of ${\Bbb Z}_+\cup\{\infty\}$-valued measures on
$(\meas ,\BB_\meas )$. Denote by $\BB_\varpi$ the smallest $\s$-field on
$\varpi$ such that $\varpi\ni \mu\mapsto \mu(B)$, $B\in \BB_\meas $, are
measurable.

We define a class of measure-valued random variables.
\begin{definition}
The $(\varpi, \BB_\varpi)$-valued random variable $N$ on $(\os,
\fffs, P_{\rm P})$ is a \emph{Poisson random measure} on $(\meas,
\BB_\meas)$ whenever the conditions below are satisfied:
\begin{itemize}
\item[(1)]
$P(N(A)=n) = e^{-\Lambda(A)}\Lambda(A)^n/n!$, $A\in\BB_\meas $,
where $\Lambda (A):= \EEEE[N(A)]$,
\item[(2)]
if $A_1,...,A_n\in \BB_\meas $ are pairwise disjoint, then
$N(A_1),..., N(A_n)$ are independent.
\end{itemize}
\end{definition}
$\Lambda (A)$ is called the {\it intensity} of $N(A)$, and
$\EEEE[e^{-\alpha N(A)}]= e^{\Lambda (A)(e^{-\alpha}-1)}$ holds.

Fix a measurable space $({\cal M}, \BB_{\cal M})$. By an ${\cal
M}$-valued  point function $p$ we mean a map $p: D(p)\rightarrow
{{\cal M}}$, where the domain $D(p)$ is a countable subset of
$(0,\infty)$. Define the counting measure $N_p(dtdm)$ on the measure
space $((0,\infty)\times {{\cal M}}, \BB_{(0,\infty)}\times
\BB_{{\cal M}})$ by
$$
\npt := N_p((0,t]\times U)=\#\{s\in D(p)\,|\,s\in (0,t], p(s)\in U\},
\quad t>0,\; U\in \BB_{{\cal M}},
$$
where $\BB_{(0,\infty)}$ is the Borel $\s$-field on
$(0,\infty)$. Let $\Pi({{\cal M}})$ denote the set of all point
functions on ${{\cal M}}$, and $\BB_{\Pi({{\cal M}})}$ be the
smallest $\s$-field on $\Pi({{\cal M}})$ with respect to which
$p\longmapsto \npt$,  $t>0$, $U\in \BB_{{\cal M}}$, are measurable.
\begin{definition}
A $(\Pi({{\cal M}}), \BB_{\Pi({{\cal M}})})$-valued random
variable $p$ on $(\os,\fffs,P_{\rm P})$ is called an ${\cal M}$-valued
point process on $(S,\Sigma,P_{\rm P})$.
\end{definition}
The point process $p$ is called a {\it stationary point process} if
and only if $p(\cdot)$ and $p(s+\cdot)$ have the same law for all
$s\geq 0$, with $D(p(s+\cdot)) = \{t \in (0,\infty)\,|\, s+t\in
D(p)\}$.

\begin{definition}
An   ${{\cal M}}$-valued   point process $p$ on  $(\os, \fffs,
P_{\rm P})$ is called a \emph{Poisson point process} if and only if
the counting measure $N_p(dtdm)$ is a Poisson random measure on
$((0,\infty)\times {{\cal M}}, \BB_{(0,\infty)} \times \BB_{{\cal M}})$.
\end{definition}
 It is known that a Poisson
point process $p$ is stationary if and only if its intensity measure
is of the form
 \eq{nm}
 \EEEE[N_p(dtdm)]=dt n(dm)
 \en
for some measure $n$ on $({\cal M},\BB_{\cal M})$. An ${\cal M}$-valued
point process $p$  on $(\os, \fffs, P_{\rm P})$ is called
\emph{$(\fffs_t)$-adapted} if for every $t>0$ and $U\in \BB_{\cal M}$,
$\npt $ is $\fffs_t$ measurable for all $t>0$. It is called
\emph{$\s$-finite} if there exists $U_n\in \BB_{\cal M}$, $n=1,2,...$,
such that $U_n\uparrow{\cal M}$ and $\EEEE[N_p(t, U_n)]<\infty$, for
all $t>0$ and $n=1,2,...$ Let $p$ be a $(\fffs_t)$-adapted, $\s$-finite
point process. When $\EEEE[\npt]<\infty$, $\forall t>0$, there exists a
natural integrable increasing process $(\hat N_p(t,U))_{t\geq0}$ on
$(S,\Sigma, P_{\rm P})$ such that
$$
\npt -\hat N_p(t,U) := \tilde N_p(t,U)
$$
is a martingale. $\hat N_p(t,U)$ is called the {\it compensator} of
point process $p$.

\begin{definition}
An ${\cal M}$-valued  point process $p$ on $(S,\Sigma, P_{\rm P})$
is called a \emph{$(\fffs_t)$-Poisson point process} if it is an
$(\fffs_t)$-adapted, $\s$-finite Poisson point process such that the
increments
$$
\{N_p(t+h, U)-N_p(t, U): \; {h>0,\, U\in \BB_{\cal M}}\}
$$
are independent of $\fffs_t$.
\end{definition}
Let $p$ be a $(\fffs_t)$-Poisson point process. Then if $t\mapsto
\EEEE[\npt]$ is continuous,  it holds that $\hat N_p(t,U) =
\EEEE[\npt]$.  In particular, a stationary $(\fffs_t)$-Poisson point
process has the compensator $\hat N_p(t,U) =t n(U)$, where $n$ is
that of \kak{nm}, and for a disjoint family of $U_i$ in $\fffs$,
$i=1,...,N$,
$$
\EEEE\left[e^{-\sum_{i=1}^ N\alpha_i  N_p((s,t]\times U_i)}\right]=
\exp\lk (t-s)\sum_{i=1}^ N (e^{-\alpha_i }-1)n(U_i)\rk .
$$
We give an example.
\begin{example}
Poisson point processes can be constructed through $d$-dimensional L\'evy processes.
Let $(\eta_t)_{t\geq0}$ be an $\RR^d$-valued stationary L\'evy process on probability
space $(S,\Sigma, P)$ with the natural filtration $\Sigma_t=
\s(\eta_s,s\leq t)$. Define the jump process $p(s)=p(s,\tau)=
\eta_s(\tau)-\eta_{s-}(\tau)$ for each $\tau\in S$. Let $D(p)=
\{s\in (0,\infty)\,|\,p(s)\not=0\}$. Then $p:D(p)\rightarrow
\RR^d\setminus\{0\}$,  $s\mapsto p(s)$, is 
an $\RR^d\setminus\{0\}$-valued
$(\Sigma_t)$-Poisson point process and $P(N_p(t,U)=n)=
(\nu(U)t)^n e^{-\nu(U)t}/n!$ holds, where $\nu(U)$ is
the L\'evy measure given by $\nu(U)={\Bbb E}_{\rm P}[N_p(1,U)]$
for $U\in \BB_{\RR^d\setminus\{0\}}$. Moreover, its compensator is
$\hat N_p(t,U)=t\nu(U)$.
\end{example}
Fix  a stationary $(\fffs_t)$-Poisson point process $p$ on $(\os,
\fffs, P_{\rm P})$ with values in ${{\cal  M}}$. In Section~3 we set
$(\wo, \BB_\wo, P_\wo) := (W\times \os, \BB_W\times \fffs,
P_W^0\otimes P)$ and $\bn:=w\times \ttt\in W\times \os=\wo$. Let
$\calb$ be the smallest $\s$-field on $[0,\infty)\times{\cal M}
\times\wo$ such that all $g$ having the properties below are
measurable:
\begin{itemize}
\item[(1)]
for each $t>0$, $(m,\bn)\mapsto g(t,m,\bn)$ is $\BB_{\cal M}
\times \wo_t$ measurable,
\item[(2)]
for each $(m,\bn)$, $t\mapsto g(t,m,\bn)$ is left continuous.
\end{itemize}

 \bd{pred}
We call a $\calb $-measurable function $h: [0,\infty) \times{\cal M}
\times \wo\rightarrow \RR$ \emph{$(\wo_t)$-predictable} and denote
their set by $\wo_{\rm pred}$.
 \ed
Write
 \eqn &&
 {{\Bbb F}} := \left\{f\in \wo_{\rm pred} \,|\,
 \int_0^{t+}\int_{\cal M} |f(s,m,\bn)|N_p(dsdm) < \infty\
 \;\mbox{for $t>0$, a.e. $\bn$} \right\},\\
 &&
 {{\Bbb F}}^2 := \left\{f\in \wo_{\rm pred}\,|\,
 \EEp\left[ \int_0^{t} \int_{{\cal M}}|f(s,m,\bn)|^2\hat
 N_p(dsdm)\right] < \infty\ \, \mbox{for $t>0$}
 \right\}
 \enn
and
$$
{{\Bbb F}}^{\rm 2,loc} := \left\{f\in \wo_{\rm pred} \,|\, \exists
\, \tau_{n} \;\; (\wo_t)\!-\!{\rm stopping \ times}: \,
\tau_{n}\uparrow \infty\ \,\mbox{and}\, 1_{[0,\tau_n]}(t)
f(t,m,\bn)\in {{\Bbb F}}^2\right\}.
$$
Let $f^i(t,\bn)$ and $g^i(s,\bn)$
 be adapted with respect to $(\wo_t)$,
$\EEp[\int_0^t |f^i(s,\cdot)|^2 ds ]<\infty$ and $g^i(\cdot,\bn)\in
L_{\rm loc}^1(\RR)$ for a.e. $\bn\in \wo$. Furthermore, take $h^i_1
\in {{\Bbb F}}$ and $h^i_2\in{{\Bbb F}}^{\rm 2,loc}$. Define the
semi-martingale $X_t =(X^1_t,...,X^d_t)$ on $(\wo, \BB_\wo,
P_\wo)$  by
 \eqnn
 \label{12}
 \lefteqn{
 X^i_t =
 \int _0^t f^i(s,\bn) dB_s^i + \int _0^t g^i(s,\bn) ds } \\
 &&
 \hspace{0.8cm}\non
 + \int_0^{t+}\int_{{\cal M}} h_1^i(s,m,\bn) N_p(dsdm)+
 \int_0^{t+}\int_{{\cal M}} h_2^i(s,m,\bn)\tilde N_p(ds dm).
 \ennn
Here $\tilde N_p(dsdm)=N_p(dsdm)-dsn(dm)$.

 \bp{2}
Let $F\in C^2(\RR^d)$ and $X_t=(X_t^1,...,X_t^d)$ be given by
\kak{12}. Suppose $h_1^i\in{\Bbb F}$, $h_2^j\in{\Bbb F}^{2,loc}$,
and $ h_1^ih_2^j=0$ for $i,j=1,...,d$. Then $F(X_t)$ is a
semimartingale and the following It\^o formula holds:
 \eqn
 \lefteqn{
 dF(X_t)= \sum_{i=1}^d \muu
 \int_0^t\frac{\partial F(X_s)}{\partial x_i}
 f_\mu ^i(s,\bn) dB_s^\mu} \\
 &&
 + \sum_{i=1}^d \int_0^t\frac{\partial F(X_s)}{\partial x_i}
 g^i(s,\bn) ds +\half \sum_{i,j=1}^d \int_0^t\frac{\partial^2 F(X_s)}
 {\partial x_i\partial x_j }f^i(s,\bn)f^j(s,\bn) ds\\
 &&
 + \int_0^{t+}\int_{{\cal M}}\left(F(X_{s-} + h_1(s,m,\bn))-
 F(X_{s-})\right) N_p(dsdm)\\
 &&
 +\int_0^{t+}\int_{{\cal M}}\left( F(X_{s-}+h_2(s,m,\bn))- F(X_{s-})
 \right) \tilde N_p(dsdm)\\
 &&
 +\int_0^{t}\int_{{\cal M}}\left( F(X_s+h_2(s,m,\bn))-F(X_s)-
 \sum_{i=1}^d h_2^i(s,m,\bn)\frac{\partial F(X_s)}{\partial x_i}
 \right)\hat N_p(dsdm),
 \enn
where $\hat N_p(dsdm) = dsn(dm)$.
 \ep
 \proof
 See, e.g., \cite[Theorem 5.1]{iw}.
 \qed
Write \kak{12} as $dX^i = f^i dB^i + g^i dt + \int_{\cal M}h_1^i dN
+ \int_{\cal M} h_2^i d\tilde N$ in concise notation.
 Let $d=1$,
  $B_{t}^1=B_t$ and
 \eqn
 &&
 dZ=u_Zdt+v_ZdB+\int_{\cal M} f_Z dN+\int_X g_Zd\tilde N,\\
 &&
 dY=u_Ydt+v_YdB+\int_{\cal M} f_Y dN+\int_X g_Yd\tilde N
 \enn
with $f_Zg_Z=0$, $f_Z g_Y=0$, $f_Yg_Y=0$ and $f_Y g_Z=0$. Then by
Proposition \ref{2} we have the product rule
 \eqn
 &&
 d(ZY) = Z_s u_Yds+Z_s  v_Y dB_s + \int_{{\cal M}} Z_{s -}
 f_Y
 N_p(dsdm) + \int_{{\cal M}} Z_{s-} g_Y \tilde N_p(dsdm) \\
 &&\hspace{1cm}
 + Y_s u_Zds+ Y(s) v_Z dB_s+ \int_{{\cal M}} Y_{s-}
 f_Z  N_p(dsdm) +
 \int_{{\cal M}} Y(s-) g_Z \tilde N_p(dsdm) \\
 &&\hspace{1cm}
 + v_Zv_Y ds+ \int_{{\cal M}} (f_Zf_Y+g_Zg_Y) N_p(dsdm).
 \enn
This formula is written as $d(ZY)=dZ\cdot Y +Z\cdot dY +dZ\cdot dY$
in the concise notation.

Suppose $n({{\cal M}})=1$ and set $N_t := N_p((0,t]\times {{\cal
M}})$ and $dN_t := \int_{{\cal M}} N_p(dtdm)$ as mentioned in
Section 3.2. Then the  compensator of $p$ is given by $\hat N_p
(t,{\cal M}) = t$ and $ \EEp[e^{-\alpha N_t}]=e^{t(e^{-\alpha}-1)}$.
Moreover,
$$
\EEp\left[\int_0^{t+}\int_{{\cal M}} f(s,\bn, m)N_p(dsdm)\right]=
\EEp\left[ \int_0^t\int_{{\cal M}} f(s,\bn,m)dsn(dm)\right].
$$
Hence we have for $f=f(s,\bn)$ independent of $m\in {{\cal M}}$,
  \eq{8}
  \EEp\left[\int_0^{t+} f(s,\bn)dN_s\right]
  = \EEp\left[\int_0^tf(s,\bn)ds\right].
  \en
Furthermore, Proposition \ref{2} gives
 \bp{40}
Suppose $h^i\in{\Bbb F}$, $i=1,...,d$, are independent of $m\in
{\cal M}$.
 Let $dX^i=f_\mu^i dB^\mu +g^i dt+ h^i dN$, $i=1,...,d$,  and $F\in
C^2(\RR^d)$. Then
 \eqn
 dF(X_t)
 &=&
 \sum_{i=1}^ d \muu \int_0^t\frac{\partial F(X_s)}
 {\partial x_i}
 f_\mu^i (s,\bn) dB_s^\mu \\
 &&
 + \sum_{i=1}^d \int_0^t\frac{\partial F(X_s)}{\partial x_i}
 g^i(s,\bn) ds +\half \sum_{i,j=1}^d \int_0^t\frac{\partial^2 F(X_s)}
 {\partial x^2_i\partial x_j} f^i(s,\bn) f^j(s,\bn) ds\\
 &&
 +\int_0^{t+}\left(F(X_{s-}+h(s,\bn))-F(X_{s-})\right) dN_s.
 \enn
 \ep

\bigskip
\noindent {\bf Acknowledgments:} We thank V. Betz, M. Gubinelli and
I. Sasaki for useful discussions. This work was partially done at
Warwick University, Coventry, and at Erwin Schr\"odinger Institute,
Vienna, both of whom we thank for kind hospitality. J.L. is grateful
to Kyushu University for a travel grant and warm hospitality. This
work is financially supported by Grant-in-Aid for Science Research
(C) 17540181 from JSPS.


\bigskip\bigskip

{\footnotesize
\begin{thebibliography}{99}
\bibitem[AGG04]{agg}
L. Amour, B. Grebert and  J.C. Guillot, The dressed nonrelativistic
electron in a magnetic field, {\it Math. Methods Appl. Sci.} {\bf 29}
(2006), 1121-1146

\bibitem[ALS83]{als}
G. F. De Angelis, G.J. Lasinio and M. Sirugue, Probabilistic
solution of Pauli type equations, {\it J. Phys. A: Math. Gen.}
{\bf 16} (1983), 2433--2444.

\bibitem[Ara83a]{ar4}
A.  Arai, Rigorous theory of spectra and radiation for a model in
quantum electrodynamics, {\it  J.  Math.  Phys.  }{\bf 24}  (1983),
1896--1910.

\bibitem[AHS78]{ahs}
J. Avron, I. Herbst and B. Simon, Schr\"odinger operators with
magnetic fields. I. General interactions, {\it Duke Math. J.} {\bf
45} (1978), 847--883.

\bibitem[BFS98a]{bfs2}
V.  Bach,   J.  Fr\"ohlich and   I.  M.  Sigal, Renormalization
group analysis of spectral problems in quantum field theory, {\it
Adv. Math.} {\bf 137}  (1998),   205--298.

\bibitem[BFS98b]{bfs1}
V.  Bach,   J.  Fr\"ohlich and    I.  M.  Sigal, Quantum
electrodynamics of confined non-relativistic particles, {\it  Adv.
Math. } {\bf 137}  (1998),   299--395.

\bibitem[BFS99]{bfs}
V. Bach,  J. Fr\"ohlich and I.M.  Sigal, Spectral analysis for
systems of atoms and molecules coupled to the quantized radiation
field, {\it Commun. Math. Phys.} {\bf 207}  (1999), 249--290.

\bibitem[BFP05]{bfp}
V. Bach, J. Fr\"ohlich and  A. Pizzo, Infrared-finite algorithms in
QED: I.~The groundstate of an atom interacting with the quantized
radiation field, {\it Commun. Math. Phys.} {\bf 264} (2006),
145--165.

\bibitem[BCFS06]{bcfs}
V.  Bach, T. Chen, J. Fr\"ohlich and I.M.  Sigal, The  renormalized
electron mass in non-relativistic quantum electrodynamics, {\it J.
Funct. Anal.} {\bf 243} (2007), 426--535.

\bibitem[BDG04]{bdg} J. M. Barbaroux, M. Dimassi and J.  C.
Guillot, Quantum electrodynamics of relativistic bound states with
cutoffs, {\it J. Hyperbolic Diff. Eq.} {\bf 1} (2004), 271--314.

\bibitem[BH07]{bh}
V. Betz and F. Hiroshima, Measures with double stochastic integrals
on a path space, preprint 2007.

\bibitem[BHLMS02]{bhlms}
V. Betz, F. Hiroshima, J. L\H{o}rinczi,   R.A. Minlos and H. Spohn,
Ground state properties of the Nelson Hamiltonian --- a Gibbs
measure-based aproach, {\it Rev. Math. Phys.},  {\bf 14} (2002),
173--198.

\bibitem[BS05]{bs}
V. Betz and H. Spohn, A central limit theorem for Gibbs measures
relative to Brownian motion, {\it Prob. Theory Rel. Fields}, {\bf
131} (2005),  459-478.

\bibitem[CH04]{caha}
I. Catto and C. Hainzl, Self-energy of one electron in
non-relativistic QED, {\it J. Funct. Anal.} {\bf 207} (2004),
68--110.

\bibitem[Che01]{th}
T. Chen, Operator-theoretic infrared renormalization and
construction of dressed 1-particle states in non-relativistic QED,
ETH Dissertation 2000.

\bibitem[Che06]{th2}
T. Chen, Infrared renormalization in non-relativistic QED for the
endpoint case, arXiv:math-ph/0601010v2 (2006).

\bibitem[DV07]{dv}
D.J. Daley and D. Vere-Jones, \emph{An Introduction to the Theory of
Point Processes}, vols. 1-2, Springer, 2002, 2007

\bibitem[Fro74]{fr}
J.  Fr\"ohlich,   Existence of dressed one electron states in a
class of persistent models,   {\it Fortschritte der Physik} {\bf 22}
(1974),   159--198.

\bibitem[Fef96]{fe}
C. Fefferman, On the electrons and nuclei in magnetic field, {\it
Adv. Math.} {\bf 124} (1996), 100--153.

\bibitem[FFG97]{ffg}
C.  Fefferman,  J.  Fr\"ohlich and  G. M. Graf, Stability of
ultraviolet-cutoff quantum electrodynamics with non-relativistic
matter,  {\it Commun.   Math.   Phys.  } {\bf 190} (1997), 309--330.

\bibitem[FGS01]{fgs1}
J.  Fr\"ohlich,  M.  Griesemer and B.  Schlein,  Asymptotic
electromagnetic fields in a mode of quantum-mechanical matter
interacting with the quantum radiation field, {\it Adv. Math.} {\bf
164} (2001), 349--398.

\bibitem[Ger00]{ge}
C.  G\'erard, On the existence of ground states for massless
Pauli-Fierz Hamiltonians, {\it  Ann.  Henri  Poincar\'e} {\bf 1}
(2000), 443--459, {\it and} A remark on the paper: ``On the existence
of ground states for Hamiltonians", mp-arc 06-146 (2006).

\bibitem[GJ68]{gj1}
J. Glimm and A. Jaffe, The $\lambda (\phi^4)_2$ quantum field theory
without cutoffs I, {\it Phys. Rev.} {\bf  176}  (1968), 1945--1951.

\bibitem[GLL01]{gll}
M.  Griesemer,   E.  Lieb and M.  Loss,   Ground states in
non-relativistic quantum electrodynamics, {\it Invent. Math.}
{\bf 145} (2001), 557--595.

\bibitem[Gro72]{gr}
L.  Gross,   Existence and uniqueness of physical ground states,
{\it J. Funct. Anal.} {\bf 10} (1972), 52--109.

\bibitem[GL07a]{GL07a}
M. Gubinelli and J. L\H{o}rinczi, Gibbs measures on Brownian
currents, to appear in \emph{Commun. Pure Appl. Math.} (2008)

\bibitem[GL07b]{GL07b}
M. Gubinelli and J. L\H{o}rinczi: Ultraviolet renormalization of
Nelson's model through functional integration, preprint (2007)

\bibitem[Hab98]{ha}
Z.  Haba,   Feynman integral in regularized nonrelativistic quantum
electrodynamics,  {\it J. Math. Phys.  } {\bf 39} (1998), 1766--1787

\bibitem[HaHe06]{hh}
D. Hassler and I. Herbst, Absence of ground states for a class of
translation invariant models in nonrelativistic QED, preprint (2006)

\bibitem[Hik99]{hi1}
M.  Hirokawa, An expression of the ground state energy of the
Spin-Boson model, {\it J.   Funct.   Anal.  } {\bf 162}  (1999),
178--218.

\bibitem[Hik01]{hi2}
M. Hirokawa, Remarks on the ground state energy of the spin-boson
model: an application of the Wirner-Weisskopf model, {\it Rev. Math.
Phys.} {\bf 13}  (2001),  221-251.

\bibitem[HH07] {hihi}
M. Hirokawa and F. Hiroshima, Poisson point process and spin-boson
models, preprint (2007)

\bibitem[Hir97]{h4}
F. Hiroshima, Functional integral representations of quantum
electrodynamics, {\it Rev. Math. Phys.} {\bf 9}  (1997), 489--530.

\bibitem[Hir00a]{h10}
F.  Hiroshima, Ground states of a model in nonrelativistic quantum
electrodynamics II, {\it J.    Math.   Phys.  }  {\bf 41}   (2000),
661--674.

\bibitem[Hir00b]{h12}
F.  Hiroshima, Essential self-adjointness of translation-invariant
quantum field models for arbitrary coupling constants, {\it Commun.
Math. Phys.} {\bf 211}   (2000),   585--613.

\bibitem[Hir02]{h19}
F.  Hiroshima, Self-adjointness of the Pauli-Fierz Hamiltonian for
arbitrary values of coupling constants, {\it Ann. Henri  Poincar\'e}
{\bf 3} (2002), 171--201.

\bibitem[Hir06]{h27}
F. Hiroshima, {Multiplicity of ground states  in quantum field
models: applications of asymptotic fields}, {\it J. Funct. Anal.}
{\bf 224} (2005), 431--470.

\bibitem[Hir07]{h26}
F. Hiroshima, Fiber Hamiltonians in nonrelativistic quantum
electrodynamics, {\it J. Funct. Anal.} {\bf 252} (2007), 314--355.


\bibitem[HL07]{hilo}
F. Hiroshima and J. L\H{o}rinczi, Localization of the ground state
in the Pauli-Fierz model at weak coupling, in preparation.

\bibitem[HI07]{hk}
F. Hiroshima and K. R. Ito, Mass renormalization in non-relativistic
quantum electrodynamics with spin 1/2, {\it Rev. Math. Phys.} {\bf
19} (2007), 405--454.

\bibitem[HS01]
{hisp2} F.  Hiroshima and H.  Spohn, Ground state degeneracy of the
Pauli-Fierz model with   spin, {\it Adv. Theor. Math. Phys.} {\bf 5}
(2001), 1091--1104.

\bibitem[HS05]{hisp3}
F.  Hiroshima and H.  Spohn, Mass renormalization in nonrelativistic
QED, {\it J. Math. Phys.} {\bf 46} (2005), 042302-42328.

\bibitem[IW81]{iw}
N. Ikeda and S. Watanabe, {\it Stochastic Differential Equations and
Diffusion Processes}, North-Holland/Kodansha, 1981.

\bibitem[KM78]{km}
T. Kato and K. Masuda, Trotter's product formula for nonlinear
semigroups generated by the subdifferentials of convex functionals,
{\it J.  Math. Soc. Japan} {\bf 30} (1978), 169-178.

\bibitem[LL00]{lilo2}
E. Lieb and M. Loss, Self-energy of electrons in non-perturbative
QED, In: {\it Differential Equations and Mathematical Physics}, eds.
R. Weikard and G. Weinstein, Cambridge, MA, AMS, 2000, 279--293.

\bibitem[LL03]{lilo} E.   Lieb and M.   Loss,   Existence of
atoms and molecules in non-relativistic quantum electrodynamics,
{\it Adv. Theor. Math. Phys. }{\bf 7} (2003), 667--710.

\bibitem[LM01]{LM01}
J.  L\H{o}rinczi and  R.A. Minlos: Gibbs measures for Brownian paths
under the effect of an external and a small pair potential, {\it J.
Stat. Phys.} {\bf 105} (2001), 605-647

\bibitem[LMS02a]{LMS02a}
J.  L\H{o}rinczi,   R.A.  Minlos and H.  Spohn, The infrared
behaviour in Nelson's model of a quantum particle coupled to a
massless scalar field, {\it  Ann. Henri  Poincar\'e} {\bf 3} (2002),
1--28.

\bibitem[LMS02b]{LMS02b}
J.  L\H{o}rinczi,   R.   A.  Minlos and H.  Spohn, Infrared regular
representation of the three dimensional massless Nelson model, {\it
Lett. Math. Phys.} {\bf 59} (2002), 189--198.

\bibitem[LMS06]{lms}
M. Loss, T. Miyao and H. Spohn, Lowest energy states in
nonrelativistic QED: Atoms and ions in motion, {\it  J. Funct.
Anal.} {\bf 243} (2006), 353-393.

\bibitem[Nel73]{ne}
E. Nelson, The free Markoff field, {\it J. Funct. Anal.} {\bf 12}
(1973), 211-227.

\bibitem[RS75]{rs2}
M. Reed and B. Simon, {\it Methods of Mathematical Physics II},
Academic Press, 1975.

\bibitem[Sas06]{sa}I.
Sasaki, Ground state of a model in the relativistic quantum
electrodynamics with a fixed total momentum, mp-arc 05-433 (2005).

\bibitem[Sim74]{si}
B. Simon, {\it The $P(\phi)_2$ Euclidean Quantum Field Theory},
Princeton University Press, 1974.


\bibitem[Spo87]{sp5}
H.  Spohn, Effective mass of the polaron: A functional integral
approach, {\it Ann.   Phys.  } {\bf 175}  (1987),   278--318.

\bibitem[Spo89]{sp3}
H.  Spohn, Ground state(s) of the spin-boson Hamiltonian,   {\it
Commun.   Math.   Phys.  } {\bf 123}    (1989),   277--304.

\bibitem[Spo97]{sp97}
H.  Spohn,  Asymptotic completeness for Rayleigh scattering, {\it J.
Math. Phys.} {\bf 38} (1997), 2281--2296.


\bibitem[Spo98]{sp98}
H.  Spohn, Ground state of quantum particle coupled to a scalar
boson field, {\it  Lett.   Math.   Phys.  } {\bf 44}   (1998),
9--16.

\bibitem[Spo04]{sp04}
H.   Spohn,   {\it Dynamics of Charged Particles and Their Radiation
Field}, Cambridge University Press, 2004.

\end{thebibliography}
}
\end{document}